\documentclass{aastex631}
\usepackage{savesym}
\savesymbol{corresponds}

\usepackage{amsmath,mathabx}%

\usepackage{graphicx}
\usepackage{subfigure}
\usepackage{diagbox,multirow}

\newcommand{\beq}{\begin{equation}}
\newcommand{\eeq}{\end{equation}}
\newcommand{\beqs}{\begin{eqnarray}}
\newcommand{\eeqs}{\end{eqnarray}}

\bibpunct{(}{)}{;}{a}{}{,}

\begin{document}

\title{Forecast of foreground cleaning strategies for AliCPT-1}

\correspondingauthor{Le Zhang, Zhi-Qi Huang, Jiazheng Dou, Hong Li}
\email{zhangle7@mail.sysu.edu.cn, huangzhq25@mail.sysu.edu.cn, doujzh@mail.ustc.edu.cn, hongli@ihep.ac.cn}

\author{Junzhou Zhang}
\affiliation{School of Physics and Astronomy, Sun Yat-sen University, 2 Daxue Road, Tangjia, Zhuhai, 519082, People's Republic of China}
\affiliation{CSST Science Center for the Guangdong-Hong Kong-Macau Greater Bay Area, Zhuhai 519082, People's Republic of China}

\author{Shamik Ghosh}
\affiliation{CAS Key Laboratory for Researches in Galaxies and Cosmology, Department of Astronomy, University of Science and Technology of China, Chinese Academy of Sciences, Hefei, Anhui 230026, People's Republic of China}
\affiliation{School of Astronomy and Space Science, University of Science and Technology of China, Hefei 230026, People's Republic of China}

\author{Jiazheng Dou}
\affiliation{CAS Key Laboratory for Researches in Galaxies and Cosmology, Department of Astronomy, University of Science and Technology of China, Chinese Academy of Sciences, Hefei, Anhui 230026, People's Republic of China}
\affiliation{School of Astronomy and Space Science, University of Science and Technology of China, Hefei 230026, People's Republic of China}

\author{Yang Liu}
\affiliation{Key Laboratory of Particle Astrophysics, Institute of High Energy Physics, Chinese Academy of Sciences, 19B Yuquan Road, Beijing 100049, People’s Republic of China}

\author{Siyu Li}
\affiliation{Key Laboratory of Particle Astrophysics, Institute of High Energy Physics, Chinese Academy of Sciences, 19B Yuquan Road, Beijing 100049, People’s Republic of China}

\author{Jiming Chen}
\affiliation{CAS Key Laboratory for Researches in Galaxies and Cosmology, Department of Astronomy, University of Science and Technology of China, Chinese Academy of Sciences, Hefei, Anhui 230026, People's Republic of China}
\affiliation{School of Astronomy and Space Science, University of Science and Technology of China, Hefei 230026, People's Republic of China}

\author{Jiaxin Wang}
\affiliation{Department of Astronomy, School of Physics and Astronomy, Shanghai Jiao Tong University, Shanghai, 200240, People's Republic of China}

\author{Zhaoxuan Zhang}
\affiliation{School of Physics and Astronomy, Sun Yat-sen University, 2 Daxue Road, Tangjia, Zhuhai, 519082, People's Republic of China}
\affiliation{CSST Science Center for the Guangdong-Hong Kong-Macau Greater Bay Area, Zhuhai 519082, People's Republic of China}

\author{Jacques Delabrouille}
\affiliation{CNRS-UCB International Research Laboratory, Centre Pierre Bin\'etruy, IRL2007, CPB-IN2P3, Berkeley, USA}

\author{Mathieu Remazeilles}
\affiliation{Instituto de F\'isica de Cantabria (CSIC-UC), Avenida de los Castros s/n, 39005 Santander, Spain}

\author{Chang Feng}
\affiliation{CAS Key Laboratory for Researches in Galaxies and Cosmology, Department of Astronomy, University of Science and Technology of China, Chinese Academy of Sciences, Hefei, Anhui 230026, People's Republic of China}
\affiliation{School of Astronomy and Space Science, University of Science and Technology of China, Hefei 230026, People's Republic of China}

\author{Bin Hu}
\affiliation{Department of Astronomy, Beijing Normal University, Beijing 100875, People's Republic of China}

\author{Hao Liu}
\affiliation{School of Physics and optoelectronics engineering, Anhui University, 111 Jiulong Road, Hefei, 230601, People's Republic of China}
\affiliation{Key Laboratory of Particle Astrophysics, Institute of High Energy Physics, Chinese Academy of Sciences, 19B Yuquan Road, Beijing 100049, People’s Republic of China}

\author{Larissa Santos}
\affiliation{Center for Gravitation and Cosmology, College of Physical Science and Technology, Yangzhou University, Yangzhou, 225009, People's Republic of China}

\author{Pengjie Zhang}
\affiliation{Department of Astronomy, School of Physics and Astronomy, Shanghai Jiao Tong University, Shanghai, 200240, People's Republic of China}
\affiliation{Division of Astronomy and Astrophysics, Tsung-Dao Lee Institute, Shanghai Jiao Tong University, Shanghai, 200240, People's Republic of China}

\affiliation{Shanghai Key Laboratory for Particle Physics and Cosmology, Shanghai, 200240, People's Republic of China}

\author{Wen Zhao}
\affiliation{CAS Key Laboratory for Researches in Galaxies and Cosmology, Department of Astronomy, University of Science and Technology of China, Chinese Academy of Sciences, Hefei, Anhui 230026, People's Republic of China}
\affiliation{School of Astronomy and Space Science, University of Science and Technology of China, Hefei 230026, People's Republic of China}

\author{Le Zhang}
\affiliation{School of Physics and Astronomy, Sun Yat-sen University, 2 Daxue Road, Tangjia, Zhuhai, 519082, People's Republic of China}
\affiliation{CSST Science Center for the Guangdong-Hong Kong-Macau Greater Bay Area, Zhuhai 519082, People's Republic of China}
\affiliation{Peng Cheng Laboratory, No.2, Xingke 1st Street, Shenzhen 518000, People’s Republic of China}

\author{Zhi-Qi Huang}
\affiliation{School of Physics and Astronomy, Sun Yat-sen University, 2 Daxue Road, Tangjia, Zhuhai, 519082, People's Republic of China}
\affiliation{CSST Science Center for the Guangdong-Hong Kong-Macau Greater Bay Area, Zhuhai 519082, People's Republic of China}

\author{Hong Li}
\affiliation{Key Laboratory of Particle Astrophysics, Institute of High Energy Physics, Chinese Academy of Sciences, 19B Yuquan Road, Beijing 100049, People’s Republic of China}

\author{Chao-Lin Kuo}
\affiliation{Stanford University, Stanford, CA 94305, USA}
\affiliation{Kavli Institute for Particle Astrophysics and Cosmology, Stanford, CA 94305, USA}

\author{Xinmin Zhang}
\affiliation{Theoretical devision, Institute of High Energy Physics, Chinese Academy of Sciences, Beijing 100049, People’s Republic of China}

\begin{abstract}

We report the test results of several independent foreground-cleaning pipelines used in the Ali CMB Polarization Telescope experiment (AliCPT-1), a high-altitude CMB imager in the Northern hemisphere with thousands of detectors dedicated to the search for a primordial CMB polarization $B$-mode signature. Based on simulated data from 4 detector modules and a single season of observation, which we refer to as Data Challenge 1 (DC1), we employ different and independent pipelines to examine the robustness and effectiveness of the estimates on foreground parameters and the primordial $B$-mode detection. 
The foreground-cleaning strategies used in the pipelines include the parametric method of template fitting (TF) and the non-parametric methods of the constrained internal linear combination (cILC), the analytical blind separation (ABS), and the generalized least squares (GLS). We examine the impact of possible foreground residuals on the estimate of the CMB tensor-to-scalar ratio ($r$) for each pipeline by changing the contamination components in the simulated maps and varying the foreground models and sky patches for various tests. According to the DC1 data with the simulation input value $r_{\rm true}=0.023$, the foreground residual contamination levels in the TF/ABS/cILC/GLS pipelines are well within the corresponding statistical errors at the $2\sigma$ level. Furthermore, by utilizing the tension estimator, which helps identify significant residual foreground contamination in the detection of the primordial $B$-mode signal by quantifying the discrepancy between various $r$ measurements, we conclude that the presence of small foreground residuals does not lead to any significant inconsistency in the estimation of $r$.

\end{abstract}

\section{Introduction} \label{sec:intro}

Measurements of the CMB $B$-mode polarization signal are thriving, with many space- and ground-based experiments either already observing (e.g., Planck~\citep{Planck:2018jri}, POLARBEAR~\citep{POLARBEAR:2019kzz}, SPTPol~\citep{SPT:2017ddy,SPT:2019nip}, BICEP 2 and 3~\citep{BICEP2:2018kqh,BICEP3:2016pqy}, Simons Array~\citep{Stebor:2016hgt}, ACT~\citep{Henderson:2015nzj,ACT:2020frw}, SPT-3G~\citep{SPT-3G:2014dbx}, SPIDER~\citep{SPIDER:2021ncy}) or preparing to do so (e.g., QUBIC~\citep{QUBIC:2020kvy}, Simons Observatory~\citep{SimonsObservatory:2018koc}, BICEP Array~\citep{Hui:2018cvg}, CMB-S4~\citep{CMB-S4:2016ple}, AliCPT~\citep{Li:2017drr}). There are also proposed space projects, such as CMBpol~\citep{CMBPolStudyTeam:2008rgp}, LiteBIRD~\citep{Matsumura:2013aja}, PIXIE~\citep{Kogut:2011xw}, PRISM~\citep{PRISM:2013fvg}, CORE~\citep{2018JCAP...04..014D} and PICO~\citep{NASAPICO:2019thw}. Although their scientific objectives are generally broad, the detection of B-mode polarization patterns on large angular scales is particularly compelling, as it can reveal horizon-scale primordial gravitational waves~\citep{Zaldarriaga:1996xe,Kamionkowski:1996zd,Hu:2001bc} and provide new insights into the inflationary epoch of our universe. 

To successfully measure the small primordial $B$-mode signal at the level of the tensor-to-scalar ratio of about $r \sim  0.01$ or better, one must have extraordinary instrument sensitivity and precise control of systematics. Besides this, whether foreground contamination can be accurately subtracted is another important limiting factor in $B$-mode detection. Observationally, the Planck full-sky high-sensitivity measurements across broad frequencies~\citep{Planck:2015mis,Planck:2018yye}, have revealed that diffuse Galactic synchrotron and thermal dust emissions are the two dominant foreground components, which mainly dominate the polarized sky at frequencies of $\nu<70$ GHz and $\gtrsim 70$ GHz, respectively, with exact values depending on angular scale and sky coverage~\citep{Planck:2014dmk}. Even in the cleanest sky patches, thermal dust emission is  stronger than the primordial CMB $B$-mode signal at frequencies of 95 and 150 GHz~\citep{BICEP2:2015nss}, the two most suitable microwave atmospheric windows for ground-based CMB observations. Other possible components such as spinning dust~\citep{Planck:2011ag} and magnetic dust~\citep{Draine:2012wr,Planck:2014skm,Hoang:2015tfa} may also have non-negligible contributions in comparison with the tiny primordial $B$-modes. This could increase the complexity of the polarized sky, even at high galactic latitudes, whereas our current knowledge of these contaminants is limited.
Thus, a good component separation algorithm  plays a key role in inflationary $B$-mode detection.

The Ali CMB Polarization Telescope (AliCPT) is a ground-based CMB experiment, whose primary objective is to precisely determine the primordial gravitational wave signal in the Northern sky~\citep{Li:2017drr,Li:2018rwc}. The AliCPT-1 telescope is a medium-sized telescope designed to make dual-band microwave measurements at frequencies of 95 and 150 GHz.  With an aperture of 72 cm and a focal plane width of 63.6 cm, it is capable of operating a receiver equipped with up to 19 detector modules, totaling 32,376 transition edge sensors (TES). Currently, this receiver is undergoing cooling tests at Stanford University~\citep{9380315,10.1117/12.2560709}. In its initial phase, AliCPT-1 will observe a 4,000 square degree ``deep patch'' of sky with a low level of dust foreground contamination in the northern hemisphere, for a period of one year, using 4 modules. To prepare for the AliCPT-1 observations, the pipeline team has conducted end-to-end simulations for the entire observing season. The sky coverage exceeding 10\% results from several factors: 1) our telescope's sensitivity is enhanced by numerous detectors and a wide field of view, enabling nearly 1\% sky coverage in a single exposure; 2) we focus on detecting the primordial $B$-mode signal, which appears at intermediate scales. However, atmospheric low-frequency perturbations can distort our observations. To counter this, we employ scanning techniques and polynomial filtering to subtract atmospheric radiation, thus expanding sky coverage; 3) mechanical limitations require the telescope to perform azimuth scanning at a fixed elevation angle, balancing effective scanning distance with minimal atmospheric interference; 4) to optimize  error checks and reduce scanning patterns, we limit the number of scanning modes, resulting in a relatively fixed elevation angle; 4) the telescope's location in a mid-latitude region (around $20^\circ$ north) means that Earth's rotation further broadens the scanning area; 5)
although the Planck satellite offers the best scan for northern sky foreground contamination, significant polarization errors remain. Hence, we conduct preliminary screenings across large areas.

This paper is part of a series of publications describing the AliCPT-1 experiment. The overview of the scientific objectives and expected performance is described in Paper I~\citep[in preparation]{paper1}, the simulation pipeline in Paper II~\citep[in preparation]{paper2}, the performance of the foreground-cleaning pipelines in constraining the tensor-to-scalar ratio $r$ in Paper IV~\citep{paper4}, the forecasts on CMB lensing
observations in Paper V~\citep{paper5}, a further step in 
characterizing foreground emissions in Paper VI~\citep[in preparation]{paper6}, and the instrumental design is described in~\citep{9380315,10.1117/12.2560709}. The main objective of this paper is to describe in detail the pipeline framework of foreground cleaning strategies in the AliCPT-1 and to report results of various adopted source separation methods for signal estimation based on current simulated sky maps, as well as to discuss impacts of possible foreground residuals on primordial $B$-mode detection. The Planck Sky Model (PSM)~\citep{PSM:2013} is used to simulate the various microwave foreground components, such as synchrotron, thermal dust, free-free, spinning dust, CO lines, and emissions from extragalactic sources, and a dedicated pipeline generates mock observations that account for instrumental effects and the the filtering of the time-ordered data (TOD). In this way, the simulated intensity and polarized sky maps are expected to realistically reflect the observations. These map sets were utilized to validate the data analysis pipelines, which we refer to as Data Challenge 1 (DC1). The primordial $BB$ amplitude is set to $r_{\rm true} = 0.023$ in our simulations, which is compatible with the latest constraint ($r_{0.05}<0.036$ at 95\% confidence) from the latest joint analysis of  Planck, WMAP, and BICEP/Keck Observations~\citep{BICEP:2021xfz}.
The lensing $B$-modes, which would contribute a small fraction of the total noise in the observed maps, are not included in the current DC1 CMB simulations. Including them would not substantially alter our conclusions on the foreground cleaning strategies. 

Simulated observations are passed through foreground cleaning pipelines to quantitatively evaluate foreground residuals and $r$ constraints with blind, non-blind, or semi-blind source separation methods. Specifically, we use the analytical blind separation (ABS) method~\citep{2016arXiv160803707Z,2018ApJS..239...36Y,2021A&A...650A..65S}, a generalized least squares (GLS) inversion approach assuming a prior for synchrotron and dust emission laws~\citep{Delabrouille2009}, a constrained ILC~\citep{Remazeilles2010} and template fitting (TF)~\citep{paper4} on reconstructing the foreground and CMB signals.

The paper is organized as follows. In Sect.~\ref{sec:mock}, we briefly describe the mock data of the AliCPT-1 and in Sect.~\ref{sec:comsep}, we outline the foreground-cleaning methods and each analysis pipeline used in this study. Foreground residual levels and their impact of residuals on the $B$-mode detection are quantified in Sect.~\ref{sec:fgre}. Finally, we summarize and conclude in Sect.~\ref{sec:con}.

\section{Mock data sets}\label{sec:mock}
We provide here a brief summary of the DC1 mock data used for validating the foreground removal pipelines of AliCPT-1. Additional details can be found in Paper I \& II~\citep[in preparation]{paper1, paper2}.  

We observe a sky patch in the Northern sky, centered at RA $=180^\circ$ and DEC$=30^\circ$, covering about $17\%$ of the sky. The observing season is from October 1 to March 31. We use simulated meteorological data at the Ali site based on the MERRA2 data from 2011 to 2017~\citep{MERRA2} and the desired instrumental parameters of AliCPT-1 to simulate the noise in the time-ordered data (TOD) for each detector as realistically as possible. Large-scale signals caused by atmospheric emissions and scan synchronous pickup are attenuated by applying two filters, a polynomial filter and a ground subtraction filter, in the time-domain signal simulations. The primary effects of the filtering are twofold: i) suppression of $EE$ and $BB$ power in observed sky maps on large angular scales, and ii) generation of $E$-to-$B$ leakage. In order to accurately estimate the filtering effects, we simulate the sky maps obtained with and without each filter separately, for a total of four cases.

Besides the data in the 95 GHz and 150 GHz bands of AliCPT-1, we also produce ``re-observed'' sky maps in the Planck HFI four bands at 100, 143, 217, 353 GHz and in the WMAP K band~\citep{2013ApJS..208...20B}. The CMB and foreground sky maps are simulated with the PSM, while the associated Planck noise maps are downloaded from the Planck Legacy Archive\footnote{http://www.cosmos.esa.int/web/planck/pla}. Additionally, the K-band noise maps with \texttt{NSIDE}=1024 are generated using the full noise covariance matrix provided with WMAP 9-year data. We coadd all sky components and noise to form sky observation maps in each band, and then re-observe them using the AliCPT scanning and map-making pipeline, with the same time-domain filtering as applied on AliCPT data. 
Two categories of DC1 data are generated: simulation data, and ancillary data, described next. 

{\bf Simulation data}: the simulation data consist of a set of observed sky maps for all seven frequency bands, including two AliCPT-1 bands (95 and 150 GHz), four Planck HFI bands (100, 143, 217, and 353 GHz), and the WMAP-K band. Each of them contains CMB, foregrounds, and noise. The tensor-to-scalar ratio $r$ was fixed to 0.023, and other standard $\Lambda$CDM cosmological parameters where drawn at random in agreement with the best-fit values in the Planck 2018 at $1\sigma$ level. Cosmological parameters were not disclosed a priori to team members participating in the DC1 data challenge. As mentioned earlier, throughout this work, the lensing $B$-modes in the simulation are set to zero, and the contribution of the lensing $B$-modes is not taken into account in the fitting procedure. In the foreground simulation, the Galactic diffuse components consist of synchrotron, free-free, thermal dust emission and anomalous dust emission, as well as the CO lines, with randomly-generated small scale fluctuations. Galactic and extra-galactic strong and faint point sources (radio and infrared) are also included in foreground maps. Noise maps for AliCPT-1 are obtained from the simulated noise TODs using the map-making procedure. The re-observed Planck HFI noise maps are generated by propagating the 87th FFP10 noise simulation through the re-observation pipeline, and the WMAP-K noise map is derived from a random noise realization according to its covariance matrix map.

 {\bf Ancillary data}: Ancillary data is used to estimate the filtering effects and the covariance matrix of the power spectra
 The Ancillary data
 include 1) 50 sets of CMB only maps of all bands, each comprised of both cases with and without filtering, where we set $r=0$ and the cosmological parameters are fixed to the best-fit values 
of the Planck 2018 result; 2) the same 50 sets of CMB only maps as above, but for 
$r=0.01$ for all bands; 3) 100 sets of noise maps for AliCPT-1 95 and 150 GHz bands, each with/without filtering, respectively, with thermal loads for noise simulations taking into account precipitable water vapor based on meterological data; 4) 50 sets of re-observed noise maps for Planck HFI bands, each with/without filtering, respectively, based on the FFP10 noise simulations of the Planck Legacy Archive; 5) 50 sets of re-observed noise maps of WMAP K band, each with /without filtering, based on the WMAP noise covariance; 6) white noise covariance matrix of AliCPT-1 95/150 GHz bands; 7) 2 sets of re-observed foreground realization maps for all bands, each consisting of both cases of with/without filtering, respectively.

The $1/f$ noise will likely be present in the actual data. However, we have confirmed that in the following scenario, such noise can be effectively subtracted, allowing us to solely focus on white noise in our simulations: a comparison is made between the operational mode of the telescope and that of BICEP, both of which feature two bolometers with perpendicular polarization orientations per focal plane pixel. The polarization signal is derived from the difference in their readings. ~\cite{2014ApJ...792...62B} (see Fig. 22) reveals that the pair difference TOD behaves as pink noise with a small spectral index, typically around 0.1--1 Hz, due to atmospheric signals being the primary source of non-polarized $1/f$ noise. Following BICEP's data processing approach, we apply a third-order polynomial filter to subtract large-scale correlations from the time-series data of the pair difference, effectively removing low-frequency components. A simulation, based on BICEP's measurements, assumes a pivot scale of 0.1 Hz and increases the white noise amplitude to the noise equivalent temperature (NET) $>400 ~\mu{\rm K} \sqrt{s}$ at 150 GHz. The filter subtracts a third-order polynomial for each half scan. Results show significant attenuation of modes below 0.1 Hz, while data above 1 Hz remains unaffected. This allows for the safe implementation of naive map-making, projecting data onto a pixelized map, albeit sacrificing uncertainty and ignoring inter-pixel correlations, inevitably losing the large-scale CMB signal. In order to quantify this loss, we calculate the suppression factor on the power spectrum, which allows us to correct for it. In order to characterize the beams of all detectors and pair-differences, we will employ far-field mapping by raster-scanning an artificial radio frequency (RF) source with light reflected by a comoving flat mirror mounted on the elevation stage.

\section{Foreground cleaning strategies}
\label{sec:comsep}
 Various approaches are applied to the multi-frequency CMB observations in order to separate the signal from the astrophysical foreground contamination. Some of the methods commonly used in the literature assume prior knowledge of the emission laws. These ``non-blind'' methods include the Wiener filtering \citep{Bunn:1994, Tegmark:1996} and the maximum entropy method \citep{Hobson:1998}. Other methods assume a parametric model for the emission laws, and estimate the values of these parameters. This is the case of Gibbs-sampling based approaches~\citep{Jewell:2004, Wandelt:2004,Larson:2007,Eriksen:2008} like  Commander~\citep{Eriksen:2004}, which implements a pixel-based parametric component separation and of ForeGround Buster~\citep{Stompor2009}, which employs an analytically derived likelihood function for the spectral parameters.
 
 

 A drawback of those methods is the fact that they rely on sometimes poorly known prior information about the foreground emission. In contrast, component separation methods that do not assume a functional form nor prior information for the emission laws as a function of frequency are referred to as ``blind'' approaches. Those methods typically use the statistical independence of the different components to learn the emission properties from the data themselves. Such blind methods include the spectral matching independent component analysis (SMICA)~\citep{Delabrouille:2003, Patanchon:2005, Betoule:2009}, SEVEM~\citep{2012MNRAS.420.2162F}, the correlated component analysis (CCA)~\citep{Bonaldi2006}, the independent component Analysis (ICA)~\citep{Baccigalupi:2004}, and variants of the internal linear combination method, such as NILC~\citep{Delabrouille:2009} and GNILC~\citep{Remazeilles:2011ze}, which have been successfully applied to both WMAP and Planck temperature and $E$-mode data~\citep{Tegmark:1996, Tegmark:2003, Bennett:2003, Eriksen:2004, Saha:2006, Basak:2012, Basak:2013}. 
 Other methods have been developed to forecast residual foreground contamination or to extend the above component separation methods, e.g. xForecast~\citep{Stompor:2016hhw,Errard:2015cxa}, BFoRe~\citep{Alonso:2016xft}, the Delta-map method accounting for spatially varying spectra to the first order in perturbation~\citep{Ichiki:2018glx}, as well as the constrained moment ILC ~\citep{Remazeilles:2020rqw}.

Each of the above methods has advantages and drawbacks, and their effectiveness depends on specifics of the data at hand (noise level, number of channels, etc.). However, high-confidence detection of CMB B-modes will be achieved when one can achieve a consistent result with blind and parametric methods. There have been many lessons learned from BICEP2~\citep{Flauger:2014qra,BICEP2:2014owc}.

Furthermore, the best method also depends on the target observation, so that there may not be a one-size-fits-all method. In the following, we will elaborate on foreground cleaning strategies and corresponding data analysis pipelines planned for AliCPT-1, which are independent to ensure their reliability and cross checking for the $r$ estimate.

\subsection{The TF method~\label{subsec:TF}}
The first method we consider in our pipeline toolbox is
a template-fitting method (TF) that performs a joint parametric fit based on a likelihood analysis in the spectral domain. 
TF employs a multi-component parametrization to characterize the band power expectation values at each frequency channels and $\ell$-bins as a combination of cosmological and foreground signals. We construct several templates for CMB and foreground components---for this very reason, we call it ``template-fitting''---to circumvent correction for filtering effects in their power spectra, which we will detail later in this section.  We also test foreground models with different parametrizations to verify the effectiveness of the estimate on the primordial $B$-mode signal and to understand the properties of foreground components.

Our CMB and foreground models consider four distinct components: one for CMB and three for foregrounds. These components include a primordial CMB $B$-mode, Galactic synchrotron and dust, and a spatially correlated component between them. Additionally, we account for the band power contributed by the noise component. Thus, the total contributions to the $BB$ auto- and cross-spectra between the $\nu_1$ and $\nu_2$ frequency maps at the $\ell$-th bin are modeled by summing over all components as

\begin{equation}
D_{\nu_1 \times \nu_2}^{BB}(\ell)=D_{\nu_1 \times \nu_2}^{BB, \rm CMB}(\ell)+D_{\nu_1 \times \nu_2}^{BB, F}(\ell) + D_{\nu_{1} \times \nu_2}^{BB, N}(\ell)
\end{equation}
where the superscripts of ``CMB'', ``$F$'' and ``$N$'' denote the CMB, foreground and noise, respectively. The measured band powers of $D_{\nu_1 \times \nu_2}^{BB}(\ell)$ are calculated by the  \texttt{NaMaster} package~\citep{2019MNRAS.484.4127A}, from which the beam and mask effects are removed unbiasedly through beam deconvolution and $B$-mode purification.

\subsubsection{Component models for the TF method}
\subsubsection*{CMB model}
The CMB contribution to the $BB$ spectrum expressed in CMB temperature units is given by 
\begin{equation}
D_{\nu_1 \times \nu_2}^{BB, \rm CMB}(\ell)=\frac{r}{0.01} \times \widebar{D}_{ r=0.01}^{\nu 1 \times \nu 2}(\ell) + \widebar{D}_{T/E~\rm only}^{\nu 1 \times \nu 2}(\ell), \label{eq:BB_scale}
\end{equation}
where $\widebar{D}_{\rm r=0.01}^{\nu 1 \times \nu 2}(\ell)$ represents a template for the CMB tensor only signal and the  $\widebar{D}_{T/E~\rm only}^{\nu 1 \times \nu 2}(\ell)$ for $E$-to-$B$ leakage from filtering. The leakage template is constructed by averaging the $BB$ power spectra of 50 $T/E$ only filtered CMB simulation maps. The input CMB spectra are generated using fiducial cosmological parameters, and notably, the $BB$ spectrum is explicitly nullified by setting $r=0$. Additionally, we find that the $B$-to-$E$ leakage is negligibly small, around $10^{-7}~\mu\text{K}^2$, as the CMB $BB$ spectrum is at the level of $10^{-3}~\mu\text{K}^2$.

Since the filtering effect (a combined effect of the polynomial filtering and the ground subtraction) is highly linear, instead of constructing filtering matrix as well as applying it on the theoretically predicted CMB power spectrum, we can build templates for the filtered shapes of various independent CMB components based on simulations. When considering the contribution of the non-diagonal elements of the filtering matrix, our fitting results for $r$ are presented in  Appendix~\ref{App:EE}. For building the $\widebar{D}_{\rm r=0.01}^{\nu 1 \times \nu 2}(\ell)$, we use the simulation pipeline to produce 50 realization maps, generated from the CMB tensor only $BB$ power spectrum with a fiducial value of $r=0.01$. These observed maps include the instrumental and filtering effects. We then compute the auto/cross band powers across all frequency pairs using~\texttt{NaMaster} to correct for the mask and beam effects. By averaging all these band powers, we finally determined the shape of the $B$-mode only template, i.e. of the spectral signature of the signal from primordial gravitational waves. In the same way, we build the template for $\widebar{D}_{T/E~\rm only}^{\nu 1 \times \nu 2}(\ell)$ by using 50 simulated $T/E$-only maps in the fiducial $\Lambda$CDM model. We can of course also construct a template for gravitational lensing induced $BB$ spectrum, but for the DC1 data, since the simulations does not include any gravitational lensing effect, we thus just assume $A_L=0$, ignoring it completely when fitting the data. An overall $B$-mode amplitude thus can be obtained by rescaling the $r$-only template, i,e., the $BB$ spectrum depending only on the tensor-to-scale ratio $r$. However, we note that although the CMB band power is ideally fixed across frequencies, the influence of filtering causes slight variations in the $r$-only template between frequencies.

In the following, the priors we choose for the TF likelihood analysis are flat with broad bounds---sufficiently large to guarantee that the priors chosen are independent of the estimates---unless they are non-physical, e.g., $r\in [0,1]$. 

\subsubsection*{Foreground models}

Besides the CMB, in the following, we use a three-component foreground model~\citep{Choi:2015xha} of the form  
\begin{equation}
\begin{aligned}
 D_{\nu_{1} \times \nu_{2}}^{B B, F}(\ell) =&A_s(\ell)f^{\nu_1}_s f^{\nu_2}_s\left(\frac{\nu_{1} \nu_{2}}{23^{2}}\right)^{\beta_{s}}+A_{d}(\ell)f^{\nu_1}_d f^{\nu_2}_d\left(\frac{\nu_{1} \nu_{2}}{353^{2}}\right)^{\beta_{d}-2}  \\
&+A_{sd}(\ell)\left[f^{\nu_1}_s f^{\nu_2}_d \left(\frac{\nu_{1}}{23}\right)^{\beta_{s}}\left(\frac{\nu_{2}}{353}\right)^{\beta_{d}-2} + f^{\nu_2}_d f^{\nu_1}_d \left(\frac{\nu_{2}}{23}\right)^{\beta_{s}}\left(\frac{\nu_{1}}{353}\right)^{\beta_{d}-2} \right]\,,
\end{aligned}
\end{equation}
with
\begin{equation}
f^{\nu}_s =\frac{g(\nu)}{g(23)}\,, \quad  f^{\nu}_d =\frac{g(\nu)}{g(353)}\frac{B_\nu\left(T_{d}\right)}{B_{353}\left(T_{d}\right)}\,,
\end{equation}
where the amplitude of the band power ($D_{\ell} \equiv \ell(\ell+1) C_{l} / 2 \pi$) is expressed in units of $\mu {\rm K^2_{CMB}}$, and the frequencies $\nu_1$ and $\nu_2$ are expressed in GHz. We have introduced the factors $g(\nu)$ which convert brightness temperature units to CMB thermodynamic temperature units ($\mu {\rm K_{RJ}}\rightarrow \mu {\rm K_{CMB}} $). The Planck function $B_\nu (T_d)$ is computed by a fixed dust temperature of $T_d=19.6$ K, determined from the fit of the dust total intensity at the high Galactic latitude~\citep{Planck:2014skm}. Here the model represents the synchrotron as a power law in frequency and the dust as a modified blackbody, $A_s$ is the synchrotron foreground amplitude, $A_d$ the dust foreground amplitude, and $A_{sd}$ represents the amplitude of the correlated component between the dust and the synchrotron. 
The parameters $\beta_s$ and $\beta_d$ represent the synchrotron and dust spectral indices, respectively.  In the following, we model the foreground with different parameterization methods in order to validate the effectiveness of our proposed foreground models and the robustness of the parameter estimates.

{\bf TF model ``p10''}: we first consider a 10-parameter model, with 9 free parameters used for modelling foreground power spectra plus one CMB parameter $r$, in which the $\ell$-dependent synchrotron and dust amplitudes of $A_s$ and $A_d$  are parametrized as
\begin{equation}
\begin{aligned}
&A_{s}(\ell)=A_{s}(\ell /\ell_*)^{\alpha_{s}+0.5\alpha_{s}^{\prime} \ln (\ell/\ell_*)}\,, \\
&A_{d}(\ell)=A_{d}(\ell /\ell_*)^{\alpha_{d}+0.5 \alpha_{d}^{\prime} \ln (\ell/\ell_*)}\,,\\
& A_{sd}(\ell)=\varepsilon \sqrt{A_s(\ell)A_d(\ell)}\,.
\end{aligned}\label{eq:p10}
\end{equation}
Here $A_{d}$ and $A_s$ denote the dust and synchrotron amplitudes in units of $\mu{\rm K^2_{CMB}}$ at a pivot angular scale of $\ell_*=80$. The dust and synchrotron components scale as power laws in $\ell$ with slopes of $\alpha_d$ and $\alpha_s$, respectively. From our simulation tests, we extend the BICEP adopted model by adding two running spectral indices of $\alpha'_d$ and $\alpha'_s$ for the synchrotron and dust, respectively, with the purpose of providing a better characterization on possible non-smooth complicated scale dependence. The correlation coefficient, $\varepsilon$, determines the level of spatial correlation between the synchrotron and dust, which for the current sensitivity for the AliCPT measurement can be set to be a constant across all $\ell$, and the $\ell$ scaling of the correlated component has the average slope of the dust and synchrotron. Physically, a negative value of $\varepsilon$ reflects an anti-correlation between these two components. 

{\bf TF model ``p16''}: a 16-parameter model, dubbed as ``p16'', is designed to capture three more detailed features that we observe from simulations: 1) the filtering-induced low-$\ell$ suppression that cannot be well captured by the running parameters $\alpha_s^\prime$ and $\alpha_d^\prime$; 2) the scale-dependence of the synchrotron-dust cross correlation, which tends to vanish at high $\ell$'s; 3) the scale-dependence of $\beta_d$ and $\beta_s$. Specifically, we model the filtering-induced low-$\ell$ suppression as an extra suppression factor,
\begin{equation}
  F_{\ell} = 1 - f_0\exp\left[-\left(\frac{\ell}{\ell_F}\right)^{\alpha_F}\right]\,, \label{eq:Fl_model}
\end{equation}
which is applied on all filtered $BB$ band powers. The uniform priors are applied on $f_0\in [0, 1]$, $\alpha_F\in [1, 8]$, $\ell_F\in [50, 100]$ in the likelihood-fitting procedure. The scale-dependence of the synchrotron-dust cross correlation is modelled by replacing the constant $\varepsilon$ in ``p10'' with a function
\begin{equation}
  \varepsilon(\ell) = \varepsilon_2\left(\frac{2}{\ell}\right)^{\alpha_\epsilon}\,.  \label{eq:eps_model}
\end{equation}
where the uniform priors are applied on $\varepsilon_2\in [-1, 1]$ and $\alpha_\epsilon\in [0, 2]$. Analogously, the scale-dependence of $\beta_d$ and $\beta_s$ is modelled by replacing the constant $\beta_d$ and $\beta_s$ in ``p10'' with 
\begin{equation}
\beta_d\rightarrow \beta_d  + \beta_d^\prime \ln\frac{\ell}{\ell_*},~\quad  \beta_s \rightarrow \beta_s  + \beta_s^\prime \ln\frac{\ell}{\ell_*}, 
\end{equation}
respectively.

{\bf TF model ``p3''}: a minimal 3-parameter model is also tested in the TF pipeline. For comparison and better understanding of the fitting results, the true shapes of foreground power spectra are assumed to be completely known. As such, the ``p3'' model is highly optimistic in that the $\ell$-dependence of each foreground components are exactly the simulation truth. This prior knowledge thus greatly reduces the number of free parameters while being able to successfully fit the data. Specifically, the band-power templates in Eq.~\ref{eq:p10} are calculated directly from the filtered DC1 foreground maps at 23 GHz and 353 GHz, 
\begin{equation}
    \begin{aligned}
    &A_{s}(\ell)= D_{23 \times 23}^{BB, \rm simF}(\ell)\,,\quad A_{d}(\ell)= D_{353 \times 353}^{BB, \rm sim F}(\ell)\,, \quad A_{sd}(\ell)=D_{23\times 353}^{BB,\rm sim F}(\ell)\,,  
    \end{aligned}
\end{equation}
where the superscript ``simF'' denotes the simulation truth. Interestingly, this optimistic model in fact fits the DC1 data as well as the other sophisticated ones, ``p10'' and ``p16'', certainly indicating the consistency and validity of these models for parameter estimation. In summary, the ``p3'' has one parameter for $r$  and only two free parameters for the foreground in terms of spectral index, $\beta_s$ and $\beta_d$.

\subsubsection*{Noise model}
In addition, using the same method as for constructing the CMB template, we estimate the mean noise band powers $D_{\nu_{i} \times \nu_j}^{BB, N}(\ell)$ from 50 noise-only filtering maps. Due to the limited number of realizations, we zero out the noise terms if their theoretical expectations in mean and covariance are zero, namely, $ D_{\nu_{i} \times \nu_j}^{BB, N}(\ell) = 0$ when $i\ne j$,  and $\langle \delta D_{\nu_{i} \times \nu_j}^{BB, N}(\ell) \delta D_{\nu_{p} \times \nu_q}^{BB, N}(\ell')\rangle \neq 0$ when $(i=p,j=q)$, or $(i=q, j=p)$. Due to the effect of masking, we do preserve the correlation (although fairly small) between the different $\ell$-bins according to the simulations. The elements of the covariance matrix that are more than two $\ell$-bins separated are too small to be reliably estimated from our limited mock samples, so we set them to zero.

\subsubsection{Gaussian likelihood for TF}
In the range of $\ell$-bins of interest (e.g. $\ell\ge 20$), the non-Gaussianity of the distribution of band power is small since the effective number of degrees of freedom per $\ell$-bin is large enough. In the course of this study, we thus use a Gaussian likelihood as a good approximation to derive the parameter estimates.  The multi-frequency multi-component likelihood function for the template fitting is written as 
\begin{equation}
-2\ln \mathcal{L}\left(\hat{\mathbf{C}}| \mathbf{C}(\theta)\right)=\left[\hat{\mathbf{C}}-\mathbf{C}(\theta)\right]^{T} \mathcal{M}^{-1}\left[\hat{\mathbf{C}}-\mathbf{C}(\theta)\right]\,,
\end{equation}
where $\mathbf{C}(\theta)$ represents the model band power vector with parameters $\theta$, and $\hat{\mathbf{C}}$ is the data band power vector. The full covariance matrix of the auto-/cross-spectral band powers derived from simulations is denoted by $\mathcal{M}$.  The data band power vector is constructed from all $BB$ spectral band powers over $N_\ell$ $\ell$-bins and $N$ frequency channels, $\hat{\mathbf{C}} =\left[\mathbf{\hat{D}}^{BB}(\ell_1), \mathbf{\hat{D}}^{BB}(\ell_2), \cdots, \mathbf{\hat{D}}^{BB}(\ell_{N_\ell}) \right]$, where the band power vector for at the $i$-th $\ell$ bin is concatenated by the following frequency spectra,
\begin{equation}
\mathbf{\hat{D}}^{BB}(\ell_i)= \left[\hat{D}_{\nu_1 \times \nu_1}^{BB}, \cdots, \hat{D}_{\nu_1 \times \nu_N}^{BB},\hat{D}_{\nu_2 \times \nu_2}^{BB},\cdots,\hat{D}_{\nu_2 \times \nu_N}^{BB}, \cdots,\hat{D}_{\nu_N \times \nu_N}^{BB}\right](\ell_i)\,,
\end{equation}
thereby having total $N_\ell\times N(N+1)/2$ elements in $\hat{\mathbf{C}}$. In the TF pipeline, we set $N_\ell=9$, resulting in a dimension size of $\mathcal{M}$ being $252\times 252$ for 7 frequency channels. The model band power vector $\mathbf{C}(\theta)$ is constructed in the same way.

The full covariance matrix of these band powers is estimated from the Monte Carlo simulations together with a semi-analytic calculation. The contributions from CMB signal and noise to the covariance are estimated purely from the 50 realizations of the end-to-end simulations for each of them, generated from the fiducial CMB model ($r=0.01$+ $\Lambda$CDM model) and fiducial noise model. The band powers for CMB and noise are then computed using \texttt{NaMaster}, in the same way as for $\mathbf{\hat{D}}^{BB}(\ell)$. Hence the statistical variances of each component and filtering-induced fluctuations are fully preserved in $\mathcal{M}$ estimated from simulations, which will accurately propagate the overall uncertainty in the final parameter estimates.

It is important to mention that the foreground emissions do make a significant contribution to the covariance of the power spectrum. Fluctuations of chance correlation $\delta D_{\ell}^{F\times N}$ between a fixed foreground map and random noise realizations would be proportional to the fluctuation level of the noise $\delta D_{\ell}^{N}$. Intuitively, although the chance correlation on average is expected to be zero, $\langle D^{F\times N}_{\ell}\rangle=0$, the square of the chance correlation, corresponding the covariance of the power spectrum (equivalently 4-point correlations of $a_{\ell m}$), is absolutely non-zero, $\langle(\delta D^{F\times N}_{\ell})^2\rangle>0$. Due to the limited number of simulations (50 realizations), we cannot estimate this chance correlation with acceptable accuracy, so instead, we use a semi-analytic approach to calculate it.

The mathematical formalism implemented to calculate the full covariance contributed from the CMB, noise and foreground-noise covariance (denoted by $\mathcal{M_{\rm CMB}}, \mathcal{M}_N, \mathcal{M}_{N\times F}$, respectively)  between frequency pairs of $\nu_i \times \nu_j$ and  $\nu_p \times \nu_q$ is written as
\begin{equation}\label{eq:M}
  \mathcal{M}(i\times j,p\times q)(\ell,\ell') \equiv \langle \delta D_{\nu_i \times \nu_j}^{BB}(\ell)\delta D_{\nu_p \times \nu_q}^{BB}(\ell')\rangle \approx \mathcal{M}_{\rm CMB}+ \mathcal{M}_N +\mathcal{M}_{N\times F}\,,
\end{equation}
where we have ignored terms of $\mathcal{M_{\rm CMB\times F}}$ and $\mathcal{M_{\rm CMB\times N}}$ relating to the CMB-foreground and noise-foreground cross correlations, as these correlations are significantly smaller compared to $\mathcal{M}_{N}$ and $\mathcal{M}_{N \times F}$ (e.g., at least two orders of magnitude smaller at 95 GHz) in our fiducial model. If needed, one can consider these additional terms and extend the above formula for $\mathcal{M}_{N\times F}$ straightforwardly. The CMB and noise covariance matrices estimated from simulations read 

\begin{equation}
  \begin{aligned}
   &\mathcal{M}_{\rm CMB}(i\times j,p\times q)(\ell,\ell')= \langle \delta D_{\nu_i \times \nu_j}^{BB,\rm CMB}(\ell)\delta D_{\nu_p \times \nu_q}^{BB,\rm CMB}(\ell')\rangle\,,\\
    &\mathcal{M}_N(i\times j,p\times q)(\ell,\ell')= \langle \delta D_{\nu_i \times \nu_j}^{BB,N}(\ell)\delta D_{\nu_p \times \nu_q}^{BB,N}(\ell')\rangle\,,
  \end{aligned}
 \end{equation} 
and the foreground-noise covariance can be expressed in terms of foreground and noise band powers, 
 \begin{equation}\label{eq:cov_fn}
  \begin{aligned}   
    &\mathcal{M}_{N\times F}(i\times j,p\times q)(\ell,\ell') = \langle \delta D_{\nu_i \times \nu_j}^{BB,F\times N}(\ell)~\delta D_{\nu_p \times \nu_q}^{BB,F\times N}(\ell')\rangle \\
    &\approx \frac{\delta_{\ell,\ell'}}{N_{\rm d.o.f.}(\ell)} \left[D^{BB,F}_{\nu_i\times\nu_p}  D_{\nu_j \times \nu_q}^{BB, N} + D^{BB,F}_{\nu_i \times \nu_q} D_{\nu_{j} \times \nu_p}^{BB, N} + D^{BB,F}_{\nu_j \times\nu_p}  D_{\nu_i \times \nu_q}^{BB, N} +D^{BB,F}_{\nu_j \times \nu_q} D_{\nu_i \times \nu_p}^{BB, N}\right](\ell)\,,
  \end{aligned}
 \end{equation} 
 where we have assumed that the correlations between different $\ell$ bins are quite small, with the correlation coefficient between adjacent bins being less than 0.1, and can be safely neglected. The foreground band powers are computed from the foreground model. We have introduced a scaling factor $N_{\rm d.o.f.}(\ell)$ that describes the effective number of modes. Recall that, the sampling variance of a power spectrum is theoretically given by $(\delta C_\ell)^2 =2C^2_\ell/N_{\rm d.o.f.}(\ell)$, with $N_{\rm d.o.f.}(\ell)$ defined as $N_{\rm d.o.f.}(\ell)=(2\ell+1)\Delta \ell f_{\rm sky}$ and where $\Delta \ell$ is the bin width and $f_{\rm sky}$ is the effective fraction of sky. However, various effects from the filtering, the mask and the spatially inhomogenenous noise lead to strong fluctuations, thereby increasing  $\mathcal{M}_{N\times F}$, so that, by using such simple theoretical estimation, one would highly underestimate the variance of band powers, especially for the low-$\ell$ variance where the filtering effects on band powers become stronger. Therefore, the effective way to accurately determine the factor $N_{\rm d.o.f.}$ is to use simulations to account for all these complex effects. By averaging the noise auto spectral powers over all frequencies, one can derive the effective degrees of freedom as     
\begin{equation}
\frac{1}{N_{\rm d.o.f.}(\ell)}\approx \frac{1}{2N_\nu} \sum_i^{N_\nu}\frac{\mathcal{M}_N(i\times i, i\times i)(\ell,\ell) }{\big[D_{\nu_i \times \nu_i}^{BB, N}(\ell)\big]^2 }\,, 
\end{equation}
where the statistical variance and noise band powers are estimated from 50 DC1 simulations. Once the foreground parameters are known, we can then use Eq.~\ref{eq:cov_fn} to calculate the covariance $\mathcal{M}_{N\times F}$ analytically. However, in practice, the foreground parameters before a full exploration of the likelihood  are unknown, so that we iteratively refine the estimate on the $\mathcal{M}_{N\times F}$. We first assume $\mathcal{M}_{N\times F}=0$---which does not bias the parameter estimate but merely underestimates error bars---and use a maximum-likelihood fit of the parametric CMB and foreground models to determine the foreground parameters. Taken with their median values, we re-calculate the $\mathcal{M}_{N\times F}$ and perform the maximum-likelihood fit again to update the CMB and the foreground parameters and consequently the $\mathcal{M}_{N\times F}$. We iterate this process until all parameter posteriors converge. From our tests, we find this iterative process can provide a unbiased estimate on $r$ and foreground parameters, and basically achieves an excellent convergence after only one iteration (the change in 
$r$ is less than 0.001 upon performing another iteration). Note that the exact number of iterations needed depends on the starting point. If it is close to the truth, one iteration is usually enough. Otherwise, two or three is sufficient.

\subsection{The GLS method}
\label{sec:gls}
We model the observed sky as a linear mixture of different emission processes. We assume that the total emission of each component can be decomposed as the product of a spatial template $s_c(p)$ which varies with position on the sky but it is assumed to be independent of the frequency, and a spectral emission law for the astrophysical process that is assumed to be independent of position in the sky. Then the total emission $x_c(\nu, p)$ from the $c$-th component in band $\nu$ at pixel $p$ is given by:
\begin{equation}
    x_c(\nu, p) = a_c(\nu)s_c(p),
\end{equation}
where $a_c(\nu)$ is the mixing amplitude of component $c$ in band $\nu$, computed from the spectral emission law. Then the model for the observed data $d(\nu, p)$ can be written as:
\begin{align}
    d(\nu, p) =& \sum_c x_c(\nu, p) + n(\nu, p) \nonumber \\
    \boldsymbol{d}(p) =& \boldsymbol{A s}(p) + \boldsymbol{n}(p). 
    \label{eq:model}
\end{align}
This matrix equation represents the linear mixture model for the observed sky. Here $\boldsymbol{d}(p)$ is a $n_\nu$-dim column vector, $\boldsymbol{A}$ is a $n_\nu\times n_c$ matrix, $\boldsymbol{s}(p)$ and $\boldsymbol{n}(p)$ are $n_c$-dim column vectors. Hereafter we omit the $p$ subscripts.

In our implementation we consider the three dominant components---the CMB, synchrotron and thermal dust. We model the CMB as differential black body, synchrotron as a power law with spectral index, $\beta_{\rm sync}=-3$ for data in Rayleigh-Jeans brightness temperature units, and the thermal dust as a modified black body with dust temperature, $T_{\rm dust}=19.6$ K and dust spectral index, $\beta_{\rm dust}=-1.59$~\citep{Choi:2015xha}, based on the Planck constraints on polarized foregrounds~\citep{Planck:2014dmk,Planck:2018gnk}. 
Our simplified emission model does not consider variations in these three foreground parameters. Instead, it aims to eliminate foreground emissions using average parameters, assuming only weak frequency-decorrelation of foregrounds within our analysis patch. This approach remains effective even when the amplitude of foreground contamination varies non-uniformly across our region.
We integrate the emission laws with the bandpass function, and account for the conversion of units to $\mu{\rm K_{CMB}}$. The mixing vector of CMB is 1 (i.e., $a_{\rm cmb}(\nu)=1$) for all frequency bands.

For a system with more frequencies than components, the left pseudo-inverse of $\boldsymbol{A}$ can solve the linear system. However, for noisy data we weight the system with inverse noise variance weights, before inverting the system. The generalized least squares (GLS) solution \cite{Delabrouille2009} is the inverse noise weighted solution of the linear system defined above giving the weights:
\begin{equation}
    \boldsymbol{W}^{\rm GLS} = \left[ \boldsymbol{A}^t \boldsymbol{N}^{-1} \boldsymbol{A} \right]^{-1}\boldsymbol{A}^t \boldsymbol{N}^{-1}
\end{equation}
where $\boldsymbol{N}$ represents the noise covariance matrix. In harmonic space, $\boldsymbol{N}_{\ell m \ell' m'}$ is a $n_{\nu} \times n_{\nu}$ matrix, approximated as $\boldsymbol{N}_{\ell}$ by neglecting mode coupling. In practice, we estimate it by averaging the noise products $n_{\nu,\ell m}^* n_{\nu',\ell m}$ from one simulation over all $\ell m$'s in the multipole range of $(0.8\ell, 1.2\ell)$, and then averaging the obtained $\boldsymbol{N}_{\ell}$ over all noise simulations. $\boldsymbol{W}^{\mathrm{GLS}}_{\ell}$ is an $n_c \times n_{\nu}$-dimensional matrix applied to the frequency maps. The solution for the component template is given by:
\begin{equation}
    \hat s_{c,\ell m} = \sum_\nu W^{GLS}_{c, \nu,\ell} d_{\nu,\ell m}.
\end{equation}
Using the CMB component we get the projected CMB. The foreground components are also projected with their corresponding weights. 

\begin{figure}[htpb]
    \centering
    \includegraphics[width=0.48\textwidth]{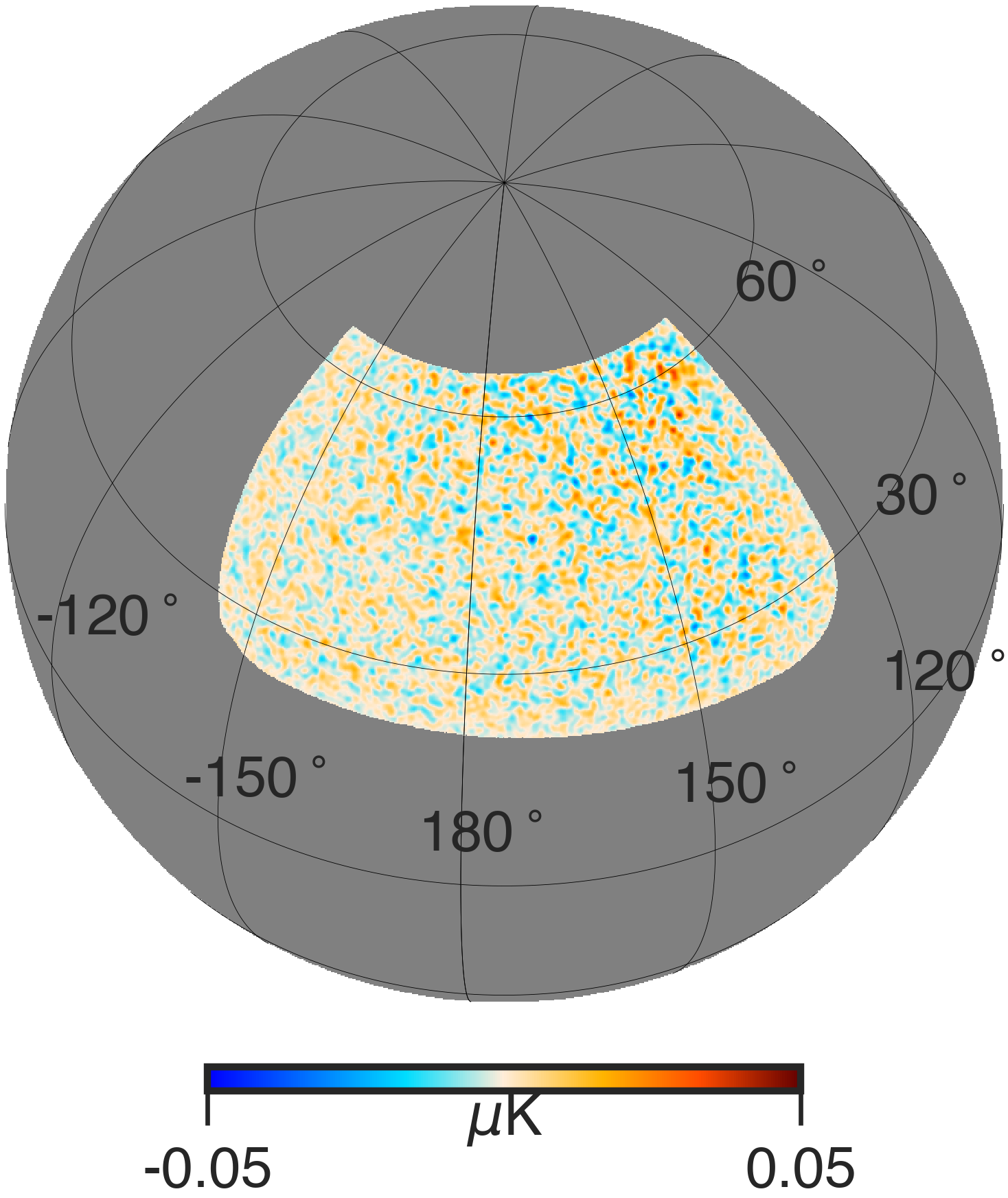}
    \includegraphics[width=0.48\textwidth]{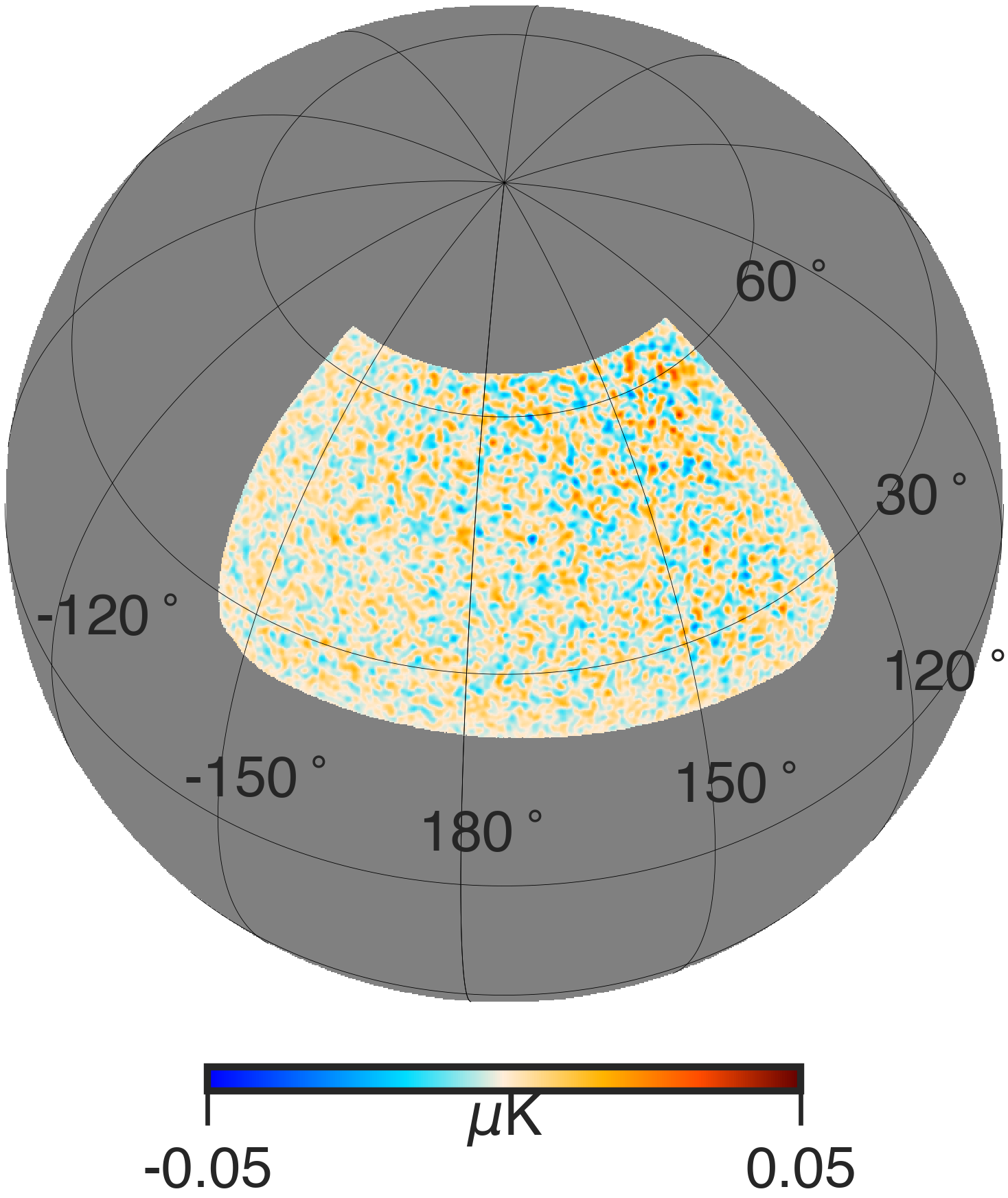}
    \caption{{\it Left}: The GLS solution for CMB for DC1 multiplied by the UNP mask. {\it Right}: The cILC cleaned CMB map for DC1 multiplied by the UNP mask.}
    \label{fig:GLS-cILC_Bmap}
\end{figure}
\emph{Mask choice:} For the AliCPT DC1 we preprocess the DC1 maps through the point source pipeline so that resolved point sources have been masked and inpainted in these maps. The preprocessing of point sources is detailed in Appendix~\ref{sec:point_source}. For our analysis we adopt a further sky mask that removes the few regions of relatively stronger foreground in the AliCPT patch. To identify these regions we use the Planck 2015 SMICA polarized thermal dust and polarized synchrotron maps. We smooth them to the common resolution of $5^\circ$ and compute the total polarized flux map. In the AliCPT patch we additionally mask all pixels where polarized flux of the foreground exceeds 0.01 ${\rm mK}_{\rm RJ}$, and all pixels above declination of $65^\circ$. This gives us a relatively uniform foreground region. Note that we estimate this region from the actual Planck polarized foreground maps. While the foregrounds in DC1 will differ from the actual sky, it should give a reasonable first estimate to eliminate brighter foreground regions from analysis. We further optimize the sky region by considering the noise variance maps of the AliCPT 150 GHz band (the noisier of the two AliCPT bands). We select pixels with noise standard deviation less than $\le 10~\mu$K-pixel, with a pixel size of 3.4 arcminute. We further weight the AliCPT map with inverse noise variance of the AliCPT 150 GHz band. We refer to this union mask of polarized foregrounds and noise as the UNP mask, shown on the right panel of Fig.~\ref{fig:masks}. 

\begin{figure}[htpb]
    \centering
    \includegraphics[width=0.48\textwidth]{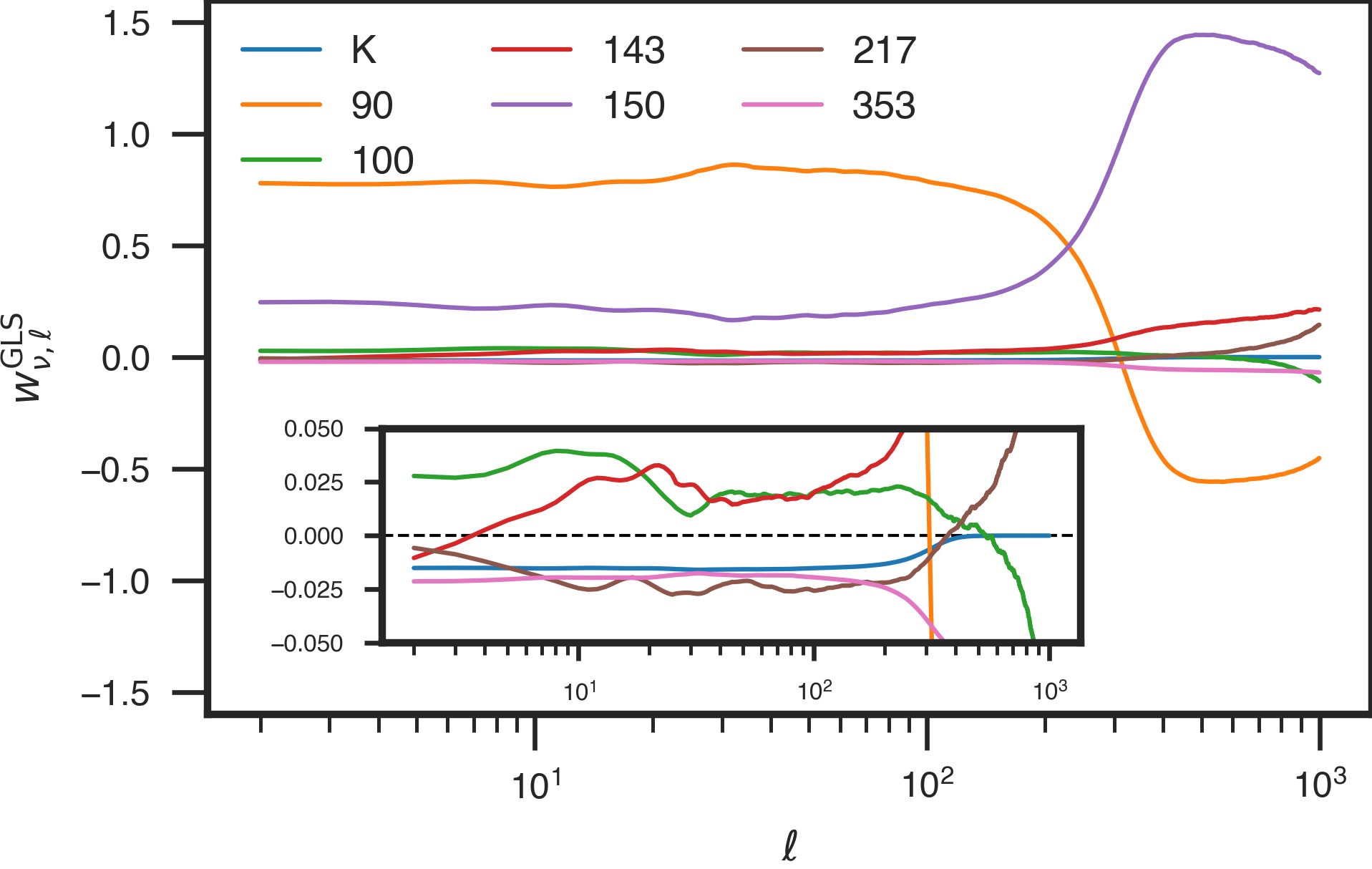}
    \includegraphics[width=0.48\textwidth]{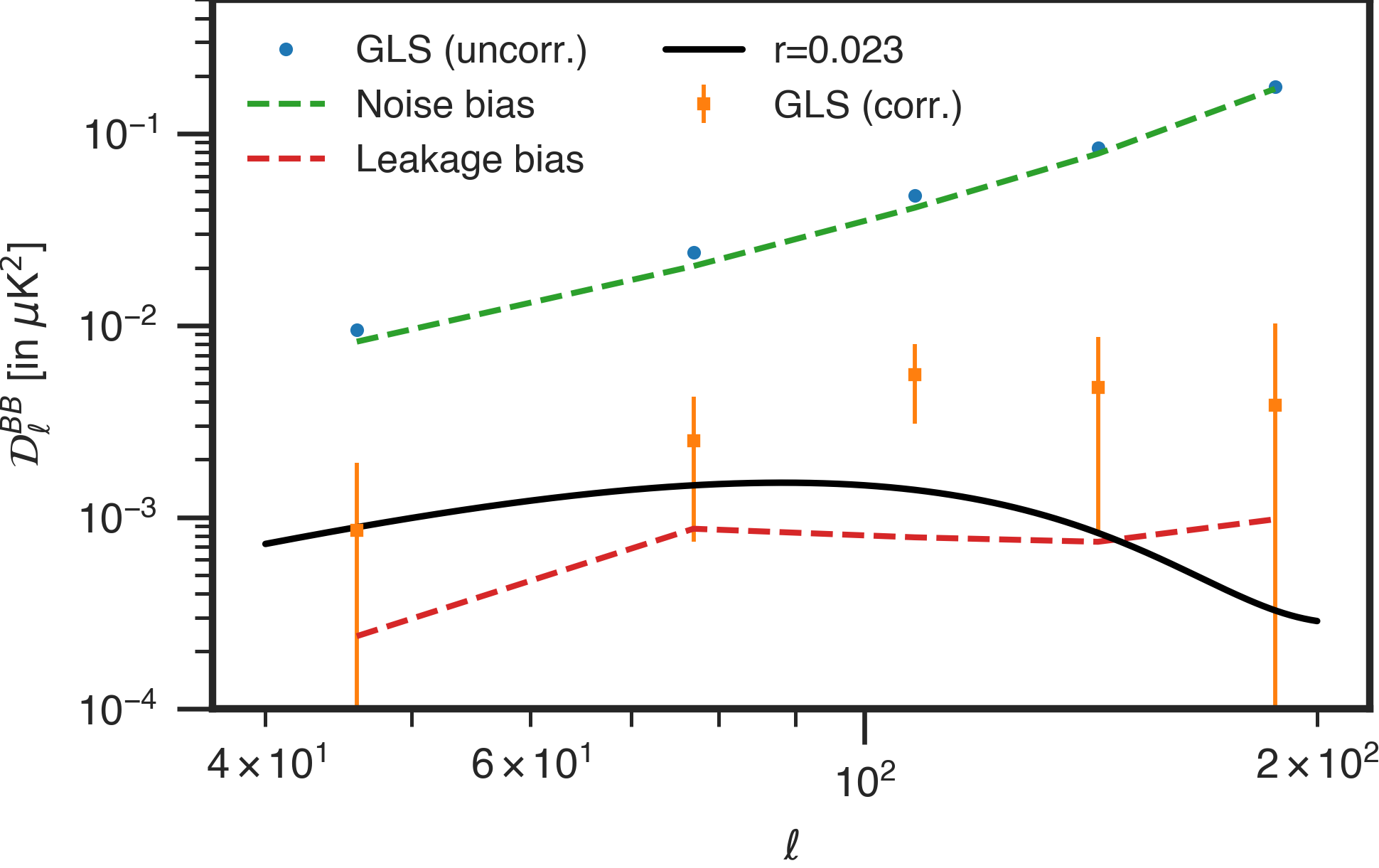}
    \caption{Left: the GLS weights for the CMB components in all 7 bands of DC1, with the values close to zero zoomed in in the sub-figure. Right: the final power spectrum recovered for DC1 with GLS method (orange bars) which has been corrected for the noise bias (green dashed) and the filtering leakage bias (red dashed). The blue points represent the GLS power spectrum before debiasing. We also show the actual input $B$-modes at $r_{\rm true}=0.023$ used in DC1 (black solid). As a comparison, the ABS-recovered $BB$ spectrum can be seen in Fig. 4 of~\cite{paper4}.}
    \label{fig:GLS-solution}
\end{figure}

We implement the GLS cleaning pipeline in harmonic space, where all spherical harmonic coefficients for a given $\ell$ are cleaned with the corresponding GLS weight. All maps are converted to the common resolution of the WMAP K band with a Gaussian beam of $52.8'$. We compute the noise covariance matrix from 50 TOD filtered noise simulations with the UNP mask, and then use the noise covariance matrix to compute the GLS weights in harmonic space. The GLS CMB weights are shown in the left panel of Fig.~\ref{fig:GLS-solution}. We can see that the AliCPT channels dominate the contribution to the CMB with the 95 GHz contributing more till the better resolution of 150 GHz takes over at higher multipoles. The pre-processed DC1 $B$-mode maps are masked with the UNP mask to obtain the $B$-mode $a_{\ell m}$s. Finally we apply the GLS CMB weights to obtain the GLS solution for the CMB $B$ modes. The CMB map from the GLS is shown on the left panel of Fig.~\ref{fig:GLS-cILC_Bmap}. Note that this map is noise dominated and has the inverse noise variance weighting from the masking. The map also contains all effects of TOD filtering which we will correct for at the stage of power spectrum calculation.

The detailed description of the power spectrum calculation for the GLS pipeline is discussed in~\citep{paper4}, and we will only summarise the main steps here.
We use the PCL-TC method~\citep{paper4} to compute the $B$-mode power spectrum. The mode coupled power spectrum is averaged with a rolling bin width of ${\rm min}(5, 0.2\ell)$, at every multipole. This step helps to average out the fluctuations as the projected noise power spectra is much larger than the CMB $B$-mode signal in the data. The GLS CMB weights are used to compute the projected noise in the CMB map. We use 50 TOD filtered noise simulations and compute the projected noise power spectrum with identical rolling bins, the average of which is the noise bias power spectrum. The leakage due to filtering is computed by taking a cross spectrum with the leakage template prepared from Planck and WMAP $E$-mode information and is detailed in~\cite{paper4}.  In short, we apply NILC on $T$ and $E$-mode maps from Planck HFI and WMAP K band, and then propagate the $T$ and $E$-mode IQU map through the AliCPT observation pipeline to obtain the filtered $T$ and $E$ mode only IQU map, from which we obtain the $B$-mode map as the leakage template. We debias the mode coupled $B$-mode spectrum from the noise and leakage biases. Finally we decouple and debeam the debiased power spectrum with \texttt{NaMaster}, which corrects for the mode coupling due to masking, the beam transfer function and the window effects. This power spectrum still contains power suppression from the TOD filtering. We correct for power suppression with the $B$-mode auto transfer function. The power spectrum with and without debiasing is shown in Fig.~\ref{fig:GLS-solution} on the right. We can see that the power spectrum of the GLS map is noise dominated. After noise debiasing the power spectrum estimates are close to the input power spectrum. 

\begin{figure}[htpb]
    \centering
    \includegraphics[width=0.48\textwidth]{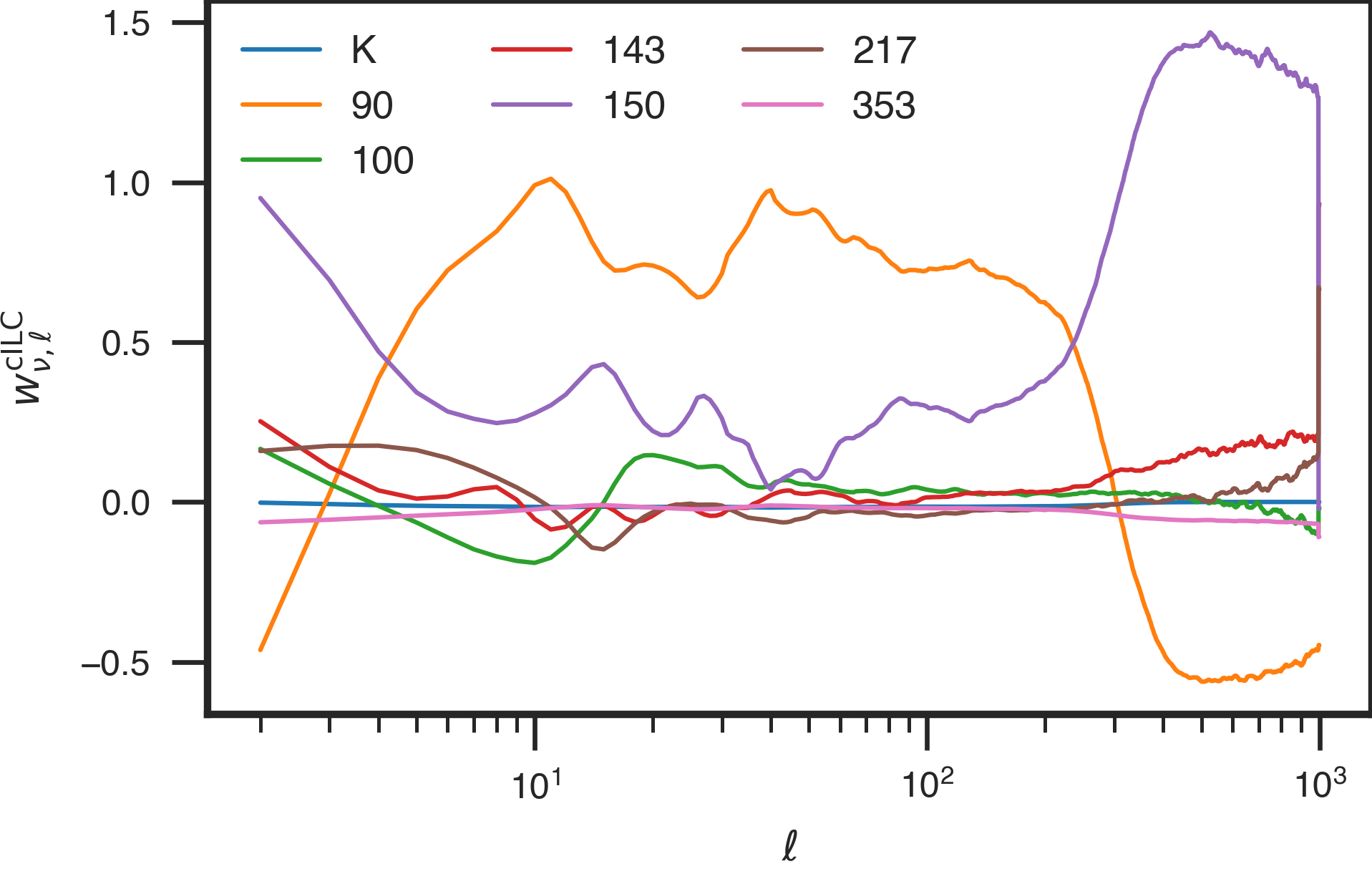}
    \includegraphics[width=0.48\textwidth]{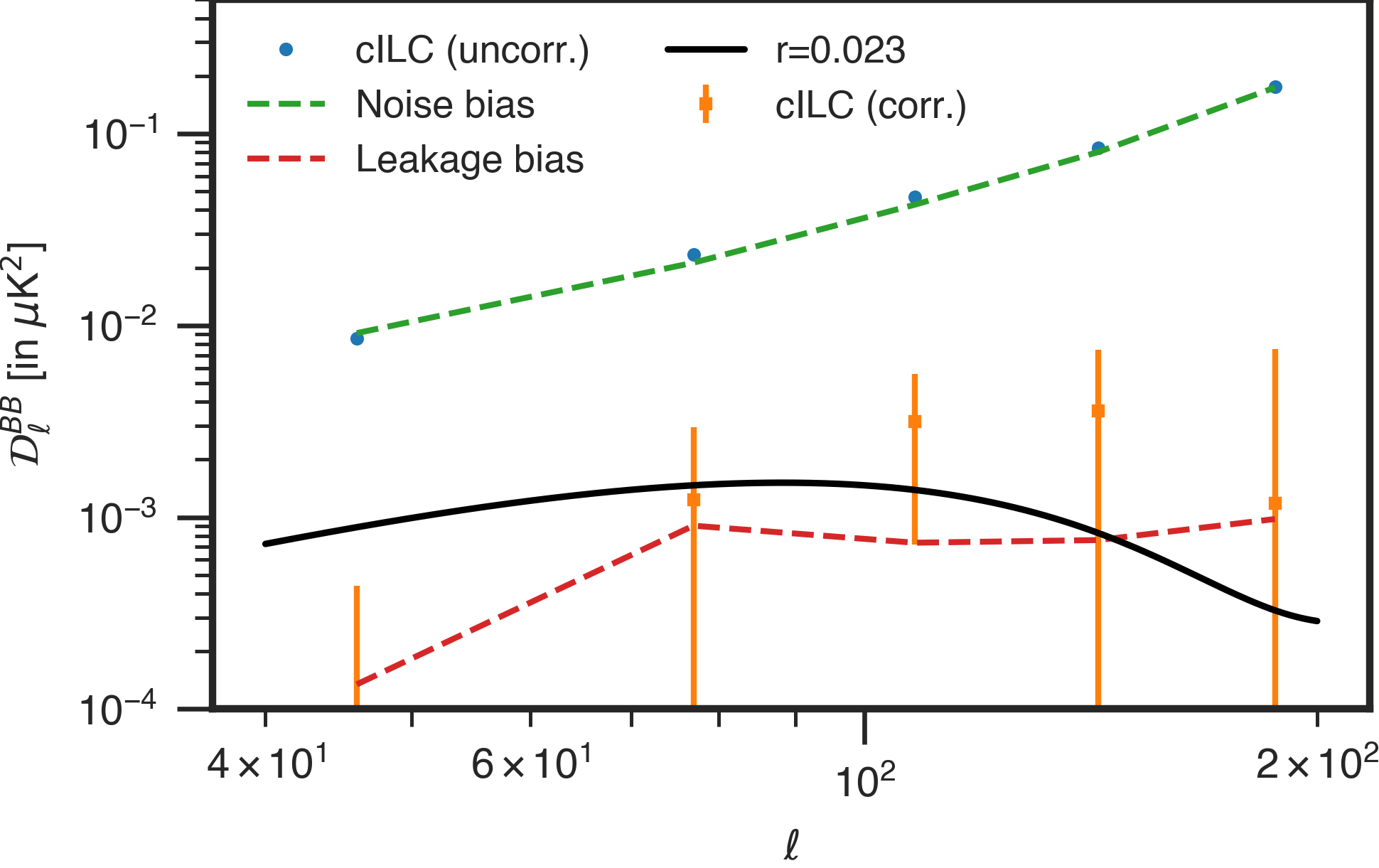}
    \caption{Same as in Fig.~\ref{fig:GLS-solution}, but for the cILC weights and corresponding $B$-mode recovery.}
    \label{fig:cILC-solution}
\end{figure}

\subsection{The cILC method}
\label{sec:cilc}
The constrained ILC (cILC) method~\citep{Chluba:2017, Remazeilles2010, Remazeilles2021} aims to eliminate selected foreground residuals by utilizing prior knowledge about foreground components, rather than minimizing global contamination variance in the ILC. The cILC method is a semi-blind foreground cleaning technique. It leverages foreground information priors to introduce additional constraints that effectively remove the average dust and synchrotron emissions captured by data modeling.

The data is modeled as in Eq.~\ref{eq:model}, where the mixing matrix has three columns corresponding to CMB, dust and synchrotron. The cILC pipleine is also implemented in harmonic space. The weights of cILC, $W^{\rm cILC}$, are computed with the constraints:
\begin{equation}
    \sum_\nu W^{\rm cILC}_{\nu,\ell} A_{\nu,c} = e_c\,,
\end{equation}
where $e_c$ is $1$ for CMB and $0$ for all foreground components modeled in $A$. The constraint in this case is computed with prior information given in Sect.~\ref{sec:gls}, so cILC is a semi-blind analysis using the priors on the main foreground components. The cILC weights are given by
\begin{equation}
    \boldsymbol{W}^{\rm cILC} = \boldsymbol{e} \left( \boldsymbol{A}^T\boldsymbol{C}^{-1} \boldsymbol{A}\right)^{-1}\boldsymbol{A}^T \boldsymbol{C}^{-1}\,,
\end{equation}
where $\boldsymbol{C}$ is the $n_\nu\times n_\nu$ covariance matrix of the data. It is estimated from data in harmonic space in the same way as the noise covariance matrix estimated in GLS from a single noise simulation. Here, $\boldsymbol{e}$ is a $1 \times n_c$ row vector, and $\boldsymbol{W}^{\rm cILC}_{\ell}$ is an $n_{\nu}$-dimensional column vector. The cILC cleaned CMB map is $\hat s_{\rm CMB,\ell m} = \sum_\nu W^{\rm cILC}_{\nu,\ell} d_{\nu,\ell m}$.

Note that the removal of noise bias from auto power spectra is no longer a viable approach due to concerns about potential bias. In modern CMB data analysis, only cross power spectra are widely used to avoid debiasing. In Appendix~\ref{app:null}, we present results for cILC and GLS, showing the utilization of cross power spectra to mitigate noise bias in $r$ estimation.

\subsection{The ABS method}
\label{sec:abs}
A new and computationally efficient method has recently been proposed by~\citep{2016arXiv160803707Z} for the blind separation of the CMB from foregrounds (ABS). The method does not require any prior assumptions or parameterization about foreground properties. Based on the measured cross band power between different frequencies, the CMB band power spectrum can be analytically solved, whilst avoiding multi-parameter fitting on foreground parameters. The ABS method has been tested for simulated Planck temperature maps~\citep{2018ApJS..239...36Y} and has also been able to recover the $E$- and $B$-mode power spectra from simulated CMB polarization observations of prospective high-precision space missions~\citep{2021A&A...650A..65S}. The main characteristics of ABS can be summarized as 1) the use of ``thresholding'' to remove noise-dominated signals and 2) the introduction of a ``shift parameter'' to shift observed multi-frequency power spectrum, especially for stabilizing the estimation of extremely faint CMB $BB$ signals. 


The formalism of the ABS method is briefly reviewed as follows. We consider observations of the CMB signal at $N_f$ different frequencies, where the observed data is a noisy linear mixture of the CMB and foreground emissions. In this scenario, the observed cross-band power spectrum between the $i$- and $j$-th frequency channels in the multipole bin $\ell$ is given by
\beq\label{eq:data}
\mathcal{D}^{\rm obs}_{ij}(\ell) = f_if_j\mathcal{D}^{\rm cmb}(\ell) + \mathcal{D}^{\rm fore}_{ij}(\ell) + \delta \mathcal{D}_{ij}^{\rm noise}(\ell)\,,
\eeq
where $i,j = 1,2\cdots N_f$ and $f_i$ collects the emission law of the CMB in which $f_i = 1$ for all channels in units of thermodynamic temperature. The CMB power spectrum, $\mathcal{D}^{\rm cmb}$, thus does not vary with frequency, and Eq.~\ref{eq:data} is applied independently to each of the $T$, $E$ and $B$ fields. Here, $\mathcal{D}_{ij}^{\rm fore}$ and $\delta \mathcal{D}_{ij}^{\rm noise}$ are the cross band power matrices of the foreground and residual instrumental noise, respectively, where the ensemble-averaged noise power spectrum is assumed to be known and has been implicitly subtracted out beforehand from the measured cross power spectrum.



Accounting for instrumental noise, the ABS solution in general can be written as a simple form,  
\beq\label{eq:abs1}
\mathcal{\hat{D}}^{\rm cmb} = \left( \sum^{\tilde{\lambda}_{\mu}\geq \lambda_{\rm cut}} \tilde{G}^2_{\mu}\tilde{\lambda}_{\mu}^{-1}\right)^{-1} - \mathcal{S}\,.
\eeq
Here, we have introduced new variables, defined by
\beq\label{eq:noiseD}
\mathcal{\tilde{D}}^{\rm obs}_{ij}\equiv \frac{\mathcal{D}^{\rm obs}_{ij}}{\sqrt{\sigma_{\mathcal{D},i}^{\rm noise}\sigma_{\mathcal{D},j}^{\rm noise}}} + \tilde{f}_i\tilde{f}_j\mathcal{S}\,,~~~~{\rm with}~~
\tilde{f_i} \equiv \frac{f_i}{\sqrt{\sigma_{\mathcal{D},i}^{\rm noise}}}\,,~\tilde{G}_{\mu}\equiv {\bf \tilde{f}}\cdot {\bf \tilde{E}}^\mu\,.
\eeq
Here, the associated dispersion in each residual noise matrix element is given by $\sigma_{\mathcal{D}, i j}^{\rm noise} \equiv \big< \left( \delta \mathcal{D}_{i j}^{\rm noise} \right)^2 \big>$.  Additionally, ${\bf \tilde{E}}^\mu$ and $\tilde{\lambda}_\mu$ are the $\mu$-th eigenvector and corresponding eigenvalue of $\mathcal{\tilde{D}}^{\rm obs}_{ij}$, respectively. The instrumental noise thus can lead to nonphysical (i.e., noise dominated) eigenmodes with eigenvalues of $|\tilde{\lambda}_\mu| \lesssim1$ in $\mathcal{\tilde{D}}^{\rm obs}_{ij}$. The ABS method thresholds the eigenvalues $\tilde{\lambda}_\mu$, only keeping those signal-dominated modes (when $0.5\leq\tilde{\lambda}_{\rm cut}\leq1$). In practice, we choose $\tilde{\lambda}_{\rm cut} = 1$ for $TT$, $EE$ and $BB$ spectra as suggested in~\citep{2018ApJS..239...36Y}. In Eq.~\ref{eq:abs1}, the introduced free parameter $\mathcal{S}$ corresponds to shifting the amplitude of the input CMB power spectrum from $\mathcal{D}^{\rm cmb}$ to $\mathcal{D}^{\rm cmb} +\mathcal{S}$. In practice, a positive and large {\it shift} parameter plays a crucial role in stabilizing the computation, ensuring CMB modes in the subspace spanned by the eigenvectors are not discarded in the thresholding process. This ``shift'' strategy becomes particularly important for low signal-to-noise regime, especially for the $B$-modes, with a typical value suggested to be $\mathcal{S}\approx 10\mathcal{D}^{\rm noise}$.

Note that, an incomplete sky coverage can lead to a considerable $E$-to-$B$ leakage, often well described in the mode-mode coupling matrix~\citep{Bunn:2002df, Chon:2003gx, Brown:2004jn, Smith:2005gi}. This effect is particularly strong at larger scales, significantly biasing $B$-mode estimate. For unbiased estimation of the cross band power spectra of $\mathcal{\tilde{D}}^{\rm obs}$ and $\mathcal{\tilde{D}}^{\rm noise}_{ij}$ in Eq.~\ref{eq:data} from partial-sky observations, we use the Python implementation of \texttt{NaMaster} library~\citep{2019MNRAS.484.4127A} to correct the mask effects as well as the beam convolution. 

\section{Analysis results and foreground residual levels}

\label{sec:fgre}
In this study, we focus on the TF, GLS, cILC, and ABS pipelines and investigate the foreground residuals for each pipeline. From the simulations, since each foreground component map is completely known, we can quantify the residual levels by comparing the reconstructed foregrounds and simulation inputs. Due to the fact that the ABS pipeline returns the CMB $BB$ spectrum directly without any foreground estimation, we leave the study of foreground residual level for ABS and its impacts on $r$ to Sect.~\ref{sect:fgr}.

\subsection{Quantifying foreground residuals for TF}
\label{sect:res_TF}
In the following, we will present the results for different foreground models and parameters in the TF pipeline. Using these derived parameters, we compare the reconstructed foreground power spectra at different frequencies with the real ones in the simulation. The reconstructed $BB$ band powers from TF are summarized in Appendix~\ref{app:BB-TF}. 

In this study, we do not discuss whether more complex foreground models (with more free parameters) are competitive in terms of fitting the data. Usually the model comparison can be achieved by some criteria, such as AIC~\citep{1100705} and BIC~\citep{1978AnSta...6..461S}, based on penalties related to the number of parameters required for the model to interpret the data. However, this is beyond the scope of this study.

\subsubsection{Foreground parameter estimate from TF pipeline}  



We compute the Gaussian likelihood of the DC1 data for the various  TF models considered in this work, using the full covariance matrix of the auto-/cross-spectral band powers as derived from the simulations and the analytical calculations. Compared to the TF pipeline used in~\cite{paper4}, the upgraded TF pipeline used in the present work further reduces point-source contamination by truncating very bright pixels (more than $50$ standard deviations off the mean)\footnote{The truncation is realized with a mapping $x\rightarrow 50\tanh (x/50)$, where $x$ is the number of standard deviations off the mean.}. This minor improvement slightly improves the reconstruction of spectral indices of the diffuse foreground (synchrotron and thermal dust), but has negligible impact on the measurement of $r$. 

We focus our discussion on the physical parameters $r$, $\beta_s$ and $\beta_d$, which can be directly compared with the input physical parameters in the DC1 simulation. Other parameters have less direct physical meaning, because they describe the power spectra with filtering-induced suppression at low $\ell$'s and point-source contamination at high $\ell$'s. 

\begin{table}[htpb]  
\centering
\begin{tabular}{crrr|r}
\hline
\hline 
{\bf TF model} & p16 & p10 & p3 & input  \\
\hline 
$r$ & $0.035 \pm 0.021$ & $0.034 \pm 0.020$ & $0.031\pm 0.019$ & 0.023\\
$\beta_d$ & $1.584 \pm 0.107$ & $1.614 \pm 0.092$ & $1.605\pm 0.055$ & $1.6\pm 0.02$  \\
$\beta_s$ & $-3.170 \pm 0.123$ & $-3.136 \pm 0.107$& $-3.088\pm 0.088$ & $-3.08$ \\
\hline
\end{tabular}
\caption{Constraints on the TF models of ``p16'', ``p10'' and ``p3'', derived from the Gaussian likelihood-based TF pipeline for the DC1 data. For each parameter, the {\it mean and the standard deviation} of posterior distribution are shown. The constraints on $r$ from all TF models are compatible with the DC1 input value $r = 0.023$ within 1$\sigma$ level.
For comparison, the corresponding input values utilized in DC1 simulations are presented, consistent with those from ~\cite{2016A&A...596A.109P}.}
\label{tab:TF}
\end{table}

\begin{figure}[htpb] 
    \centering
    \includegraphics[width=0.45\textwidth]{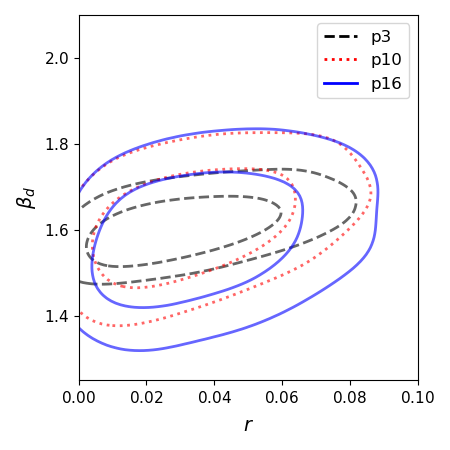}
    \includegraphics[width=0.45\textwidth]{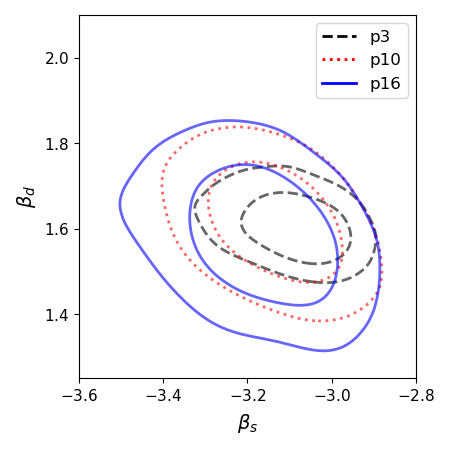}
    \caption{The marginalized $r$-$\beta_d$ (left panel) and $\beta_s$-$\beta_d$ (right panel) constraints from the TF pipeline with ``p3'', ``p10'' and ``p16'' foreground models. Inner and outer contours indicate 68.3\% CL and 95.4\% CL, respectively.}
\label{fig:mcmc}
\end{figure}


The MCMC runs converge well, with reduced $\chi^2$ (best $\chi^2$ / degrees of freedom in the data) of $218.2/231$, $230.9/242$ and $232.1/236$ for p3, p10 and p16, respectively. The marginalized posterior distributions of $r$, $\beta_s$ and $\beta_d$ are shown in Tab.~\ref{tab:TF} and  Fig.~\ref{fig:mcmc}. Notably, the complexity of foreground models does not have a significant impact on the constraint on $r$, which agrees well with the DC1 input value $r=0.023$.

For ``p10'', although spectral indices in harmonic space $\alpha_d$ and $\alpha_s$ are measured with respect to the power spectra suppressed by filtering, we may still compare them with the measurement from other experiments, which should be close to the DC1 simulation input, to gain some insights of the filtering effect and other sources of contamination. We find the dust spatial index ($D^{\rm dust}_\ell \propto  \ell^{\alpha_d}$) is $\alpha_d = -0.66 \pm 0.21$, very consistent with the BICEP/{\it Keck} and Planck derived value of $\alpha_d\approx -0.4$~\citep{Planck:2014skm,Planck:2018gnk,BICEP:2021xfz}. The spatial index for synchrotron ($D^{\rm syn}_\ell \propto \, \ell^{\alpha_s}$) is $\alpha_s=0.71\pm 0.28$, more than $2\sigma$ higher than the BICEP/{\it Keck} best fit of $\alpha_s=0$ (marginalizing over the range  of $[-1,0]$ and without a running parameter $\alpha_s^\prime$). Because what ``p10'' measures is the power spectra of TOD-filtered maps, and given the different parameterization and priors, the $\sim 2\sigma$ discrepancy here does not necessarily indicate significant inconsistency. More accurate analysis with additional information from $E$ maps, however, shows that there is indeed some minor biases in both the spatial index $\alpha_s$ and frequency index $\beta_s$ of synchrotron power. See Appendix~\ref{App:EE} and Fig.~\ref{fig:EEcosmo_contours1} for further discussion on this issue. 


For the more sophisticated model ``p16'', the estimate on $r$ and associated error bar are almost the same as those in ``p10'', giving confidence in a robust estimate of the value of $r$. We note that the frequency indices of $\beta_d$ and  $\beta_s$ are also consistent with those in ``p10'' (see Fig.~\ref{fig:mcmc}.) 
As the filtering-induced suppression effect is modeled in ``p16'', the $\alpha_s$ parameter in ``p16'' can be approximately interpreted as the spectral index of the synchrotron foreground component at the pivot scale. (Although its precise value may still be twisted by the point source contamination that is not yet modeled in ``p16''.) In this case we obtain $\alpha_s= 0.09\pm 0.52$, consistent with the BICEP/{\it Keck} + Planck best-fit value $\alpha_s=0$. Note that the filtering correction in ``p16'' is not determined by the simulations with and without filtering (see \ref{eq:Fl}), but is fitted directly from the observed power spectrum data. The transition scale of $\ell_F =74.1 \pm 11.2$ confirms the fact that the strongest suppression of the filtering lies in the low-$\ell$ bins. The derived value, $f_0=0.61\pm 0.26$, which characterizes the degree of the suppression and deviates from zero at 95\% CL, again implies a 2$\sigma$ detection of the filtering effect.  
 
We find a significant negative synchrotron-dust correlation, $\varepsilon=-0.35\pm 0.061$ for p10, deviating from zero by more than 5$\sigma$.
For p16, we obtain $\epsilon_2 = -0.66 \pm  0.20$ and  $\alpha_\epsilon = 0.181 \pm  0.094$. The $\sim 2\sigma$ hint of a positive running $\alpha_\epsilon$ indicates that the synchrotron-dust anti-correlation is more significant at low ells. 
Substituting the mean values of $\varepsilon_2$ and $\alpha_\epsilon$ into Eq.~\eqref{eq:eps_model}, we find that the mean $\varepsilon$ value obtained in ``p10'', namely $\varepsilon(\ell)=-0.35$, corresponds to a pivot scale $\ell=67\sim l_F$. This indicates that the synchrotron-dust anti-correlation measured in ``p10'' and ``p16'' could just be a filtering-induced observational artifact.

The ``p3'' model provides an estimate of the ideal case where both the synchrotron and dust spectra are perfectly known at reference frequencies (here $23\,\mathrm{GHz}$ for synchrotron and $353\,\mathrm{GHz}$ for dust). This optimistic assumption significantly reduces the uncertainties of the spectral indices on frequency ($\beta_s$ and $\beta_d$), but has negligible impact ($\sim$$10^{-3}$ shifts of mean and uncertainty) on $r$. For more ambitious target beyond AliCPT-1, say $\sigma(r)$$\sim$$10^{-3}$, however, we expect that additional observations of foreground emission at low and high frequencies would 
help with foreground modeling, and give confidence in the physical relevance of parametric models that fit the data well across an extended span of frequency channels.

\subsubsection{Quantifying foreground residuals for TF}

	\begin{figure}  
	\centering
	    \subfigure{
	        \includegraphics[trim=0cm 0cm 0.4cm 0.4cm,clip, height=0.25\textwidth]{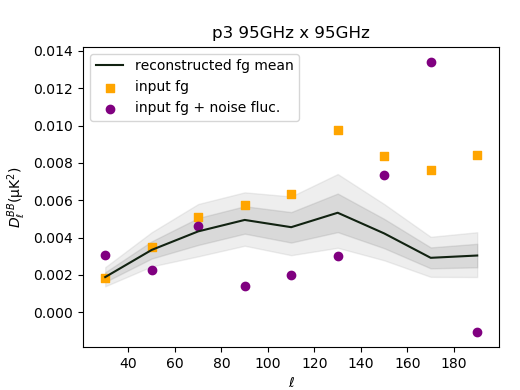}
	    }\hspace{-3mm}
	    \subfigure{
	        \includegraphics[trim=0.8cm 0cm 0.4cm 0.4cm,clip, height=0.25\textwidth]{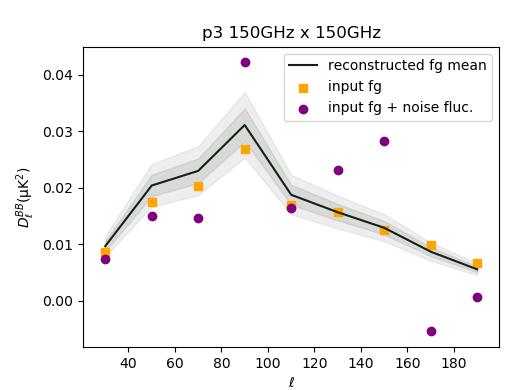}
	    }\hspace{-3mm}
	    \subfigure{
	        \includegraphics[trim=0.8cm 0cm 0.4cm 0.4cm,clip, height=0.25\textwidth]{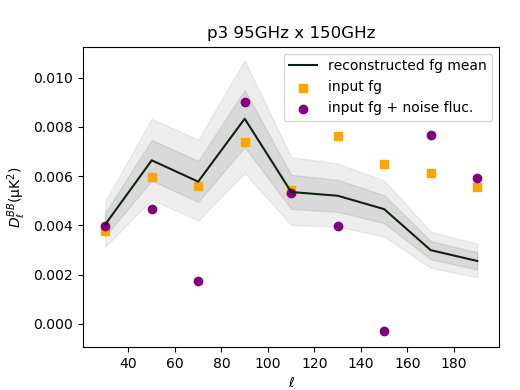}
	    }\\
	    \subfigure{
	        \includegraphics[trim=0cm 0cm 0.4cm 0.4cm,clip, height=0.25\textwidth]{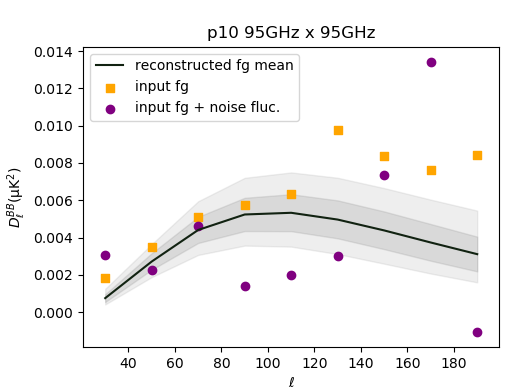}
	    }\hspace{-3mm}
	    \subfigure{
	        \includegraphics[trim=0.8cm 0cm 0.4cm 0.4cm,clip, height=0.25\textwidth]{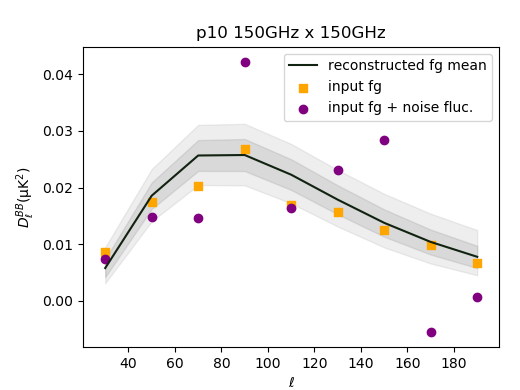}
	    }\hspace{-3mm}
	    \subfigure{
	      \includegraphics[trim=0.8cm 0cm 0.4cm 0.4cm,clip, height=0.25\textwidth]{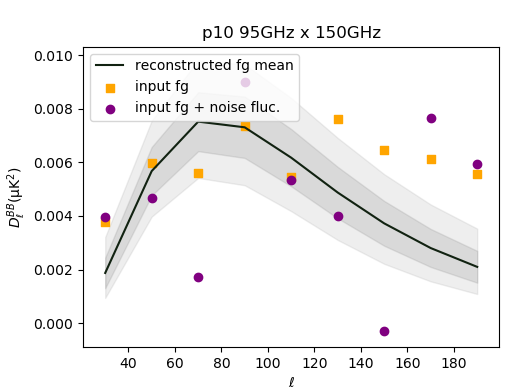}
            }\\
             \subfigure{
	        \includegraphics[trim=0cm 0cm 0.4cm 0.4cm,clip, height=0.25\textwidth]{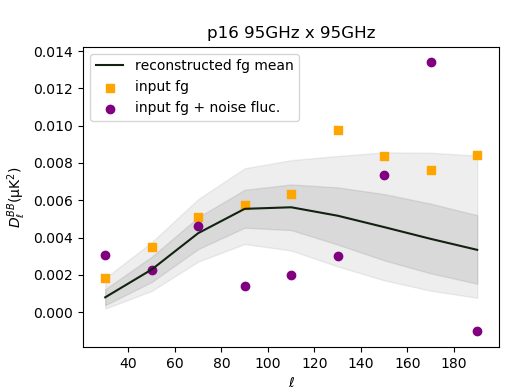}
	    }\hspace{-3mm}
	    \subfigure{
	        \includegraphics[trim=0.8cm 0cm 0.4cm 0.4cm,clip, height=0.25\textwidth]{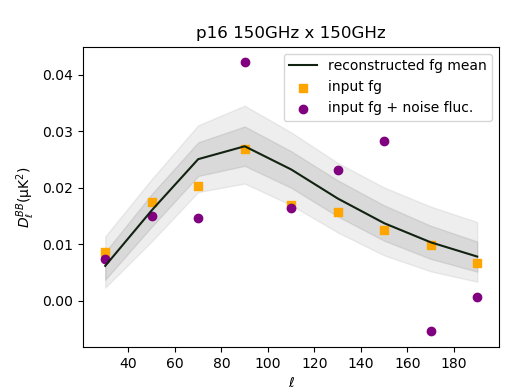}
	    }\hspace{-3mm}
	    \subfigure{
	        \includegraphics[trim=0.8cm 0cm 0.4cm 0.4cm,clip, height=0.25\textwidth]{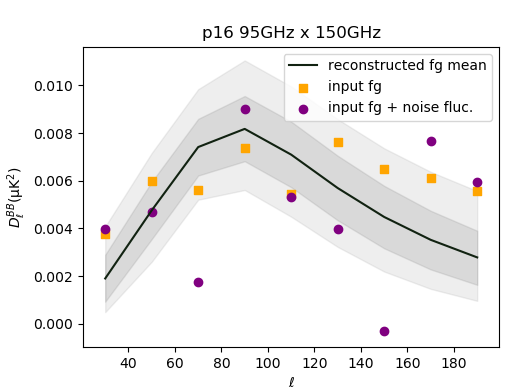}
	    }
	    \caption{Reconstructed $BB$ band powers of the foreground at 95$\times$95  (left), 150$\times$150 (middle) and 95$\times$150 (right) GHz, for the ``p3'' (top), ``p10'' (middle) and ``p16'' (bottom) mpdels, derived from the TF pipeline using the DC1 data. The solid black line is the mean of the ensemble of trajectories, whereas the dark-gray and light-gray  shaded regions indicate the 68.3\% (dark gray) and 95.4\% (light gray) confidence levels. Orange squares are the band powers of DC1 input foreground, and the purple dots shows the spectra with extra contribution from residual noise fluctuations. The reconstructed foreground band powers agree well with the DC1 input ones at $\ell \lesssim 120$ for 95$\times$95 and 95$\times$150, and $\ell\lesssim 200$ for 150$\times$150. The disagreement of 95$\times$95 and 95$\times$150 band powers at $\ell \gtrsim 120$ are due to the contribution from point sources that are not modeled in the TF pipeline. If we ignore the degeneracy between foreground and CMB, which is very subdominant here, the TF pipeline typically fits the foreground models to the combination of DC1 foreground and residual noise fluctuations, shown as purple dots. }
	\label{fig:fg_rec}
	\end{figure}

In practice, by reason of foreground complexity, limited frequency coverage, and contaminating instrumental noise, foreground cleaning always leaves a non-zero amount of foreground residuals in the reconstructed CMB $B$-mode, which may bias the estimate on $r$. In order to quantify the impact of foreground residuals on the estimate of $r$, we first examine the reconstructed foreground power spectra from the TF models, as compared with the simulation truth.

In the case of TF likelihoods for the DC1 data, joint probability distributions of $r$ and foreground parameters are derived by running the Markov Chain Monte Carlo (MCMC), and foreground power spectra are then computed from the corresponding parameters at each step of the final MCMC chain. Results are displayed in Fig.~\ref{fig:fg_rec} for the 95$\times$95, 150$\times$150 and 95$\times$150 GHz auto and cross spectra, for the ``p3'', ``p10'' and ``p16'' TF models. We focus on these three frequency pairs because the constraining power on $r$ mainly originates from these low-noise channels. 
Note that since the TF pipeline fits the total band powers, the deviation of the particular noise realization of DC1 from the simulation  mean, namely the residual noise fluctuations, may have an impact on the reconstructed foreground band powers.  
In all cases, the foreground band powers are well reconstructed at the large angular scales $\ell \lesssim 120$ that are relevant for the measurement of $r$. The good agreement indicates that residual noise fluctuations, although noticeable, do not have a significant impact on the fitting of a smooth component.
 At smaller angular scales ($\ell \gtrsim 120 $), the reconstructed band powers at 95$\times$95 and 95$\times$150 GHz channels appear to be lower than the input ones. The disagreement at smaller scales and lower frequencies are due to the contribution from radio and infrared point sources, which are not yet modeled in the TF pipeline\footnote{In the current version of TF pipeline we smooth very bright pixels that are more than 50 standard deviations off the mean, but most of the point sources remain untouched in the TF pipeline.} and typically have different spectral indices from the diffuse components. 

 As the model complexity increases, the scattering of the reconstructed foreground band powers increases from ``p3'' to ``p16''. The most flexible p16 model is capable of capturing some features of the high-$\ell$ rise in 95$\times$95 and 95$\times$150 GHz channels. This is because ``p16'' allows spectral running in both frequency and scale ($\ell$), which partially captures the property of point-source contribution. We leave further extensions of the TF model to include point source contamination to a future implementation of the component separation pipeline.

\subsection{Quantifying foreground residuals for GLS and cILC}\label{sect:fre_gc} 
\begin{figure}[htpb]
    \centering
    \includegraphics[width=0.7\textwidth]{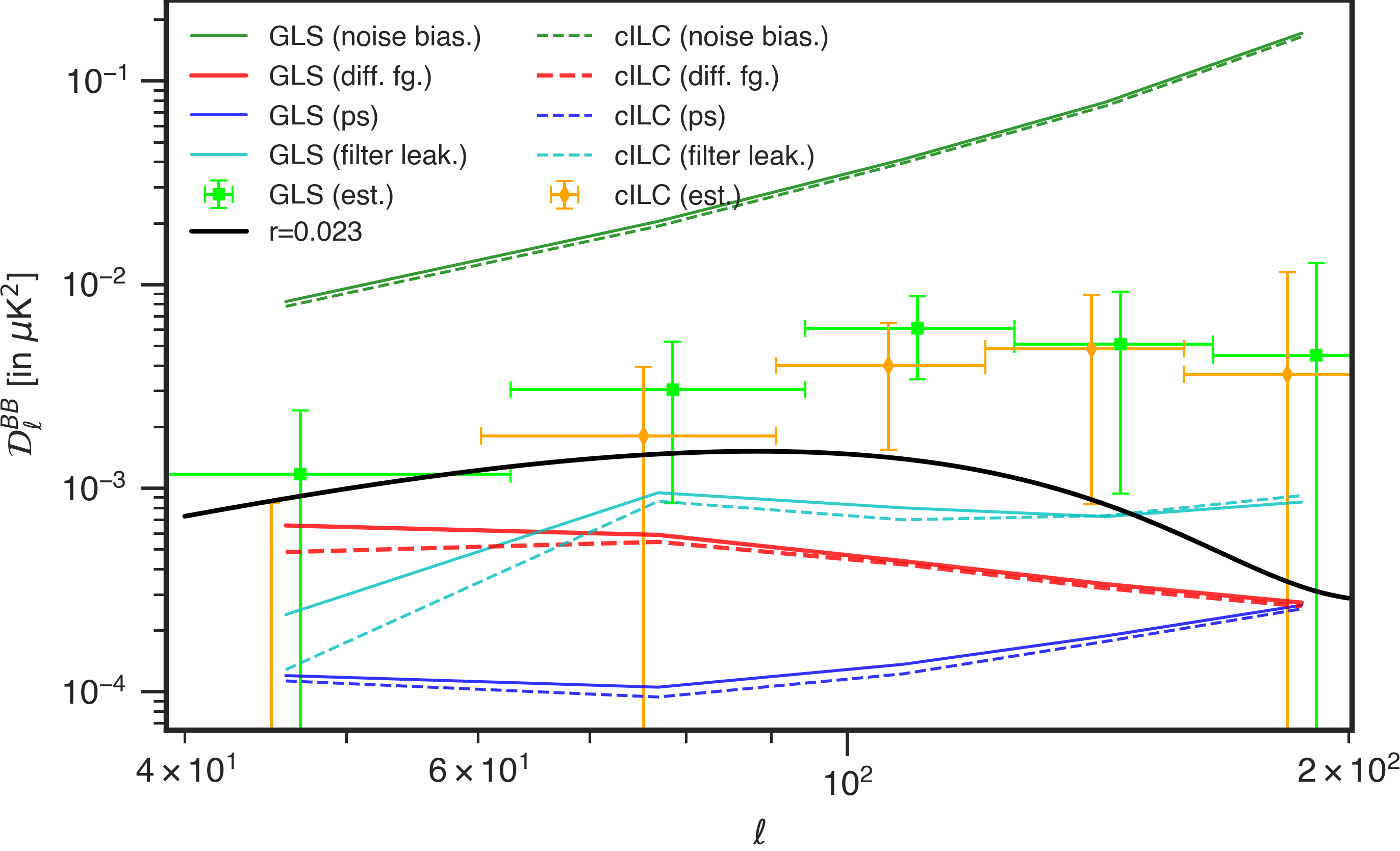}
    \caption{Comparison of residual contamination levels for DC1 $B$-maps obtained with GLS (solid) and cILC (dashed) for DC1 data, where the residuals are sourced by several components, including the noise bias (green), the diffuse foreground (red), the point sources (blue), and the $E$-to-$B$ filtering leakage (cyan). Among them, the noise and the filtering leakage are properly debiased by ancillary noise simulations and by the leakage template, respectively, while other contaminations are blind for the data. The GLS (and cILC with a slight shift on the $x$-axis for a clear display) debiased power spectrum is plotted with $x$-error bars denoting the $\ell$-bin width, and $y$-error bars denoting the uncertainty of the noise bias. Note that the residual levels are very comparable between the GLS and cILC methods, with the cILC yielding slightly lower biases for all residuals.}
    \label{fig:GLS-cILC_residuals}
\end{figure}

To quantitatively evaluate the bias level of foreground residuals for the cILC and GLS methods, a comparison of contamination levels in $BB$ band powers is performed. The possible components of contamination include the residual diffuse foreground, the residual point source, the residual noise, and the filtering leakage. These components can be categorized into two sets: residual diffuse and point source foreground emission is hard to evaluate for real data, while residual noise and E-to-B leakage from filtering and part-sky coverage can, in principle, be estimated given prior knowledge of the noise properties and of polarization $E$-modes, e.g., from Planck and WMAP measurements (see \cite{2021JCAP...02..036G} for a method minimizing $E$-to-$B$ leakage using ancillary full-sky E-mode maps).

The residual maps, which include the residual diffuse foreground, the residual point source, and the residual noise, are obtained using the same processing method as the DC1 maps. However, instead of having multi-frequency maps with all components combined, the input sky maps contain only each individual foreground or noise component. We preprocess these component maps by first masking out the point sources and then inpainting the masked spots. Subsequently, we apply the ``UNP'' mask to obtain the $B$-mode $a_{lm}$'s.

To obtain the foreground-cleaned residuals, we project out these $a_{lm}$'s for all frequencies by multiplying them with the frequency-dependent weights precomputed from the cILC/GLS pipelines for DC1 data. Finally, with the filtering correction and \texttt{NaMaster} estimation, we calculate the $BB$ band power of each residual component map in the range of $\ell \in [30, 200]$, as shown in Fig.~\ref{fig:GLS-cILC_residuals}.

The noise bias is estimated by averaging 50 residual noises whose input maps are from 50 ancillary noise simulations. The filtering leakage, however, is debiased by the cross-spectrum of the cILC/GLS map and the leakage template (see~\cite{paper4} for details). Since the residual noise is much larger than the filtering leakage, the uncertainty of the debiased power spectrum is approximately the standard deviation of 50 residual noise realizations, as plotted in the $y$-error bars of Fig.~\ref{fig:GLS-cILC_residuals}.

To evaluate the effect of the residual contaminations on the estimation of $r$, we fit $r$ with the residual band powers assuming a Gaussian likelihood: 
\begin{equation}
    - 2 \ln \mathcal{L}(r) = \sum_{\ell_b \ell_{b'}} \left[C^{\rm res}_{\ell_b} - rC_{\ell_b}^{r=1} \right]\left[M^{-1}_{\rm fid}\right]_{ \ell_b \ell_{b'}} \left[C^{\rm res}_{\ell_{b'}} - rC_{\ell_{b'}}^{r=1}\right]\,.
    \label{eq:gauss_llh}
\end{equation}
Here, $\boldsymbol M_{\rm fid}$ represents the covariance matrix of band powers computed from 300 fiducial simulations with $r_{\rm fid}=0.03$. The $r$-values fitted from the residual diffuse foreground and the residual point source are shown in Tab.~\ref{tab:cls-gls_res}, indicating that these two components contribute to a total bias of 0.012 on the estimated $r$-value for cILC and 0.014 for GLS. 

For GLS, the debiased components do not introduce biases on average but result in an uncertainty of $r$ determined by the uncertainty of the debiased power spectrum. For cILC, the estimated noise bias power spectrum is slightly higher than the actual noise bias on average, while the error has been found to be negligible compared to the uncertainty~\citep{Dou:2023vqg}. Therefore, we also fit $\sigma_{\rm noise}(r)$ using the standard deviation of the residual noise band powers with the same likelihood. We fit $\sigma_{\rm leak}(r)$ using the deviation of the estimated leakage bias from the true leakage power derived from the filtered $T/E$-only CMB maps. The total uncertainty of $r$ is computed as: $\sigma_{\rm tot}(r) = \sqrt{\sigma^2_{\rm noise}(r) + \sigma^2_{\rm leak}(r)}$. All the fittings are done with a prior of $r\in [-1, 1]$, as the effect of residuals on $r$ might be negative. As seen in Tab.~\ref{tab:cls-gls_res}, the total bias is smaller than the total uncertainty, implying that our estimation of $r$ is reliable even with these residuals.

\begin{table}[htpb]
    \centering
    \setlength{\tabcolsep}{1.1mm}
    \begin{tabular}{c|c c c c c c} 
        \hline \\[-1ex]
         \textbf{Component} &  $\Delta r_{\rm fg}$ &  $~\Delta r_{\rm ps}$   &  $~\Delta r_{\rm tot}$ & $\sigma_{\rm noise}$ & $\sigma_{\rm leak}$ & $\sigma_{\rm tot}$  \\[1ex] 
         \hline \\[-1ex]
          cILC & 0.010 & 0.002 & 0.012 & 0.035  & 0.004 & 0.035  \\[1ex] 
         \hline \\[-1ex]
          GLS  & 0.012 & 0.002 & 0.014 & 0.036  & 0.004 & 0.036   \\[1ex]  
        \hline
    \end{tabular}
    \caption{Summary of the impact of different component residuals on $r$ for the cILC and GLS pipelines in DC1 data, estimated through cILC/GLS likelihood evaluations for the band powers of residuals across all 5 $\ell$-bins in the range of $\ell \in [30, 200]$ (as shown in Fig.~\ref{fig:GLS-cILC_residuals}). The total biases ($\Delta r \equiv r - r_{\rm true}$) induced by the diffuse foregrounds (fg) and point sources (ps) are $\Delta r_{\rm tot}= 0.012/0.014$ for cILC/GLS, respectively. On the other hand, the debiased components, including the filtering leakage (leak) and the noise, result in total statistical uncertainties in $r$ of $\sigma_{\rm tot} = 0.035/0.036$ for cILC/GLS. Note that these fitting procedures were performed with a prior of $r \in [-1, 1]$. As seen, the total biases are well below the total uncertainties $\sigma_{\rm tot}$.
      }
    \label{tab:cls-gls_res}
\end{table}

\subsection{Comparing \texorpdfstring{$r$}{r}-values derived with and without foregrounds \label{sect:fgr}} 

In the previous analysis, we focused on analyzing various residual levels. Here, we will further examine the impact of foreground residuals on $r$ for our pipelines, including TF, cILC, GLS, and ABS,  by comparing the measured $r$-values from the DC1 data with and without foreground.


To do so, a test with foreground-free DC1 data was performed, in which all the map-level foreground components at each frequency were entirely removed, while keeping the noise and CMB sky maps unchanged. As such, the derived $r$ in the foreground-free case might be reasonably considered to be the expected value in the case of perfect foreground cleaning. Comparing its value with that from the fiducial DC1 data allows us to further quantify the impact of foreground residuals.

Using the Gaussian likelihood-based TF pipeline and adopting the same covariance matrix as used in the fiducial DC1 (for a fair comparison), we find a mean value of $r$ for foreground-free DC1 data of $r_{\rm free}\simeq 0.011\pm 0.02$ at 68\% confidence level. Foreground models are not considered in the likelihood fit for this estimate, as no foreground emission is present.  Here we used one realization of the CMB map in DC1. Due to significant noise fluctuations, there is a large statistical uncertainty of approximately $0.02$ for the fitted $r$. Thus, deviating from $r_{\rm true} = 0.023$ by $0.01$ is highly plausible. Note that while this value of $r_{\rm free}$ is derived from TF, other pipelines obtain almost the same value in the foreground-free case ($r_{\rm free}=0.013$--$0.015$ for the other pipelines), so we use this value as a benchmark to discuss the contamination of the foreground residuals for estimates with all foreground-cleaning methods.


Let us go back and re-examine the results for the fiducial DC1 data in the presence of the foreground. As seen in Tab.~\ref{tab:fg_res}, all the pipelines give compatible estimates for $r$, varying from 0.031 (``p3'') to 0.036 (ABS) with statistical uncertainties at the level of $\sigma(r)\approx0.02$. Compared with $r_{\rm free}$ derived in the foreground-free case, we find that the derived $r$-values in the DC1 data from the different pipelines deviate from $r_{\rm free}$ by $\Delta r\in$ [+0.014, +0.025]. Thus, the foreground residuals from the different foreground cleaning methods would lead to positive contributions to $BB$ band powers (at the level of $\Delta r\sim +0.01$), slightly increasing the $r$ estimate. The results obtained from the TF, GLS, and cILC pipelines are consistent with the findings presented in Sects.~\ref{sect:res_TF} and~\ref{sect:fre_gc}.

We observe that the relative deviations with respect to the statistical errors, $(r-r_{\rm free})/\sigma(r)$ are in the range of $[+0.88, +1.15]$ for all pipelines (see the last row of  Tab.~\ref{tab:fg_res}). Hence, foreground residuals slightly bias the $r$ estimate at the $1\sigma$ level for the current DC1 simulation. This bias is lower than the standard deviation of the estimator, and thus not very sisgnificant statistically. We conclude that our foreground cleaning pipelines are reasonably robust and effective for the AliCPT-1.

\begin{table}[htpb]
    \centering
    \setlength{\tabcolsep}{1.1mm}
    \begin{tabular}{c|c c c c c c c} 
        \hline \\[-1ex]
         \textbf{Method} &  TF (p3)&  TF (p10) &  TF (p16) & ABS & GLS & cILC  \\[1ex] 
         \hline \\[-1ex]
         $r/\sigma(r)$ & $0.031/0.019$ & $0.034/0.020$ & $0.035/0.021$& $0.036/0.025$& $0.030/0.020$ &$0.025/0.016$      \\[1ex] 
         \hline \\[-1ex]
         $ \left(r-r_{\rm free}\right)/\sigma(r)$ & $+1.05$ &$+1.15$&$+1.14$ &$+0.99$&$+0.95$&$+0.88$     \\[1ex]  
        \hline
    \end{tabular}
    \caption{ {\it Top:} the mean and associated statistical uncertainty of $r$ derived from each pipeline for the fiducial DC1 data.  {\it Bottom:} the relative deviation calculated based on $r$-values of DC1 and foreground-free data, $(r-r_{\rm free})/\sigma(r)$, where $r_{\rm free}=0.011$ is derived from the TF pipeline in the foreground-free case. The relative deviations fall within the range of $[0.88, 1.15]$, which indicates that the level of residual contamination is not statistically significant.}
    \label{tab:fg_res}
\end{table}

\subsection{A further examination: impact of varying sky patch on \texorpdfstring{$r$}{r}  estimates}

\begin{figure}
    \centering
    \includegraphics[width=0.4\textwidth]{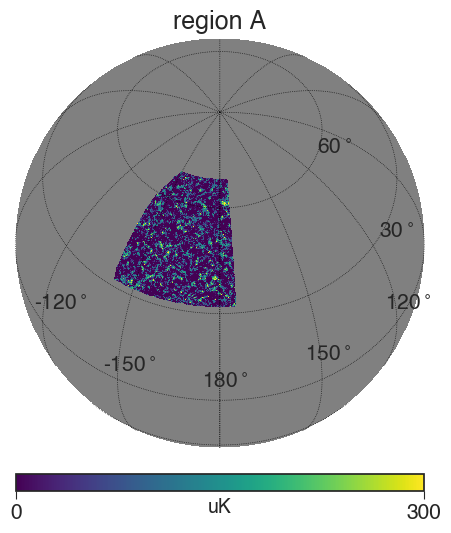}
    \includegraphics[width=0.4\textwidth]{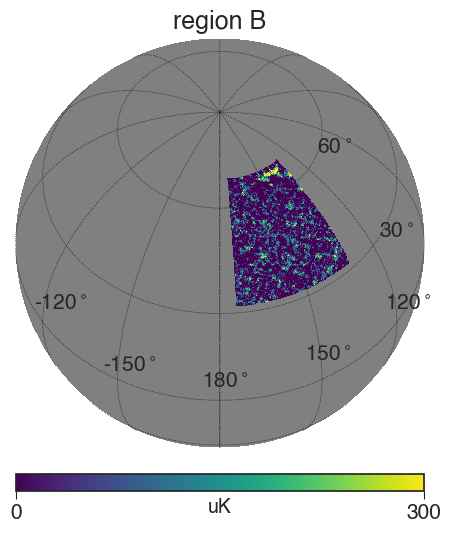}
    \includegraphics[width=0.4\textwidth]{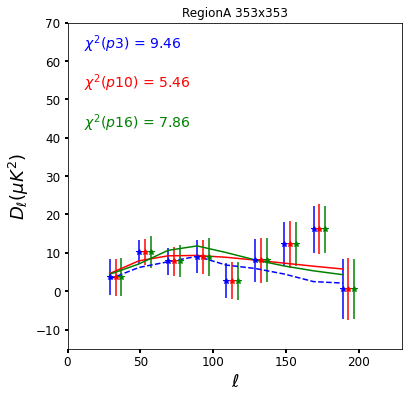}
    \includegraphics[width=0.4\textwidth]{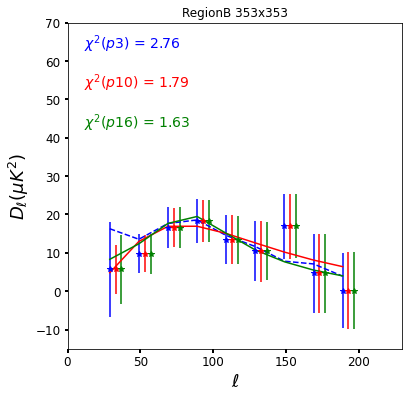}
    \caption{{\it Top}: the observed intensity (Stokes $I$) sky at 353 GHz from the simulation for the ``region A'' (left) and ``region B'' (right), separated by splitting the fiduical TF mask along RA$\simeq 170^{\circ}$, both of which have almost equal sky coverage and geometry, ensuring a fair comparison between the two. {\it Bottom}: the corresponding $BB$ band powers and associated $1\sigma$ uncertainties for the ``region A'' (right) and ``region B'' (left) derived from the best-fit models of ``p3'' (dotted blue) , ``p10'' (solid red) and ``p16''(solid green) in $50 \leq \ell \leq 200$ at $353\times 353$ GHz.}
    \label{fig:maskab}
\end{figure}

A further examination of the foreground residuals was performed by analyzing $r$-values measured in two distinct sky patches, each with distinct foreground contamination. The process involved creating two masks by dividing the fiducial TF mask along RA$\simeq 170^{\circ}$. Subsequently, the resulting two sub-patches (denoted by ``region A'' and ``region B'') have nearly equal areas and similar geometries, as depicted in the top panels of Fig.~\ref{fig:maskab}.

As observed from the splitting, the standard deviations of the intensity (Stokes $I$) fluctuations in the DC1 353 GHz maps for ``region A'' and ``region B'' are $\sigma_{I}^A=107.3~\mu$K and $\sigma_I^B=111.9~\mu$K, respectively. This suggests that ``region B'' exhibits slightly higher foreground contamination. For further validation, we have also examined the intensity fluctuations of the pure foreground components in these two sub-patches to eliminate contributions from noise and CMB. The results show $\sigma^{I}_A=57.2~\mu$K and $\sigma^I_B=68~\mu$K for ``region A'' and ``region B'', respectively. These findings indeed indicate that ``region B'' is affected by slightly higher dust contamination than ``region A''. 

Since intensity and polarization both depend on the dust column density, they are 
not independent. 
It is reasonable to expect that the dust contamination level in ``region B'' is also stronger in polarization, and the DC1 pure foreground simulation confirms this expectation. The fluctuation levels of $Q$ and $U$ maps at 353 GHz are $\sigma_Q^A=3.7~\mu$K and $\sigma_U^A=4.1~\mu$K for ``region A'', while ``region B'' exhibits higher values of $\sigma_Q^B=5.7~\mu$K and $\sigma_U^B=4.9~\mu$K. This clearly indicates that ``region A'' has approximately 34--17\% lower polarization contamination compared to ``region B''.

Note that the noisy Planck polarization data at 353 GHz themselves do not allow us to derive a reliable result about the dust contamination level in practice. Therefore, to determine the degree of foreground contamination, the choice of these two regions is mainly based on the level of intensity fluctuations rather than the polarization.

The TF, ABS, GLS, and cILC pipelines were re-run for ``region A'' and ``region B'' using the full multi-frequency DC1 data. The predicted band powers from the best-fit TF models are shown in the bottom panels of Fig.~\ref{fig:maskab} as an example. As expected, the band powers in ``region B'' appear to be higher than those in ``region A'', further confirming the stronger dust contamination in the former. For example, the ratio of band powers $D^{A}_\ell/D^{B}_\ell$ is approximately $10/15$ when $\ell\in [50,100]$.

\begin{table*}[htb]
    \footnotesize
    \centering
    \caption{Mean values of $r$ and their corresponding $1\sigma$ standard deviations estimated from the TF, ABS, GLS, and cILC pipelines for ``region A'' and ``region B''.} 
    \label{table:r-maskab}
    \begin{tabular}{ccccccc}
\hline
{\diagbox{sky}{$r$}{model}}  &p3& p10 & p16& ABS & GLS & cILC \\
\hline
\\[-1em]
\multirow{1}{*}{{region A}} & $0.037\pm 0.027$ & $0.048\pm0.030$  &   $0.046\pm 0.032$  & $0.023\pm 0.020$ & $0.021\pm 0.016$ &$0.024\pm 0.020$ \\
\\[-1em]
\hline
\\[-1em]
\multirow{1}{*}{{region B}} & $0.042 \pm 0.031$  & $0.068 \pm0.038$  &  $0.080\pm 0.041$ &  $0.070\pm 0.039$ &  $0.033\pm 0.025$ & $0.039\pm 0.028$  \\
\\[-1em]
\hline
\end{tabular}
\end{table*}

The derived $r$-values are listed in Tab.~\ref{table:r-maskab}. These values were obtained from the analysis using independent pipelines and foreground models for the two regions. We find that: 1) the relatively higher dust contamination in ``region B'' shifts $r$ towards higher values while slightly increasing its statistical uncertainty; 2) when compared with $\sigma(r)$ in Tab.~\ref{tab:fg_res}, the statistical error for each TF model (``p3'', ``p10'', and ``p16'') is roughly increased by a factor of $\sqrt{2}$ due to the approximately halved area of the sky patch; 3) the statistical errors derived from ABS/GLS/cILC for ``region A'', however, remain unchanged compared to the case of the full-sky patch.

The main reason why errors remain unchanged for ABS/GLS/cILC is that the estimated values of $r$ shift towards zero, which is smaller than those in the case of the full patch, as seen in Tab.~\ref{tab:cls-gls_res}. Since the combination of the likelihood function is left-shifted (towards zero), and a positive prior on $r$ would effectively shift the posterior distribution towards values greater than zero, shrinking the confidence interval. This leads to the observed statistical errors remaining almost unchanged or even slightly smaller. This can be confirmed quantitatively from the results in Tab.~\ref{tab:fg_res}, where the statistical errors are based on the $r\in [-1,1]$ prior. Therefore, the errors are all larger than the results of the positive prior (see Tab.~\ref{tab:cls-gls_res}). However, if the estimate remains the same, then the error in smaller patch does increase compared to the full patch case, as can be seen from the results of cILC. The derived $r$ value for cILC in ``region A'' is only 0.001 smaller than that in the full patch, resulting in a 25\% increase in the statistical error.

\begin{figure}[htpb]
    \centering
    \includegraphics[width=0.55\textwidth]{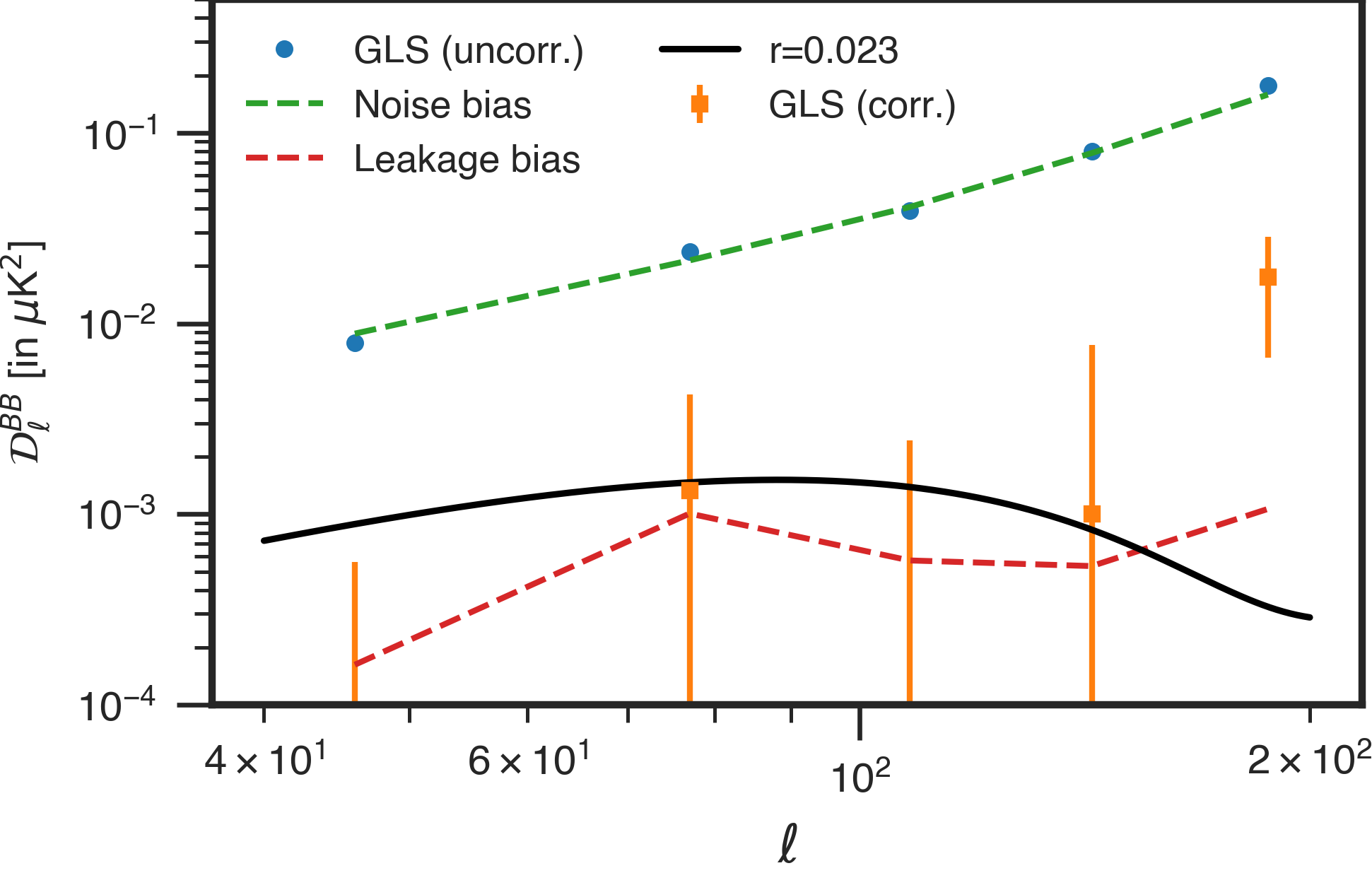}
    \includegraphics[width=0.35\textwidth]{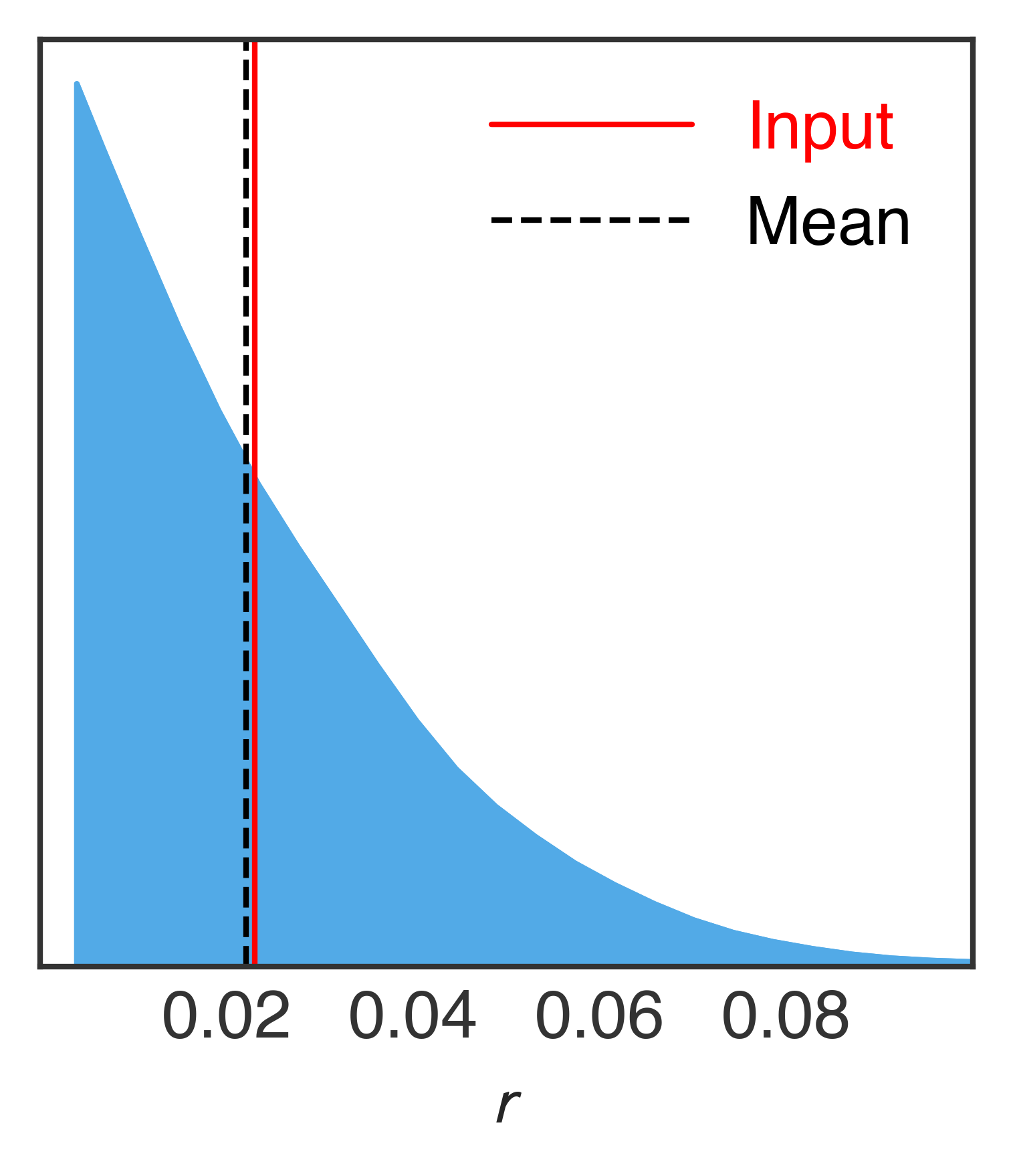}
    \caption{Power spectra (left) and the posterior distribution (right) of the tensor-to-scalar ratio $r$ obtained using the GLS pipeline for the sky sub-patch labeled as ``region A''. {\it Left}: the blue markers represent the power spectrum without noise or leakage bias correction, whereas the orange markers illustrate the final power spectrum after debiasing for both noise and leakage bias.
    For comparison, the noise bias is represented by the green dashed line, while the red dashed line indicates the leakage bias. {\it Right}:
    the solid red line represents the input $r$-value of 0.023 for DC1 data, while the dashed black line illustrates the mean value ($r=0.021$) of the posterior distribution (blue shaded) derived from the likelihood analysis under a positive prior on $r$.}
    \label{fig:GLS-patchA}
\end{figure}

To provide further clarification, we explain why the uncertainty of GLS in ``region A'' is smaller than that in the all-sky case. The band powers estimated by GLS for ``region A'' and the posterior probability distribution of $r$ are shown in Fig.~\ref{fig:GLS-patchA}. 
Due to the substantial statistical uncertainty of the noise in DC1 data, the band powers obtained by GLS appear with negative amplitudes even after correcting (subtracting) the mean power spectra of the noise and leakage. Incorporating these negative data points into the likelihood analysis while enforcing a positive prior on $r$ can cause the posterior distribution of $r$ to shift towards 0, consequently shrinking the $1\sigma$ confidence interval. For the fiducial patch with lower noise uncertainties, the GLS-recovered $BB$ band powers are all positive, resulting in a broader confidence interval compared to that in ``region A''. Additionally, we employed a different apodization for the mask, using the 6$^\circ$ \texttt{C2} apodized mask for ``region A'' instead of the inverse noise variance mask used in the GLS pipeline to alter the noise uncertainty. However, the negative amplitudes still appear, and the posterior distribution of $r$ remains unchanged.

To obtain a qualitative assessment of the tensions in $r$ arising from the foreground residuals, we use the following estimator~\citep{Lin:2017ikq}---under a Gaussian statistical assumption---to quantify the tension (measured in the number of $\sigma$'s) between measurements in two different cases, denoted as $i$ and $j$, 
\begin{equation}\label{eq:r-tension}
T_r(i,j)= \frac{|\bar{r}_i-\bar{r}_j|}{\sqrt{\sigma^2_i+\sigma^2_j}}\,,
\end{equation}
where $\bar{r}$ and $\sigma$ represent the mean and variance of the marginalized posterior of $r$, respectively. Qualitatively, values of $T_{r}<1.4$ are usually interpreted as indicating no significant tension, while values of $1.4<T_{r}<2.2$ and $2.2<T_{r}<3.1$, as well as $T_{r}>3.1$, indicate weak, moderate, and strong tensions, respectively.

\begin{table*}[htb]
    \footnotesize
    \centering
    \caption{Tensions of $r$, denoted as $T_r(i,j)$ as defined in Eq.~\ref{eq:r-tension}, between measurements in various cases. The cases include using independent foreground-cleaning pipelines and choosing different foreground models (``p3'', ``p10'', ``p16'') and sky patches (``region A'' and ``region B''). The tension matrix is symmetric, so only the upper diagonal elements are mentioned here, and the corresponding lower ones are replaced by ``*''.}
    \label{table:t-maskab}
    \scalebox{0.8}{
    \begin{tabular}{lcccccccccccc}
\hline
{{Case}}  &p3+A & p3+B & p10+A &p10+B & p16+A &p16+B & ABS+A& ABS+B & cILC+A& cILC+B & GLS+A& GLS+B 
\\
\hline
\multirow{1}{*}{{p3+A}} &  0  & 0.12     &0.27     &  0.66   & 0.21 & 0.87 & 0.41& 0.69 & 0.39& 0.05 &0.51 & 0.11
\\
\hline
\multirow{1}{*}{{p3+B}} &*        & 0     & 0.14      &  0.53   & 0.09 & 0.73 &  0.51 & 0.56 &0.49 & 0.07 &0.60 & 0.22
\\
\hline
\multirow{1}{*}{{p10+A}} &*        & *   &  0         &  0.41   & 0.05 & 0.63 & 0.69 & 0.45 & 0.67& 0.22 & 0.79 & 0.38
\\
\hline
\multirow{1}{*}{{p10+B}} &*        & *       & *      & 0        & 0.44 & 0.21 &  1.04 & 0.04 &1.03  &0.61 & 1.13 &0.77
\\
\hline
\multirow{1}{*}{{p16+A}} & *        & *       & *     &   *      & 0     &  0.65 & 0.47 &0.61 & 0.59&0.16&0.69&0.32
\\
\hline
\multirow{1}{*}{{p16+B}} & *        & *       & *    &    *    &  *  & 0  & 1.25 &  0.18 &1.24&0.82 &1.34 &0.98
\\
\hline
\multirow{1}{*}{{ABS+A}} & *        & *       & *    &    *    &  *  & *& 0 &  1.07 & 0.04&0.46 &0.08 & 0.31
\\
\hline
\multirow{1}{*}{{ABS+B}} & *        & *       & *    &    *    &  *  & *  & *& 0& 1.06&0.64&1.16&0.79
\\
\hline
\multirow{1}{*}{{cILC+A}} & *        & *       & *    &    *    &  *  & *  & *& * &0  &0.44 &0.12 & 0.29
\\
\hline
\multirow{1}{*}{{cILC+B}} & *        & *       & *    &    *    &  *  & *  & *& *&*&0 &0.56 & 0.16
\\
\hline
\multirow{1}{*}{{GLS+A}} & *        & *       & *    &    *    &  *  & *  & *& * &*  &* & 0 & 0.40
\\
\hline
\multirow{1}{*}{{GLS+B}} & *        & *       & *    &    *    &  *  & *  & *& *&*&* & * & 0

\\
\hline
\end{tabular}
}
\end{table*}

As shown in Tab.~\ref{table:t-maskab}, the tensions between ``region A'' and ``region B'' in all cases consistently satisfy the requirement of $T_r<1.4$, regardless of the chosen pipeline. This indicates that there is no significant tension among the estimates of $r$, and they remain consistent within $1\sigma$. In other words, using different sky patches and choosing various foreground models have no statistical impact (within 1$\sigma$ consistency) on the cosmological result, specifically the constraint on $r$. This strong consistency in the estimations is evident.

Consequently, we conclude that, under the nominal sensitivity in the DC1 data, the proposed foreground-cleaning pipelines, TF/ABS/GLS/cILC, effectively subtract the foreground contamination and provide consistent results without introducing any noticeable foreground residuals that could impact the estimate of $r$.

\section{Conclusion}
\label{sec:con}
Separating the foreground from the cosmological signal, particularly on large angular scales, is one of the most challenging tasks in CMB data analysis. The precise measurement of the CMB primordial $B$-modes relies on successfully cleaning foreground contamination. To achieve this, we have developed several independent analysis pipelines for the AliCPT-1 experiment, each utilizing different foreground-cleaning strategies. This paper focuses on validating each pipeline and ensuring consistency while studying the impacts of foreground residuals on CMB band powers measured from various tests. For a detailed investigation of the estimate on $r$, please refer to~\citep{paper4}.

In this study, we present in detail the template fitting (TF) pipeline, which effectively separates CMB signals from foreground contamination in a parametric manner. We describe the component modeling and each computation step involved in the TF analysis. The TF pipeline employs a Gaussian likelihood estimator to constrain the parameters for both CMB and foreground components, based on observed multi-frequency auto/cross $BB$ band powers. In the TF pipeline, we use three different foreground parameterization methods (``p3'', ``p10'', and ``p16''), which range from simple to complex, to model the frequency and spatial distributions of the synchrotron and dust components. To account for filtering effects in the CMB, we construct the CMB template using end-to-end simulations and utilize ``forward modeling''. By fitting the DC1 data to these templates, we can accurately determine the tensor-to-scalar ratio, $r$.

We find that increasing the number of TF foreground parameters leads to slight changes in the estimated values of $r$. However, all of these estimates are in good agreement with the true value within  1$\sigma$ errors. Moreover, we find that the TF models effectively capture the spectral and spatial shapes of the diffuse synchrotron and dust components. The band-power reconstruction at each frequency pair from these models agrees well with the inputs at the scale of interest of $r$. This indicates that the presence of foreground residuals does not significantly bias the measured value of $r$. Consequently, the estimation from the TF pipeline is reliable, and all the proposed foreground models are valid.

Next, we assess the impact of foreground residuals on the CMB $BB$ band powers for the GLS and cILC pipelines. To achieve this, we utilize the sky maps for each individual component in the DC1 simulation, including the diffuse foreground, point sources, filtering leakage, and noise. By passing these components through both pipelines, we accurately determine the contamination levels of the residuals. We find that the total bias is smaller than the total uncertainty for either GLS or cILC, indicating that our estimation of $r$ is reliable even in the presence of these residuals.
 

A test with foreground-free DC1 data was performed to quantify the impact of foreground separation and foreground residuals. For all the pipelines, we have compared the derived $r$-values between the fiducial and the foreground-free cases, and we find that the relative deviations from the true one, with respect to the statistical errors, are always in the range of $[0.88, 1.15]$ for the pipelines of TF, ABS, cILC, and GLS. Therefore, the foreground residuals can slightly shift the $r$ to a higher value, and the corresponding contamination level in $r$ is consistent with the statistical error, well within $2\sigma$.

For further testing, we have split the observed sky into two sub-regions, each with an equal area. By dividing along a particular direction, dust contamination is visibly stronger in one region than in the other at the re-observed 353 GHz map. From the TF likelihood analysis, we find that the reconstructed $BB$ power spectra from the best-fit model parameters are in good agreement with the measurements in both sub-regions, and no statistically significant anomalies are found. When comparing the results from TF/ABS/cILC/GLS pipelines for these two sub-regions, we assess the tensions in $r$ between the different cases and observe a good 1$\sigma$ consistency among all of them.

All these tests strongly support the validity of the foreground cleaning strategies, including TF/ABS/cILC/GLS, in our data analysis pipelines. The results indicate that any potential foreground residuals do not have a statistically significant bias on the estimation of $r$ when using the simulated AliCPT-1 data for one observation season. As we continue to optimize our data analysis pipelines and make advancements in the AliCPT project, such as expanding our frequency coverage and enhancing sensitivity, we are highly optimistic about achieving significant improvements in our measurements. With these advancements, we are confident that we will soon achieve a precise and accurate measurement of the extremely faint primordial $B$-mode signal.


\appendix
\section{Masks and apodization}\label{app:mask}
\begin{figure}[htpb]
    \centering
    \includegraphics[width=0.5\textwidth]{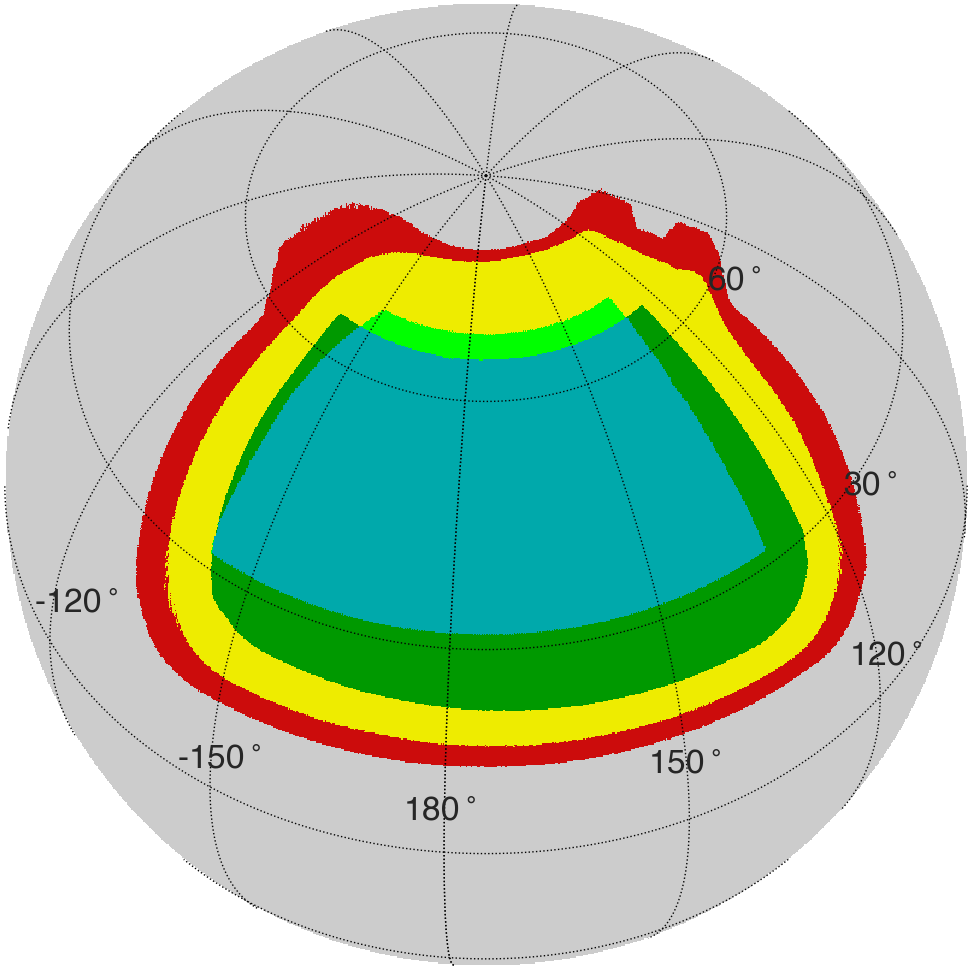}
    \caption{Summary of binary masks used for foreground cleaning in this study, including the DC1 observation patch (red), the '20$\mu$K' mask (yellow) used in the $T$- and $E$-mode analysis from the ABS and the Needlet-ILC (NILC) pipelines, and the 'UNP' mask (dark green) used in the GLS and cILC pipelines for $B$-mode analysis, as well as the 'ABSapo' mask (green) used in the ABS and TF pipelines for $B$-mode analysis. In practice, the apodization schemes of $6^\circ$ \texttt{C2}, $4^\circ$ \texttt{C2} and the inverse noise variance weighted  have  been applied to these binary  masks of  '20$\mu$K', 'ABSapo', ``UNP'', respectively.}
    \label{fig:bin_mask}
\end{figure}

Here we summarize the masks and the apodization scheme used in the study. The different pipelines in the analysis are independent, so that, for foreground cleaning and signal estimation they adopt different masks (not guaranteed to be optimal, but extensively tested). The default binary masks (unless otherwise specified) used in this paper are shown in Fig.~\ref{fig:bin_mask} and, in practice, we have applied the same \texttt{C2} scheme but with different apodization scales into the masks before computing the power spectra, in order to minimize spurious signals from a sharp edge. By convention, the \texttt{C2} function, $f$, implemented in \texttt{Namaster} tapers a mask with an given apodization scale of $\theta^*$ through  
\begin{equation}
f=\left\{\begin{array}{cc}
\frac{1}{2}[1-\cos (\pi x)] & x<1 \\
1 & \text { otherwise }
\end{array}\right.
\end{equation}
where $x=\sqrt{(1-\cos \theta) /\left(1-\cos \theta_{*}\right)}$ and $\theta$ is the angular separation of the pixel from the nearest masked pixel. Now let us outline the adopted masks as follows.

\emph{``20$\mu$K'' mask:} the accepted sky fraction is about 10\%, produced by removing any pixel with noise standard deviation above 20 $\mu$K-pixel at \texttt{NSIDE}=1024 for 150 GHz channel. It is used for estimating $TT$ and $EE$ and $TE$ spectra from the ABS and the NILC pipelines, with the apodization scale of $6^\circ$.

\emph{``UNP'' mask:} similar to the ``20$\mu$K'' mask, but further removing the sky above declination of $65^\circ$ to mitigate foreground contamination, the accepted sky fraction is about 6.7\%. It is used in the GLS and cILC pipelines for the $B$-mode estimation, with the apodization scales of $4^\circ$.

\emph{``ABSapo'' mask:} it is produced by masking the sky that has smaller than 20\% of the maximum hit count, along with removing some higher foreground areas. As such, the accepted sky fraction is about 5\%. It is used in both the ABS and the TF pipelines for estimating CMB $B$-modes, with the apodization scales of $4^\circ$.

\subsection{Point source processing} 
\label{sec:point_source}
The ABS and TF pipelines do not treat the point sources separately. In contrast, the GLS and cILC pipelines share a common pre-processing stage, wherein they mask point sources and inpaint the point source masking. The discussion of this process is as follows.

\begin{figure}[htpb]
    \centering
    \includegraphics[width=0.45\textwidth]{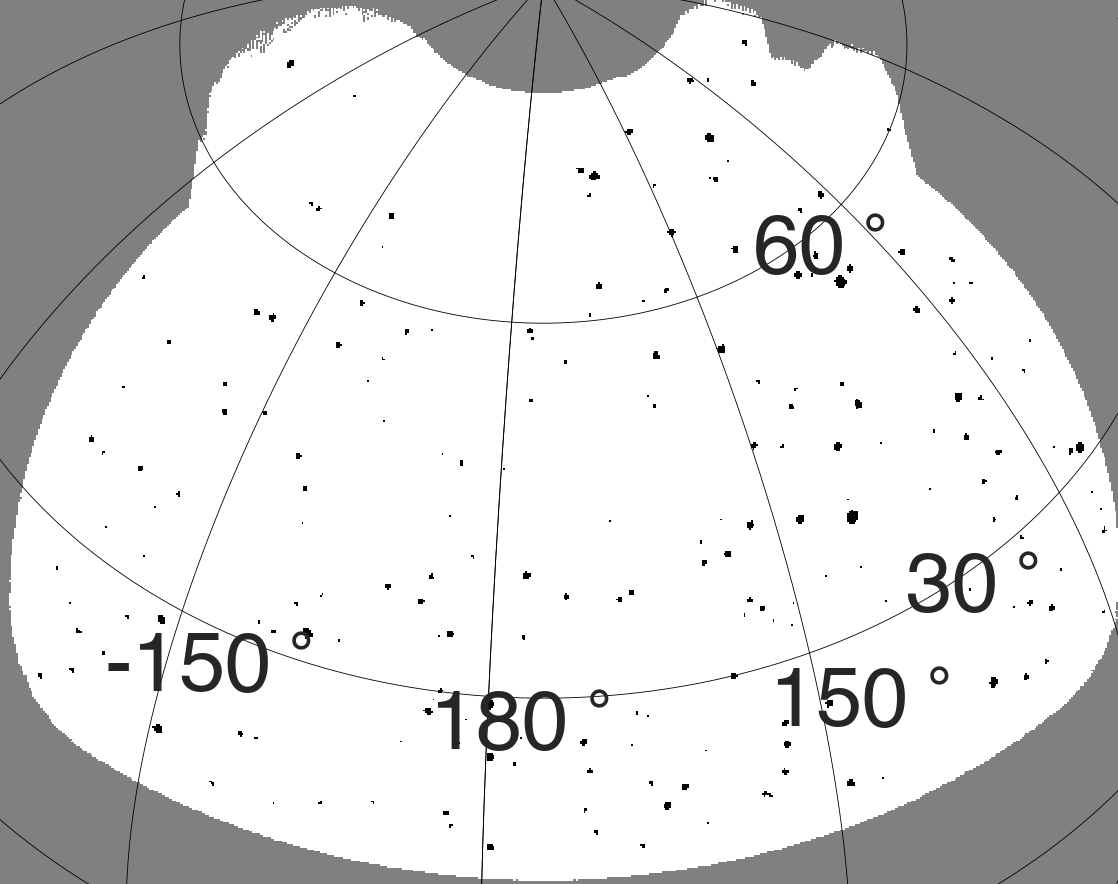}
    \includegraphics[width=0.45\textwidth]{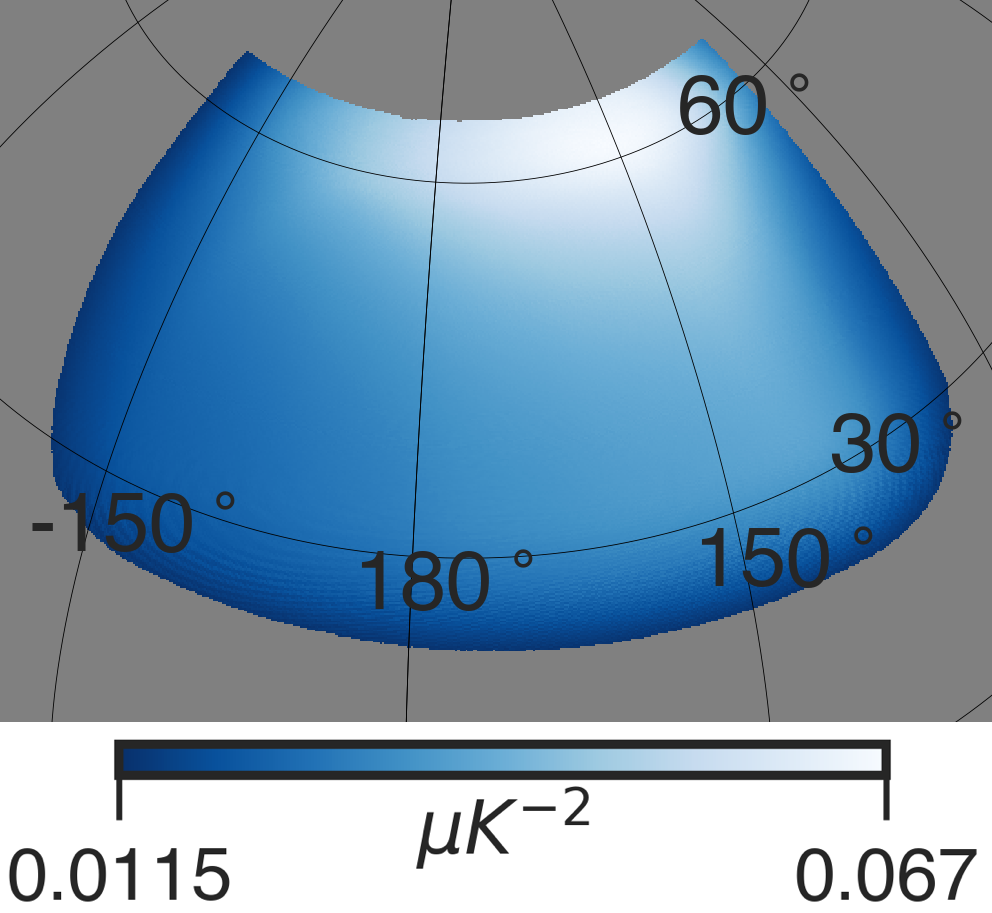}
    \caption{Left: point source mask for DC1 obtained from HFI 100 GHz temperature map with the outlier method. Right: the ``UNP'' mask for polarized foreground and noise with inverse noise variance weights.}
    \label{fig:masks}
\end{figure}

The point source mask is obtained from unfiltered HFI 100 GHz temperature  map using the outlier method~\citep{Tegmark:1998} to identify the resolved point sources. This map is without any TOD filtering. For the actual sky we can simply use the Planck point source map but in DC1 simulation we have different point source contamination. We compute the debeamed power spectrum of the diffuse components in the temperature map from the actual HFI 100 GHz map, masked with HFI point source mask. We call this $C_\ell^{\rm tot}$. Then harmonic space smoothing filter is defined as:
\begin{equation}
    S_\ell \propto \frac{1}{C_\ell^{\rm tot} B_\ell}, 
\end{equation}
where $B_\ell$ is the Gaussian beam of the HFI 100 GHz map. We obtain the filtered DC1 HFI 100 GHz temperature map after applying this filter. We then smooth the absolute valued filtered map with $0.5^\circ$ beam. We compute the mean and standard deviation of this map. We finally obtain the point source mask by masking pixels that are $3\sigma$ from the mean in the smoothed HFI map. The point source mask for the AliCPT patch is shown in Fig.~\ref{fig:masks} on the left.
We preprocess DC1 maps prior to component separation for the NILC, GLS and the cILC pipelines. For $T$ map we find point sources making contributions for $\ell > 1100$, while it does not effect the $E$-mode power spectra. However, we mask the $IQU$ maps with the point source mask and obtain $T$/$E$ maps. The pixels that are masked are then inpainted using the unmasked pixel surrounding the masked pixels. To do this we isolate the pixels around a point source masking and perform a transform to local Gnomonic coordinates. We then use the pixels that are not masked around the point source mask to obtain a least curvature interpolation surface. We use the interpolation surface to obtain inpainted values in the masked pixels. This method works better for smoother maps and with smaller point source masking. The inpainting procedure is implemented using the \texttt{iSAP} software\footnote{https://www.cosmostat.org/software/isap}\citep{starckISAPInteractiveSparse2013}.

For $B$-modes we mask the Stokes $QU$ maps maps with the point source mask, and use template cleaning method (TC)~\citep{2019JCAP...04..046L, 2019PhRvD.100b3538L,chen1,chen2} to obtain $B$-mode maps with point source masked. TC is used to clean for the $E$-to-$B$ leakage due to incomplete sky and point source masking. We then inpaint the leakage cleaned and point source masked $B$ maps. Note that the leakage from $B$-to-$E$ is negligible when comparing to the $E$-mode signal, so that we do not make any attempt to correct it. To quantify the influence of inpainting on the final NILC/GLS/cILC power spectra, we propagate the input point source map through the whole preprocessing and foreground removal pipeline, including masking and inpainting, and compute the power spectrum of the residual point source. We only find the significant point source residual in $BB$ power spectra (as shown in Fig.~\ref{fig:GLS-cILC_residuals}), which is below the diffuse foreground residual at scales of interest. This indicates that the point sources after inpainting are subdominant among the foreground contamination. We also note that, due to the TOD filtering point sources get smeared in the directions of the scans. The point source mask fails to clean the effects of filtering of point sources. After the point source processing stage we get $T$/$E$/$B$ maps for each band without resolved point sources.

\section{The likelihood for ABS and NILC method}
\subsection{The likelihood for ABS}
\label{app:likelihood}
In the ABS analysis, we employ a Gaussian likelihood to estimate the tensor-to-scalar ratio, with using 9 $\ell$-bins in the multipole range $20\le \ell \le 200$. The likelihood has the form of $-2\ln \mathcal{L}\propto  \Delta \boldsymbol{C}^T\boldsymbol{M}^{-1}_{\rm fid\,}\Delta \boldsymbol{C}$. Here $\Delta \boldsymbol{C}$ represents the difference between the model band power vector and the data one, $\boldsymbol{M}_{\rm fid}$ is the fiducial covariance matrix of band powers from simulations, which is fixed in the likelihood evaluation. According to the definition, $\left[\boldsymbol{M}_{\rm fid}\right]_{\ell\ell'}=\left<(D^{\rm fid}_{\ell}-\langle D^{\rm fid}_{\ell}\rangle) (D^{\rm fid}_{\ell'}-\langle D^{\rm fid}_{\ell'}\rangle)\right>$, where $D^{\rm fid}_{\ell}$ are the ABS-derived CMB $BB$ band powers. We estimate the covariance by using the ABS-derived band powers from 50 realization maps with the same generation process as for DC1 data, keeping the same foreground sky and the noise level as in DC1 data but setting the fiducial value of $r=0.01$ for $\Lambda$CDM model. These realizations are produced by combining the different component filtered maps from the ancillary data listed in Sect.~\ref{sec:mock}. Since the CMB tensor contribution to the covariance is extremely minor compared with the noise and other contributions, the choice of fiducial value of $r$ has almost no changes on the estimated covariance, as long as it does not deviate greatly from the true value.  As a result, various uncertainties contributed from the noise, CMB signal, $E$-to-$B$ leakage and TOD filtering effects as well as foreground residuals after the ABS cleaning have been appropriately included in the covariance. Moreover, due to the finite number of simulations, we use only the adjacent elements of $\boldsymbol{M}_{\rm fid}$ that are separated by two or fewer $\ell$-bins. 

The construction of Gaussian likelihood functions for GLS and cILC is similar to that of ABS, and details can be found in ~\citep{paper4}.

\subsection{The NILC method}
\label{sec:nilc}
In this study, we apply the NILC foreground-cleaning method for estimating the CMB $E$-mode map, which is used for removing the $E$-to-$B$ leakage in the GLS and cILC pipelines. In the following, we will detail this method.

The internal linear combination (ILC) method was first introduced by~\citep{Bennett:1992} for foreground cleaning of COBE-DMR multi-frequency observations. Since then ILC has been shown to be an effective and efficient method of obtaining foreground cleaned CMB maps from multi-frequency data with minimal assumptions about the CMB signal. The ILC method has been studied and implemented in both map and harmonic space \citep{Bennett:2003, Tegmark:2003, Eriksen:2004, Saha:2006, 2008PhRvD..78b3003S, Saha2011}. One of the most effective ILC implementation is implemented in the frame of needlets \citep{Delabrouille:2009, Basak:2012, Basak:2013, Planck2014XII}. Needlets are special kind of spherical wavelets that allow for localization both in pixel and harmonic space. We adopted the Needlet ILC (NILC) method for producing foreground cleaned CMB $T$- and $E$-mode maps for AliCPT-1. While $B$-mode measurement is the key science objective for AliCPT, several science objectives, like lensing, will require high-precision CMB $T$- and $E$-mode maps. 

\begin{figure}[htpb]
    \centering
    \includegraphics[width=0.6\textwidth]{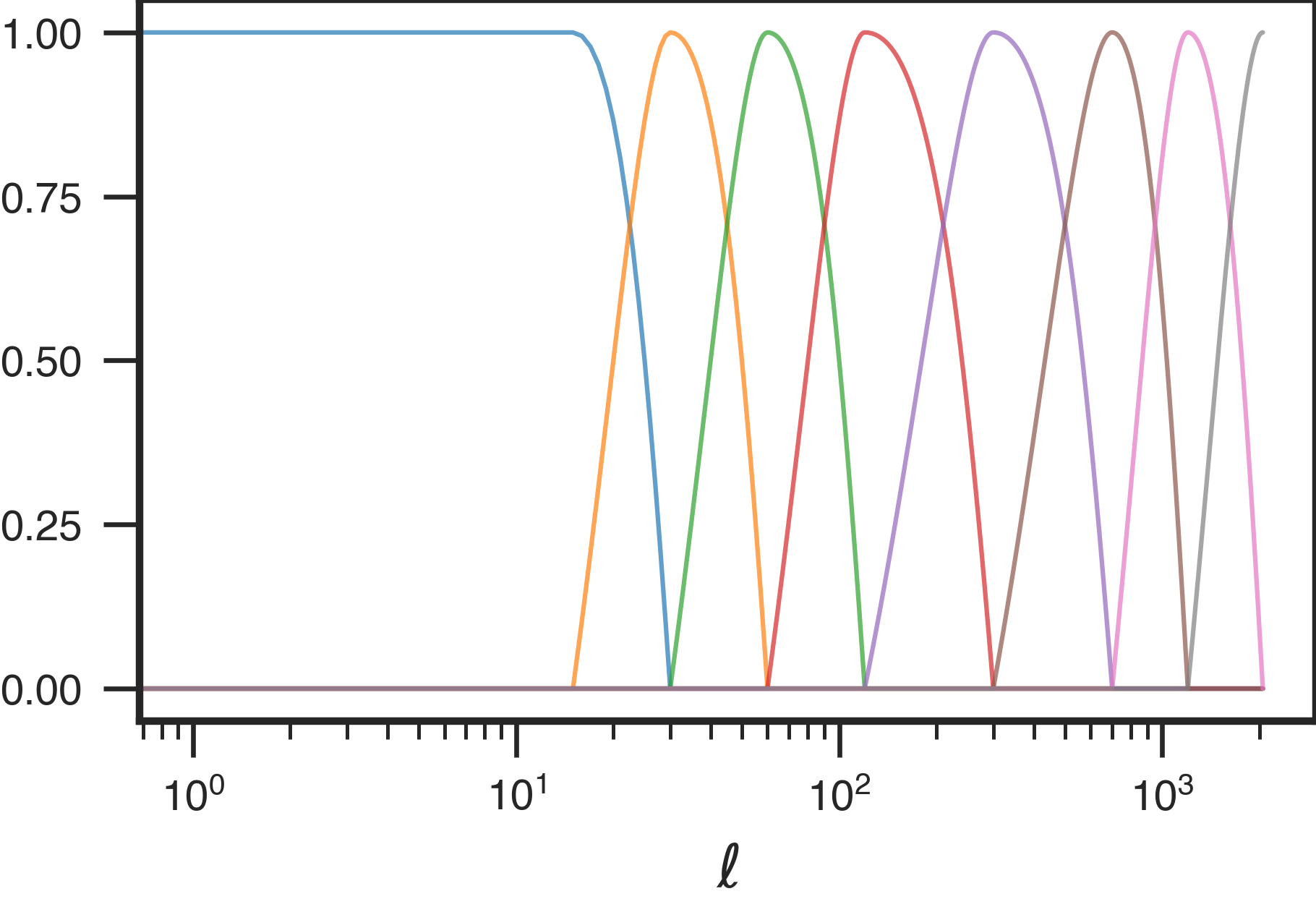}
    \caption{Cosine needlet bands used for the NILC analysis of $T$- and $E$-maps.}
    \label{fig:needlets}
\end{figure}

The observed data in frequency band $\nu$ and its spherical harmonic transformation can be written as $d(\nu, \hat n) = \sum_{\ell m}d^\nu_{\ell m}Y_{\ell m}(\hat n)$. We can convolve/deconvolve the data to the common resolution by $d^\nu_{\ell m} B_{\ell,c} / B_{\ell, \nu}$, where $B_{\ell,c}$ and $B_{\ell,\nu}$ are the beam window functions for common resolution and native map resolution respectively. For AliCPT DC1 case we set the common resolution of the final map to be $11$ arcmin, which is the best resolution of the the two AliCPT channels. Since WMAP K band and AliCPT 95 GHz have larger beam sizes we cut off the contribution of these two bands to $\ell$ of 350 and 1200, respectively. The map information contained in $d^\nu_{\ell m}$ can be decomposed into a set of filtered maps with harmonic coefficients for the $j$-th needlet band being given by:
\begin{equation}
    d^{\nu, j}_{\ell m} = h^j_\ell d^\nu_{\ell m}\,,
\end{equation}
where the needlet bands $h^j_\ell$ are responsible for localizing the information in harmonic domain and satisfy $\sum_j (h^j_\ell)^2=1$. Cosine and Gaussian needlet bands are common choices used in NILC. From our tests we find that cosine bands perform slightly better than the Gaussian bands for $E$-map resulting in smaller residual noise contamination in the AliCPT case, so we will present the analysis using cosine needlet bands. The needlet bands are shown in Fig.~\ref{fig:needlets}.

Assuming HEALPix pixelization, the spherical needlet function is defined as:
\begin{equation}
    \psi_{jk}(\hat n) = \sqrt{\frac{4\pi}{N_j}}\sum_{\ell m} h^j_\ell Y^*_{\ell m}(\hat n) Y_{\ell m}(\hat p_{jk})\,,
\end{equation}
where $N_j$ is the \texttt{NPIX} parameter, and $\hat p_{jk}$ refers to the pixel center of the $k$-th pixel of the of $j$-th needlet map. The needlet coefficient of the observed map $d(\nu, \hat n)$ is then given by:
\begin{align}
    b^\nu_j(\hat p_{jk}) &= \int d(\nu, \hat n) \psi_{jk}(\hat n) d \hat n, \\
        &= \sqrt{\frac{4\pi}{N_j}}\sum_{\ell m} h^j_\ell d^\nu_{\ell m} Y_{\ell m}(\hat p_{jk})\,.
\end{align}
The needlet transformation gives us HEALPix maps of each needlet scale, and it is also easy to see that the needlet transformation is a linear operation.  

To perform the ILC on the needlet maps we need to compute the covariance matrix for $k$-th pixel of the $j$-th needlet scale $\boldsymbol C_{jk} = C^{\nu_1\times \nu_2}_{jk} = \langle b^{\nu_1}_j(\hat p_{jk}) b^{\nu_2}_j(\hat p_{jk}) \rangle$. In practice the empirical estimate of covariance $\boldsymbol{\hat C}_{jk}$ is computed by averaging the needlet coefficient product $b^\nu_j(\hat p_{jk}) b^\nu_j(\hat p_{jk})$ over some domain of pixels in the neighbourhood of pixel $k$. This may be written as:
\begin{equation}
    \hat C^{\nu_1\times \nu_2}_{jk} = \frac{1}{n_k} \sum_{k'} w_j(k, k') b^{\nu_1}_j(\hat p_{jk'}) b^{\nu_2}_j(\hat p_{jk'})\,,
\end{equation}
where $w_j(k, k')$ are selection weights that chooses the set of pixels around pixel $k$ that will make contribution to the covariance, and $n_k$ are the number of selected pixels. Note that the weights are scale dependent, so for the needlet maps containing information on the largest angular scales, the contribution is made from nearly all pixels while for needlet maps with small scale information we choose pixels in the vicinity of pixel $k$. 

Just as with any ILC method we assume that with the data in thermodynamic temperature units, the CMB signal is frequency independent, and we set the ``mixing vector'' for the CMB, $\boldsymbol a$, to be an unit vector of length equal to the number of frequency channels. Then the NILC weights are given by:
\begin{equation}
    W^{\rm NILC}_{\nu, j}(\hat p_{jk}) = \frac{\boldsymbol {\hat C}_{jk}^{-1} \boldsymbol a}{\boldsymbol a^t\boldsymbol {\hat C}_{jk}^{-1} \boldsymbol a}\,.
    \label{eq:nilc_wgts}
\end{equation}
Then the NILC cleaned needlet maps are given by:
\begin{equation}
    b^{\rm NILC}_j(\hat p_{jk}) = \sum_{\nu} W^{\rm NILC}_{\nu, j}(\hat p_{jk}) b^\nu_j(\hat p_{jk})\,.
\end{equation}
Finally the harmonic coefficients of the NILC cleaned map, $\hat s^{\rm NILC}_{\ell m}$ are obtained by inverse needlet transform:
\begin{equation}
    \hat s^{\rm NILC}_{\ell m} = \sum_{jk}b^{\rm NILC}_j(\hat p_{jk}) \sqrt{\frac{4\pi}{N_j}} h^j_\ell Y_{\ell m}(\hat p_{jk})\,.
\end{equation}
We can recover the NILC cleaned map by doing an inverse SHT as $\hat s_{\rm NILC} (\hat n)= \sum_{\ell m} \hat s^{\rm NILC}_{\ell m} Y_{\ell m} (\hat n)$.

\begin{figure}[htpb]
    \centering
    \includegraphics[width=0.48\textwidth]{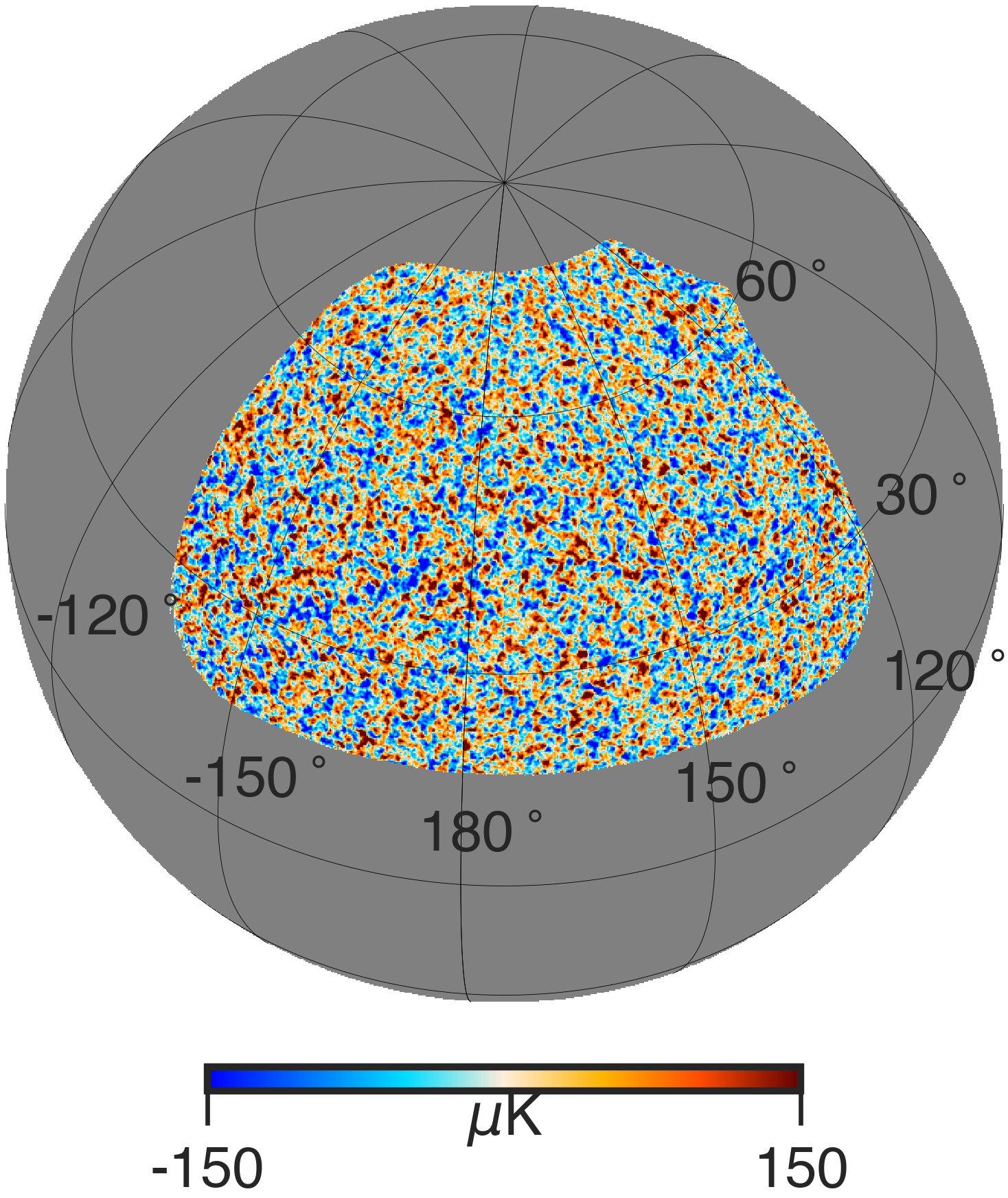}
    \includegraphics[width=0.48\textwidth]{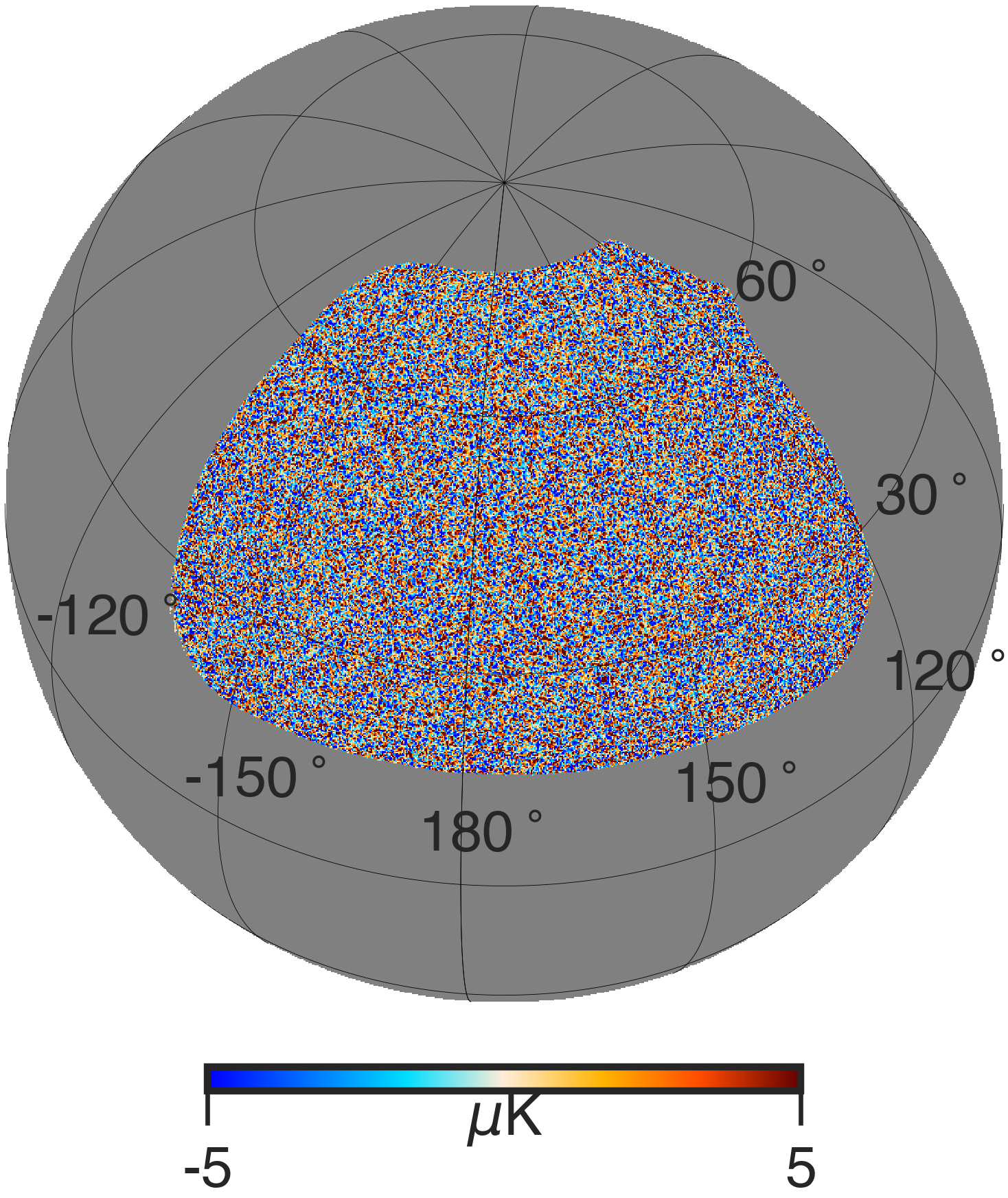}
    \caption{Left: The NILC cleaned TOD filtered $T$-map for DC1 case at 11 arcmin resolution. Right: The NILC cleaned TOD filtered $E$-map for DC1 case at 11 arcmin resolution. Note that the large and intermediate scale power is missing due to the TOD filtering of the data.}
    \label{fig:NILC_TEmaps}
\end{figure}

For DC1 data analysis we first perform point source pre-processing to obtain point source masked and inpainted $T$- and $E$-mode maps. We then proceed to perform NILC cleaning of the $T$- and $E$- maps separately. In case of $T$- and $E$-mode analysis we use a simple noise level based cut of the DC1 scan patch. This patch only keeps the region of the scan area where the 150 GHz channel has polarized noise standard deviation of 20 $\mu$K-pixel or less at \texttt{NSIDE}$=1024$. We produce NILC cleaned $T$- and $E$-maps at 11 arcmin resolution. These maps are shown in Fig.~\ref{fig:NILC_TEmaps}. We note that AliCPT improves the noise level of the $E$-maps till $\ell \sim 700$ according to its beam resolutions at 95/150 GHz. Beyond this multipole the HFI 143 GHz channel starts making increasingly larger contribution to the NILC cleaned $E$-map. Thus, the high-$\ell$ noise is close to the noise level of HFI 143 GHz rebeamed with 11 arcmin beam. In Sect.~\ref{sec:TEspectra} we have detailed the TT/EE/TE power spectra computation, filtering correction and error at power spectra level. 

\section{Recovery of CMB \texorpdfstring{$TT$/$EE$/$TE$}{TT/EE/TE} spectra}
\label{sec:TEspectra}
\begin{figure}[htpb]
    \centering
    \includegraphics[width=0.485\linewidth]{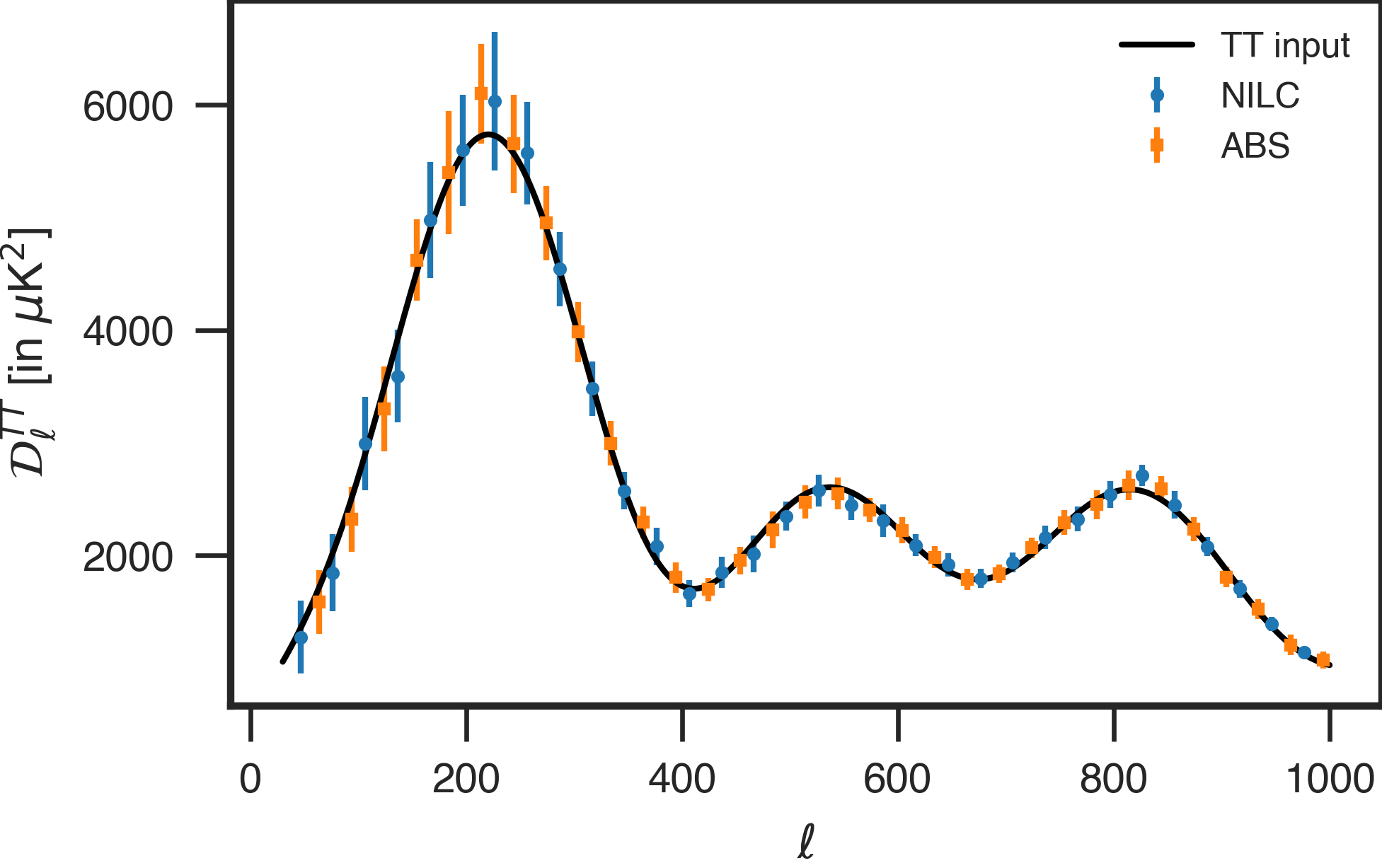}
     \includegraphics[width=0.48\linewidth]{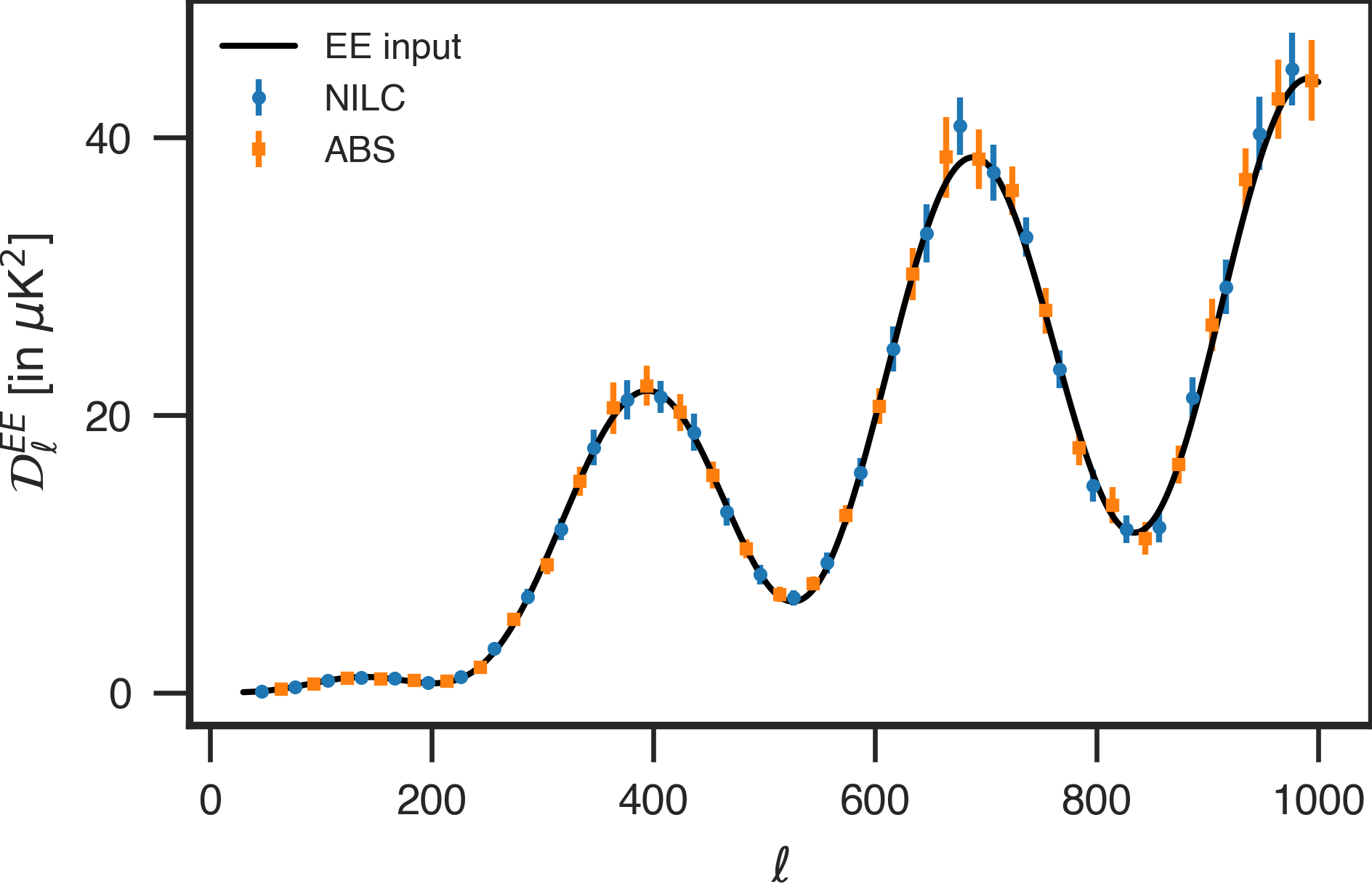}
      \includegraphics[width=0.48\linewidth]{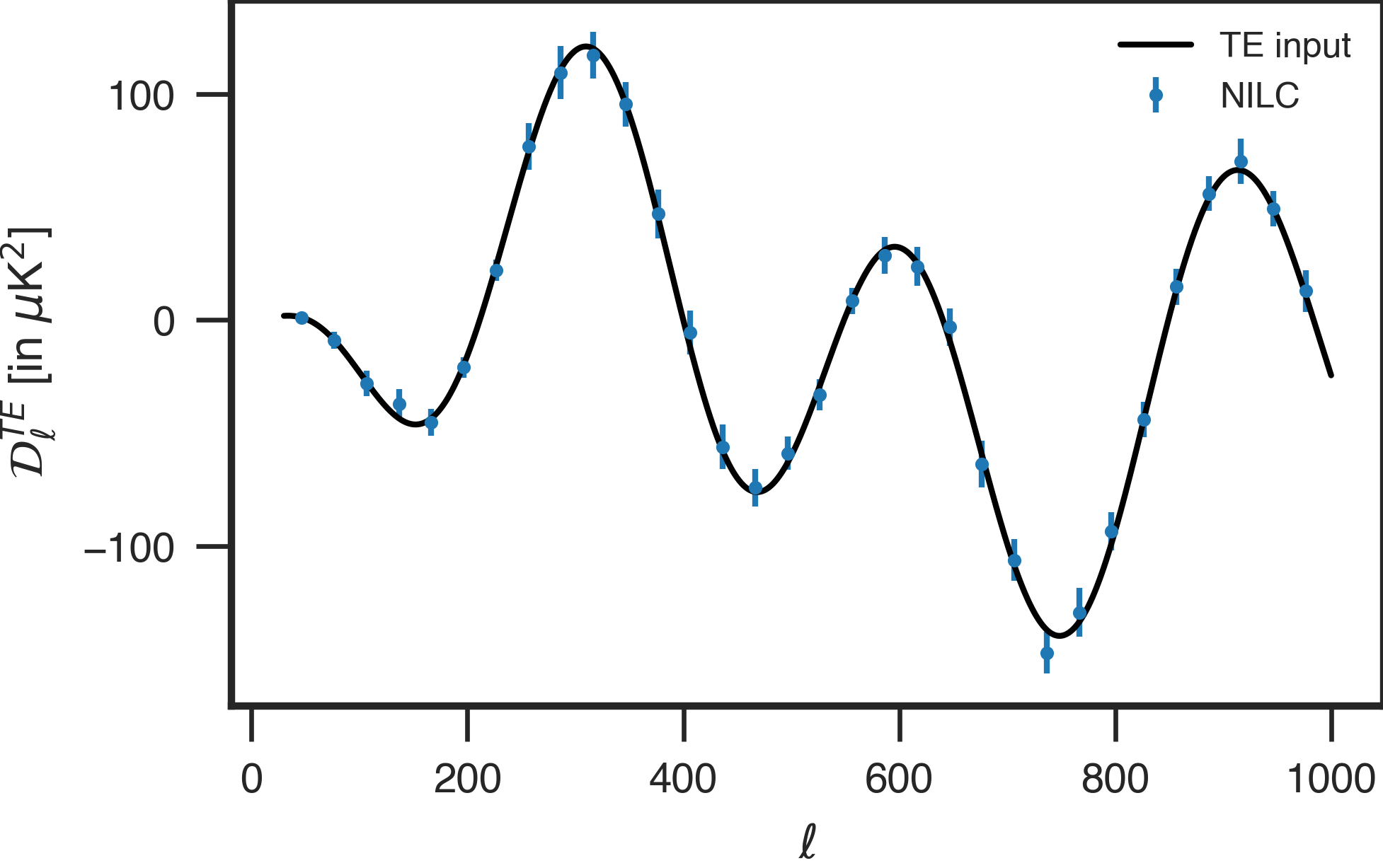}
    \caption{CMB $TT$ (top left) and $EE$ (top right) power spectra for ABS (orange) and NILC (blue) methods. The $TE$ (bottom) power spectrum for the NILC (blue) method. The multipole range is $20\le\ell\le1024$ with bin width $\Delta \ell =30$. The Planck-based theoretical spectra as the simulation inputs are plotted in solid black in each panels. The error bars show $2\sigma$ statistical uncertainties including cosmic variance, noise sampling variance and filtering effects as well as foreground residuals after cleaning, which are estimated from 50 realizations of ``observed'' multi-frequency maps with passing through the ABS and NILC pipelines. The ABS- and NILC-derived band power spectra and associated error bars are in good agreement, and all estimated spectra are consistent with the theoretical inputs within $2\sigma$ level. Note that the measured power spectrum matrix ($\tilde{\mathcal{D}}_{ij}^{\rm obs}$ in Eq. \ref{eq:noiseD}) must be symmetric during the ABS calculation to ensure real eigenvalues. However, this condition is not satisfied by the measured $TE$ data, making the ABS currently applicable only to auto-power spectra. For the applicability to cross-power spectra, a further extension is required.}
    \label{fig:TE}
\end{figure}

\begin{figure}[htpb]
    \centering
    \includegraphics[width=0.45\linewidth]{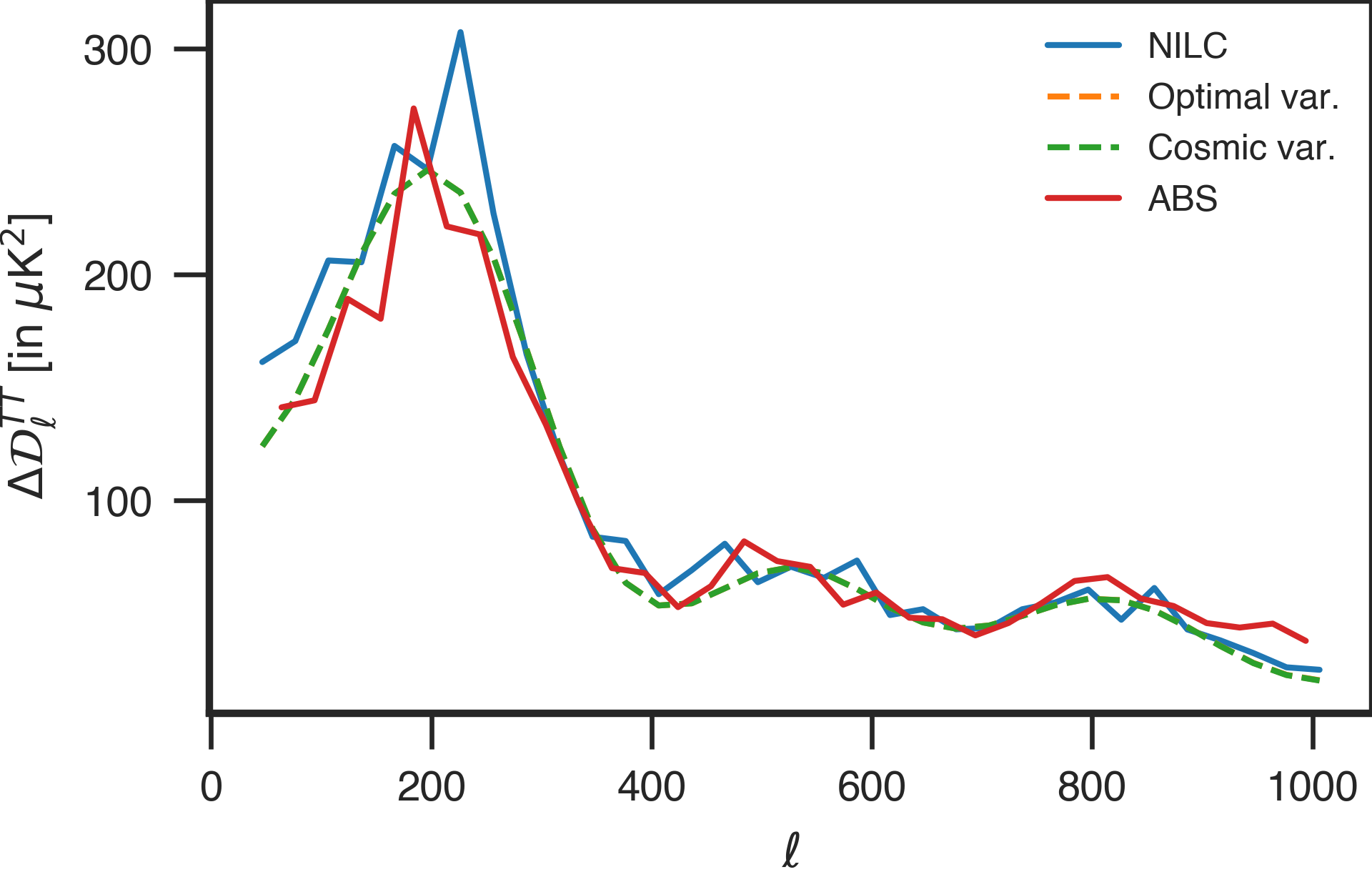}
     \includegraphics[width=0.45\linewidth]{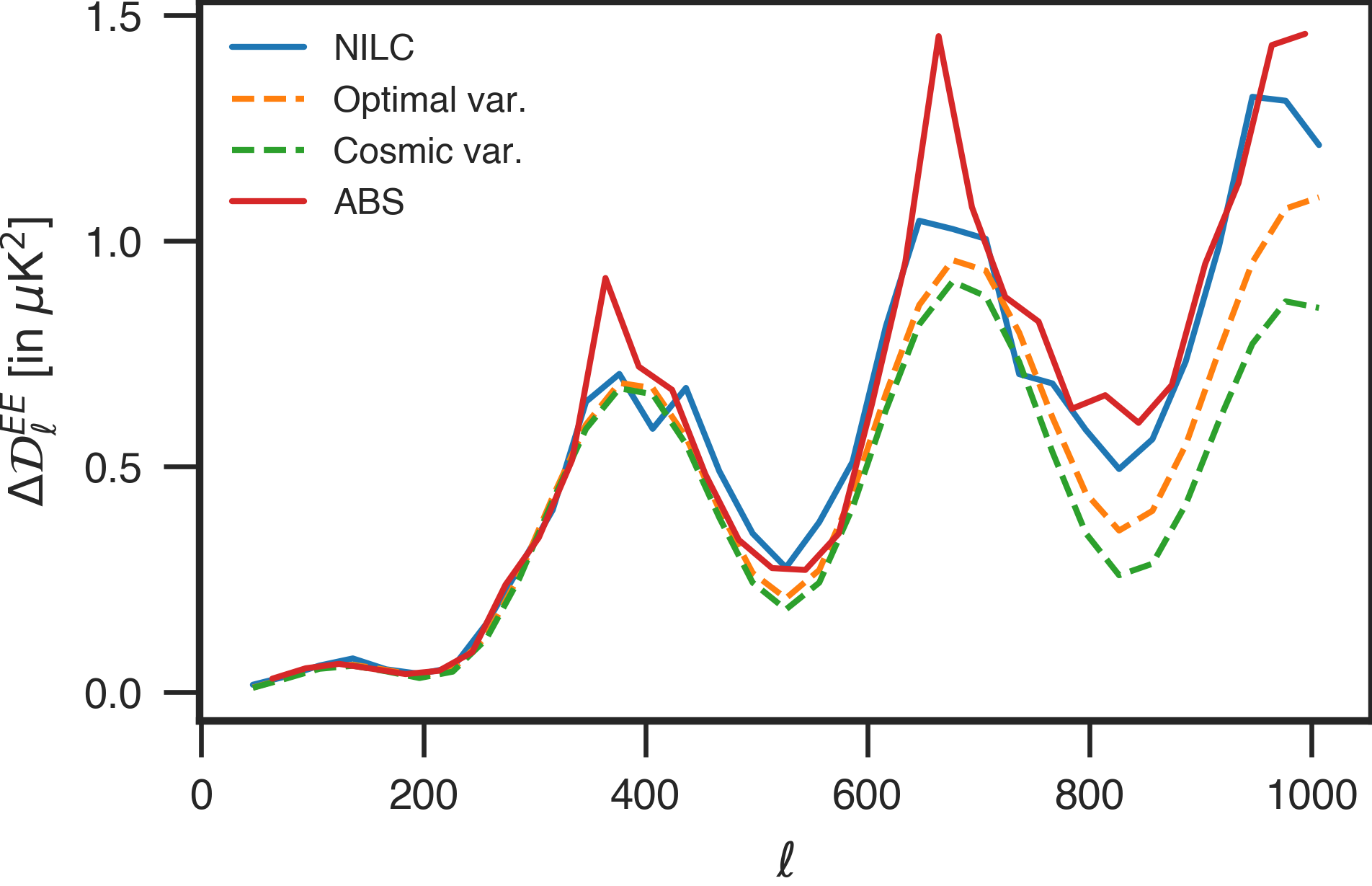}
      \includegraphics[width=0.45\linewidth]{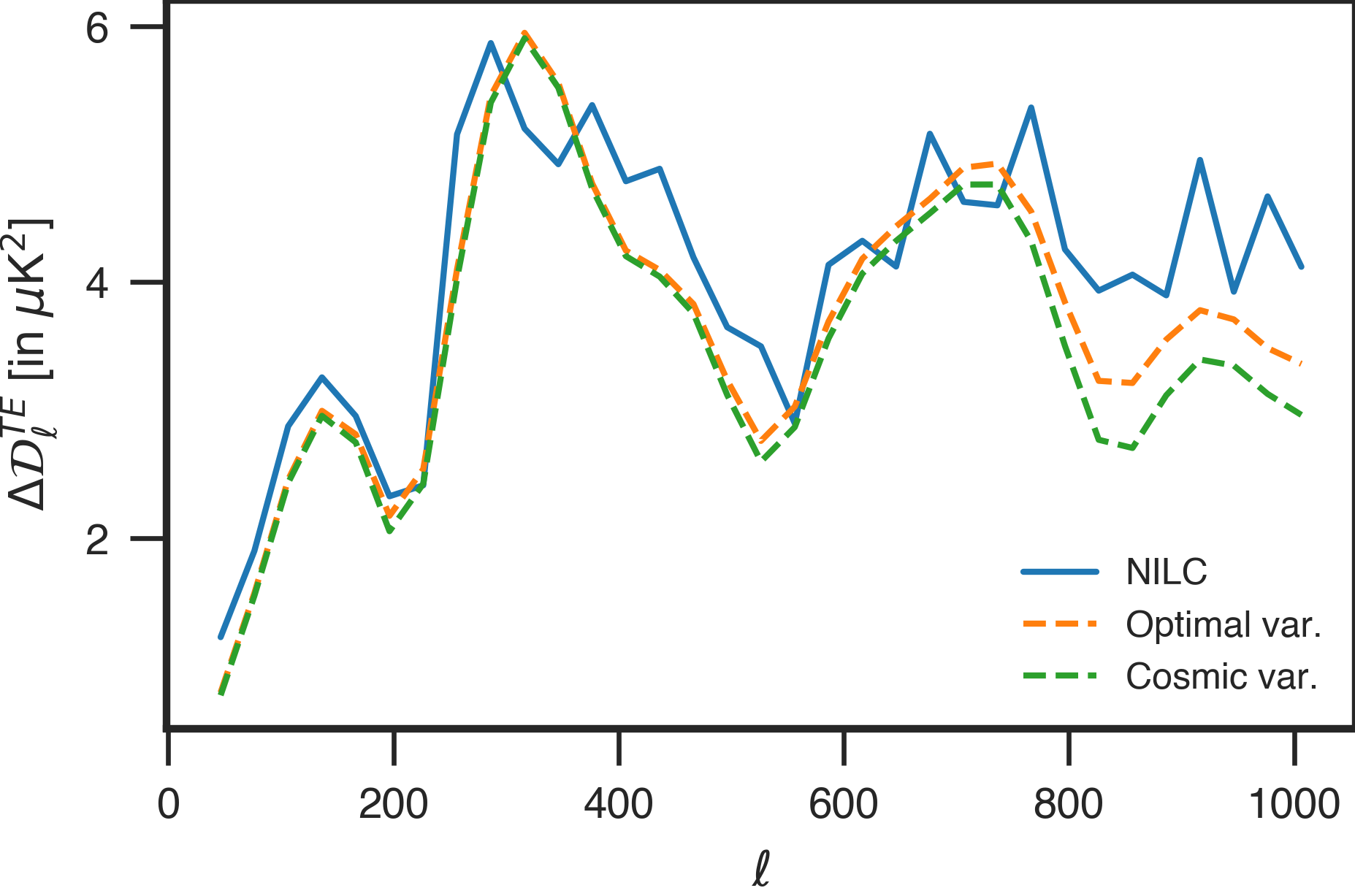}
    \caption{CMB $TT$ (top left), $EE$ (top right) and $TE$ (bottom) power spectra 1-$\sigma$ errors derived from 50 simulations of CMB and noise and propagated through the ABS (red) and NILC (blue) pipelines, respectively, in the range of $20\le\ell\le1024$ with bin width $\Delta \ell =30$. The optimal variance defined in Eq.~\ref{eq:optv} is shown in the orange dashed line, while the cosmic variance is shown with green dashed line. The errors from both ABS and NILC pipelines are very similar, with NILC having slightly smaller errors for $\ell>700$.}
    \label{fig:TE_error}
\end{figure}

For a cross-check, here we present the CMB $TT$/$EE$/$TE$ power spectra from the ABS and NILC methods, where the NILC-derived $EE$ spectrum has been used in GLS/cILC pipelines for the estimation of the $E$-to-$B$ leakage.  

We compute the CMB $TT$, $EE$ and $TE$ power spectra using \texttt{NaMaster} with $6^\circ$ C2 apodization. Due to the effect of timestream filtering the spectra computed from the maps will need further correction. For the $T$- and $E$-mode analysis of the DC1 data, we can safely ignore any leakage term due to filtering as it only affects the  $E$-modes due to leakage of the $B$-modes. However, this term is negligible compared to the $E$-mode power. The only correction we perform is for the loss of power, i.e. power suppression on large angular scales. The spectra with power suppression caused by filtering are denoted with overhead bars. The corrected spectra (without overhead bar) are written in the same way as defined in \citep{paper4}:
\begin{equation}\label{eq:Fl}
    D_\ell^{XY} = \frac{1}{F^2_{\ell, XY}}\bar{D}^{XY}_\ell\,, \quad {\rm with}\quad  F^2_{\ell, XX} = \left<\frac{\bar{D}^{XX}_\ell}{D^{XX}_\ell} \right>\,,
\end{equation}
where $XY \in [T,E]$, and the auto-transfer function for $T$- or $E$- modes is denoted by $F^2_{\ell, XX}$ and is computed from 50 $T/E$-only filtered CMB maps. The computation process is similar to the steps discussed in detail in \citep{paper4}. The auto-transfer function for the $TE$ cross spectrum is defined as $F^2_{\ell, TE} = (F^2_{\ell, TT}F^2_{\ell, EE})^{1/2}$.

In Fig.~\ref{fig:TE}, we present the recovered CMB $TT$, $EE$ and $TE$ bandpowers estimated from the ABS and the NILC foreground-cleaning pipelines for the DC1 data. The solid line in each panel is the DC1 input CMB power spectrum, consistent with Planck 2018 measurements. The error bars indicate $2\sigma$ statistical uncertainty computed from 50 realizations where the CMB and instrumental noise maps are drawn randomly based on the fiducial cosmology model but with fixed foregrounds. Here, the suppression effect caused by filtering in the CMB signal is appropriately corrected and the suppression factor for each $\ell$ bin is calculated based on the ratio of band powers with and without filtering from the 50 CMB-only simulations. We also show the associated statistical errors (i.e., standard deviation) for these spectra in each $\ell$-bin. As seen, both  ABS and NILC pipelines provide a precise foreground subtraction and all the reconstructed spectra are in complete agreement with the true ones within $2\sigma$ over the whole multipole range.

In order to quantitatively assess the statistical errors of reconstructed power spectra, we compare them with the theoretical predictions of the optimal variance defined as:
\begin{align}\label{eq:optv}
    \Delta D_{\ell, {\rm optimal}}^{TT/EE} &= \sqrt{\frac{2}{(2\ell + 1) \Delta\ell f_{\rm sky}}}\left(D^{TT/EE}_\ell +  N_\ell^{TT/EE}/B^2_\ell \right)\,,\nonumber\\
    \Delta D_{\ell, {\rm optimal}}^{TE} &= \sqrt{\frac{1}{(2\ell + 1) \Delta\ell f_{\rm sky}}}\left[\left(D^{TT}_\ell +  N_\ell^{TT}/B^2_\ell\right)\left(D^{EE}_\ell + N_\ell^{EE}/B^2_\ell\right) + \left( D^{TE}_\ell+ {N_\ell^{TE}/B^2_\ell}\right)^2\right]^{1/2}\,,
\end{align}
where $N_\ell^{TT/EE}$ is the inverse noise variance combination of the noise spectra for the seven DC1 frequency bands. The quantity $N_\ell^{TE}/B^2_\ell$ can be safely ignored since the expectation value of $N_\ell^{TE}$ is zero. In the particular noise realization of the DC1 data, we have confirmed that the ratio of $N_\ell^{TE}/B^2_\ell$ to $D^{TE}_\ell$ approaches zero. The inverse noise variance combination represents the lowest possible noise variance achievable through map combination. The cosmic variance is obtained by dropping the noise terms in the above definitions for optimal variance. We show the comparison of the $1\sigma$ errors with the optimal and cosmic variances in Fig.~\ref{fig:TE_error}. As seen, both ABS- and NILC-derived errors in $TT$ and $EE$ appear quite comparable and are very close to the optimal variance. The optical variance is a theoretical estimate of the lower bound on errors. Due to the limited number of simulations, our results fluctuate around these bounds in these spectra.

\section{Reconstructed $BB$ band powers from TF pipeline}
\label{app:BB-TF}
The derived $r$-values from different TF models have been detailed in~\citep{paper4}. Therefore, in this section, we are providing supplemental information about the reconstructed $BB$ band powers.

\begin{figure}[htpb]
    \centering
    \includegraphics[width=\linewidth]{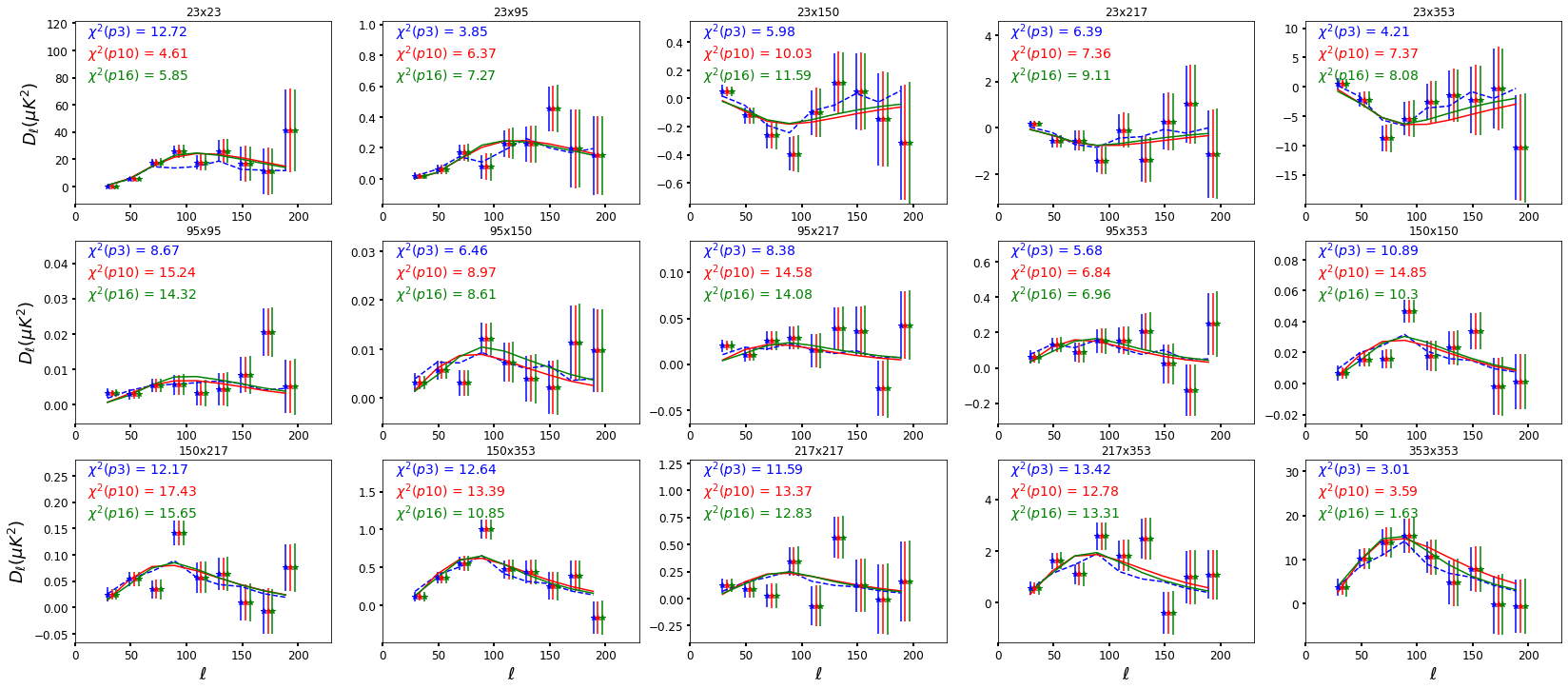}
    \caption{$BB$ auto- and cross-spectra band powers $D_\ell$ between the AliCPT 95\&150 GHz maps and bands of WMAP (23 GHz) and Planck (217\&353 GHz), in units of the CMB temperature $\rm \mu K^2_{CMB}$. In all cases the dots show the simulated data, while three solid lines show the expectation values from TF ``p3''(blue), ``p10'' (red) and ``p16'' (green), each of them using the median to predict the band powers. The error bars correspond to the 1$\sigma$ standard deviations, given by the diagonal components of the covariance matrix $\mathcal{M}$ for each frequency pairs. The $\chi^2$s (for 9 degrees of freedom in each case) for a comparison of the data with different models are listed in each plot. The agreement with the spectra and error bars as well as the $\chi^2$s is therefore a validation of these models and the TF pipeline.}
    \label{fig:mcmc_f2}
\end{figure}

We make a comparison of the $BB$ auto- and cross-spectra of the AliCPT 95\&150 GHz maps and the Planck 217\&353 GHz and WMAP K band (23 GHz) maps in simulations. Based on the median values of model parameters estimated from the TF likelihood spanning the range $\ell\in[20,200]$, the recovered spectra for ``p10'' (red; 10 parameters), ``p16'' (green; 16 parameters), and ``p3'' (blue; 3 parameters) are shown in Fig.~\ref{fig:mcmc_f2} (the complete set, including Planck 143\&100 GHz, is not shown), where the dots represent the data with error bars approximated by the square root of the diagonals of $\mathcal{M}$ (defined in Eq.~\ref{eq:M}) for each frequency pairs.

As seen, although the foreground-noise contribution to the covariance depends on the specific foreground model, 
we would like to illustrate here that the overall error bars predicted by the different models in the band-powers are essentially unchanged, especially for the bins $\ell\gtrsim50$ that dominate constraints on $r$. Moreover, the error bars at $20\le \ell<40$ may depend slightly on the models. However, these low-$\ell$ band powers suffer from the strong filtering effects and have relatively larger uncertainties, thus providing less information on CMB and foreground.

To roughly quantify how well our models fit the data, we calculate a $\chi^2$. Note that the $\chi^2$s for the separate auto-/cross-spectral band powers are obtained from the corresponding submatrices of the full covariance, and therefore the sum of the individual $\chi^2$s is not the total computed from the full one.  We see that all the TF models provide good overall fits among these 15 frequency pairs, and all the fits are consistent with each data points within about 2$\sigma$. In addition, the complex spatial shapes of the $BB$ powers of ``p3'' are apparent. This is because their templates are constructed directly from the sky maps of pure DC1 foreground simulations, which are not intrinsically smooth in \texttt{PSM}. Based on the $\chi^2$s in $23\times23$ and $353\times353$, the synchrotron and dust models appear to be very successful in reconstructing the data. We also see modest evidence for detection of negative synchrotron-dust correlation in $23\times353$ at 1$\sigma$ level, and our models are able to fit the data well within the statistical uncertainties.

\section{The impact of $EE$ power spectrum and other cosmological parameters \label{App:EE}}

\begin{figure}[!h]
    \centering
    \includegraphics[width=0.6\textwidth]{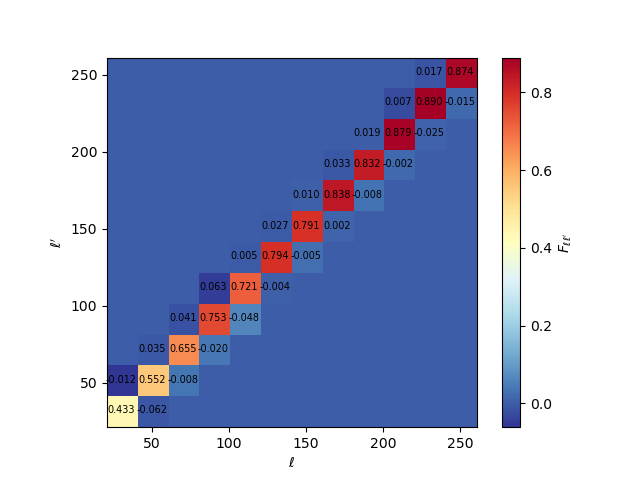} %
    \caption{ The filtering matrix $F_{\ell\ell'}$ in the TF analysis, estimated through a least-square fitting to the simulated noise power spectra before and after filtering. Due to the limited number of simulations, the estimation was restricted to those belonging to the same or neighboring $\ell$-bins.}
    \label{fig:fllinTF}
\end{figure}

The TF method in subsect.~\ref{subsec:TF} focuses on the dependence of the $BB$ power spectrum on the $r$ parameter. The impact of the $EE$ power spectrum and other cosmological parameters has been ignored. In particular, the lensing contribution to the $BB$ power spectrum, which is cosmology-dependent, was assumed to be known and neglected in the simulations and analyses. These simplified assumptions may lead to overly optimistic estimates. In this section, we extend our analysis by including the $EE$ power spectrum and varying all cosmological parameters. The CMB power spectra for a set of cosmological parameters $\theta = \{\Omega_bh^2, \Omega_ch^2, \theta, \tau, \ln\left(10^{10}A_s\right), n_s\}$ and $r$ are given by:
\begin{eqnarray}
D^{EE}_\ell(\mathbf{\theta}, r) &=& \widebar{D}_{\ell}^{EE}(\mathbf{\theta}_{\rm fiducial}, r=0) + \sum_{\ell'} F^{EE}_{\ell \ell'} \left[D^{EE, {\rm theory}}_{\ell'}(\mathbf{\theta}, r) - D^{EE,{\rm theory}}_{\ell'}(\mathbf{\theta}_{\rm fiducial}, r=0)\right], \label{eq:EEcosmo_EE} \\
D^{BB}_\ell (\mathbf{\theta}, r) &=& \widebar{D}_{\ell}^{BB}(\mathbf{\theta}_{\rm fiducial}, r=0) + \sum_{\ell'} F^{BB}_{\ell \ell'} \left[D^{BB, {\rm theory}}_{\ell'}(\mathbf{\theta}, r) - D^{BB,{\rm theory}}_{\ell'}(\mathbf{\theta}_{\rm fiducial}, r=0)\right]\,.
\label{eq:EEcosmo_BB}
\end{eqnarray}
In the time-domain filtering, we approximate it using multiplicative transfer functions $F_{\ell,\ell'}^{EE}$ and $F_{\ell,\ell'}^{BB}$, obtained through least-square fitting to the simulated noise power spectra before and after filtering. Due to the limited number of simulations, we retain $F_{\ell, \ell'}$ only for the same or neighboring $\ell$-bins. The resulting filtering matrix is displayed in Fig.~\ref{fig:fllinTF}, which is employed in the TF analysis.

The simulation-based $\widebar{D}_{\ell}^{EE}(\mathbf{\theta}_{\rm fiducial}, r=0)$ and $\widebar{D}_{\ell}^{BB}(\mathbf{\theta}_{\rm fiducial}, r=0)$ terms in Eqs.~(\ref{eq:EEcosmo_EE}-\ref{eq:EEcosmo_BB}), where $\mathbf{\theta}_{\rm fiducial}$ represents the fiducial cosmology for DC1 $r=0$ simulations, already include the time-domain filtering effect for the base model $(\theta_{\rm fiducial}, r=0)$ and therefore do not require multiplication by the transfer functions. Eq.~\ref{eq:EEcosmo_BB} propagates the uncertainties in cosmological parameters to the uncertainties in the lensed $BB$ spectrum, but the cosmic variance of the lensed $BB$ spectrum is still absent in this analysis. A comprehensive treatment of the lensing effect would necessitate a new set of simulations, which is beyond the scope of this work focused on the analysis of DC1 data.

The analysis here, dubbed ``EE+cosmo run'' , contains 26 parameters as listed in Tab.~\ref{tab:EE+cosmo}. With the transfer functions $F_{\ell,\ell'}$ estimated from simulations, we no longer use the phenomenological parameters in Eq.~\ref{eq:Fl_model} to describe the time-domain filtering effect. We have also assumed that the foreground $E$ maps and $B$ maps share the same frequency dependence ($\beta_d$ and $\beta_s$). Finally, for the six $\Lambda$CDM parameters $\mathbf{\theta} = \{\Omega_bh^2, \Omega_ch^2, \theta, \tau, \ln\left(10^{10}A_s\right), n_s\}$, we apply a Planck+BAO prior
\begin{equation}
\chi^2_{\rm Planck +BAO} = (\mathbf{\theta}-\mathbf{\theta}_{\rm fiducial})^T\mathrm{Cov}^{-1}(\mathbf{\theta}-\mathbf{\theta}_{\rm fiducial}),
\end{equation}
where $\mathrm{Cov}$, the Planck TTTEEE + lowE + lensing + BAO posterior covariance matrix of $\theta$, is available in the Planck legacy archive~(\url{https://www.cosmos.esa.int/web/planck/pla}). See also \citet{Planck:2018vyg} for the detailed description of the  products in the Planck legacy archive. 

In the ``EE+cosmo'' run we also apply a dynamic covariance matrix for the power spectra data, and add its log-determinant to the total $\chi^2$. For $r\le 0.01$ the CMB covariance matrix is linearly interpolated between the $r=0$ simulations and $r=0.01$ simulations, and for $r>0.01$ we simply use the covariance matrix from the $r=0.01$ simulations (We do not do extrapolation, as this might break the positive definiteness of the covariance matrix). These refinements do not lead to significant changes in the results, indicating that the systematics of the iteration approach in the ``p3'', ``p10'' and ``p16'' runs is negligible.

\begin{table}
    \caption{Parameters in the ``EE+cosmo'' run \label{tab:EE+cosmo}}
    \centering
    \begin{tabular}{lll}
    \hline
    \hline
    parameter & definition & prior \\
    \hline
    $r$ &  tensor-to-scalar ratio & $[0, 0.2]$ \\
    $\Omega_bh^2$ & baryon density & $[0, 0.1]$\\
    $\Omega_ch^2$ & cold dark matter density & $[0, 0.5]$ \\
    $100\theta_{\rm MC}$ & angular size of the sound horizon on the last scattering surface & $[0.5, 1.5]$ \\
    $\tau$ &  reionization optical depth & $[0, 0.2] $ \\
    $\ln\left(10^{10}A_s\right)$ & primordial scalar amplitude & $[2, 4]$ \\
    $n_s$ & primordial scalar index & $[0.8, 1.2]$ \\
    \hline
    $T_{\rm MBB}$ & dust MBB temperature parameter in $\mathrm{K}$ & $[19, 20.5]$ \\  
    $\beta_d$ & dust polarization frequency spectral index & $[1, 2.5]$ \\  
    $\beta_s$ & synchrotron polarization frequency spectral index & $[-4, -2]$ \\  
    $A_{\rm d, EE}$ & dust $EE$ spectral amplitude in $\mu \mathrm{K}^2$ & $[0, 100]$ \\
    $\alpha_{\rm d, EE}$ & dust $EE$ spatial spectral index & $[-4, 0]$ \\
    $\alpha^\prime_{\rm d, EE}$ & dust $EE$ spatial spectral running & $[-4, 4]$ \\
    $A_{\rm s, EE}$ & synchrotron $EE$ spectral amplitude in $\mu \mathrm{K}^2$ & $[0, 100]$ \\
    $\alpha_{\rm s, EE}$ & synchrotron $EE$ spatial spectral index & $[-4, 0]$ \\
    $\alpha^\prime_{\rm s, EE}$ & synchrotron $EE$ spatial spectral running & $[-4, 4]$ \\
    $\varepsilon_{\rm 2, EE}$ & synchrotron-dust EE correlation parameter defined in Eq.~\eqref{eq:eps_model} & [-1, 1] \\
    $\alpha_{\rm \epsilon, EE}$ & synchrotron-dust  EE correlation parameter defined in Eq.~\eqref{eq:eps_model}& [0, 6] \\    
    $A_{\rm d, BB}$ & dust $BB$ spectral amplitude in $\mu \mathrm{K}^2$ & $[0, 100]$ \\
    $\alpha_{\rm d, BB}$ & dust $BB$ spatial spectral index & $[-4, 0]$ \\
    $\alpha^\prime_{\rm d, BB}$ & dust $BB$ spatial spectral running & $[-4, 4]$ \\
    $A_{\rm s, BB}$ & synchrotron $BB$ spectral amplitude in $\mu \mathrm{K}^2$ & $[0, 100]$ \\
    $\alpha_{\rm s, BB}$ & synchrotron $BB$ spatial spectral index & $[-4, 0]$ \\
    $\alpha^\prime_{\rm s, BB}$ & synchrotron $BB$ spatial spectral running & $[-4, 4]$ \\
    $\varepsilon_{\rm 2, EE}$ & synchrotron-dust BB correlation parameter defined in Eq.~\eqref{eq:eps_model} & [-1, 1] \\
    $\alpha_{\rm \epsilon, EE}$ & synchrotron-dust BB correlation parameter defined in Eq.~\eqref{eq:eps_model}& [0, 6] \\    
    \hline
    \end{tabular}
\end{table}

\begin{figure}[!h]
    \centering
    \includegraphics[width=0.35\textwidth]{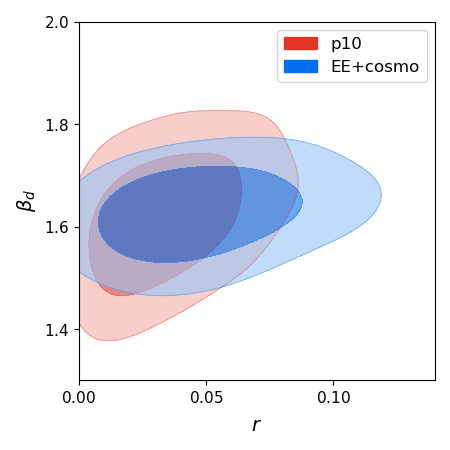} %
    \includegraphics[width=0.35\textwidth]{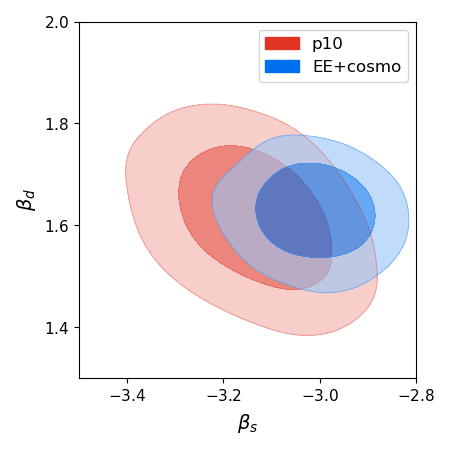}
    \caption{Constraints on $r$, $\beta_s$ and $\beta_d$ from the ``EE+cosmo'' run, compared with the ``p10'' analysis.}
    \label{fig:EEcosmo_contours1}
\end{figure}

Because ``EE+cosmo'' and ``p10'' use the same parameterization for the $BB$ power spectrum, we compare the results of the two runs in Fig.~\ref{fig:EEcosmo_contours1}. Additional information in the $EE$ power spectrum significantly improves the constraint on the spectral indices $\beta_s$ and $\beta_d$ of polarization maps, giving $\beta_d=1.627\pm 0.061$ and $\beta_s = -3.011\pm 0.081$, respectively.  

\begin{figure}[!h]
    \centering
    \includegraphics[width=0.35\textwidth]{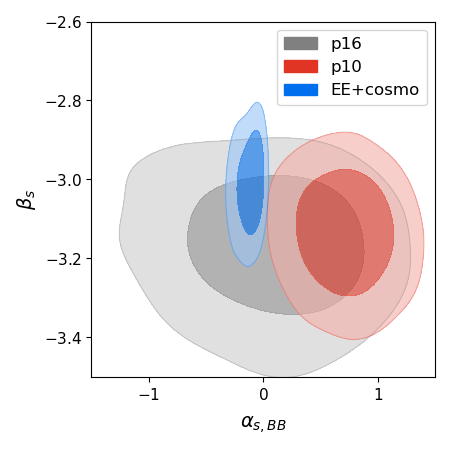} %
    \includegraphics[width=0.35\textwidth]{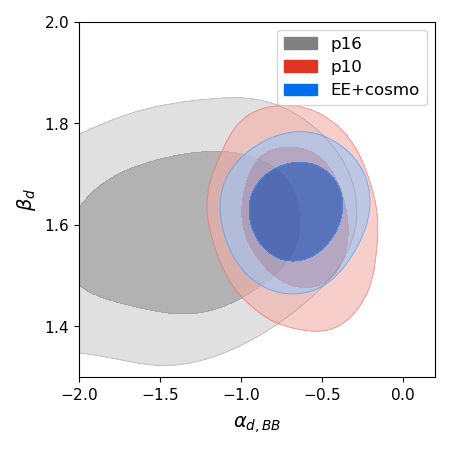}
    \caption{Comparison of foreground parameters between ``EE+cosmo'', ``p10'' and ``p16''.}
    \label{fig:EEcosmo_contours3}
\end{figure}

Fig.~\ref{fig:EEcosmo_contours3} compares the reconstructed spatial and frequency spectral indices for ``EE+cosmo'', ``p10'' and ``p16''. While the dust spectral indices are very stable among different analyses, the synchrotron spectral indices are biased in ``p10'', if we trust that ``EE+cosmo'' is more accurate and reliable. The bias in ``p10'', which only uses $B$ maps, seems to be due to over-fitting the synchrotron model when there is essentially only one synchrotron-dominated frequency band with very large noise in $B$ map.

\begin{figure}
    \centering
    \includegraphics[width=\textwidth]{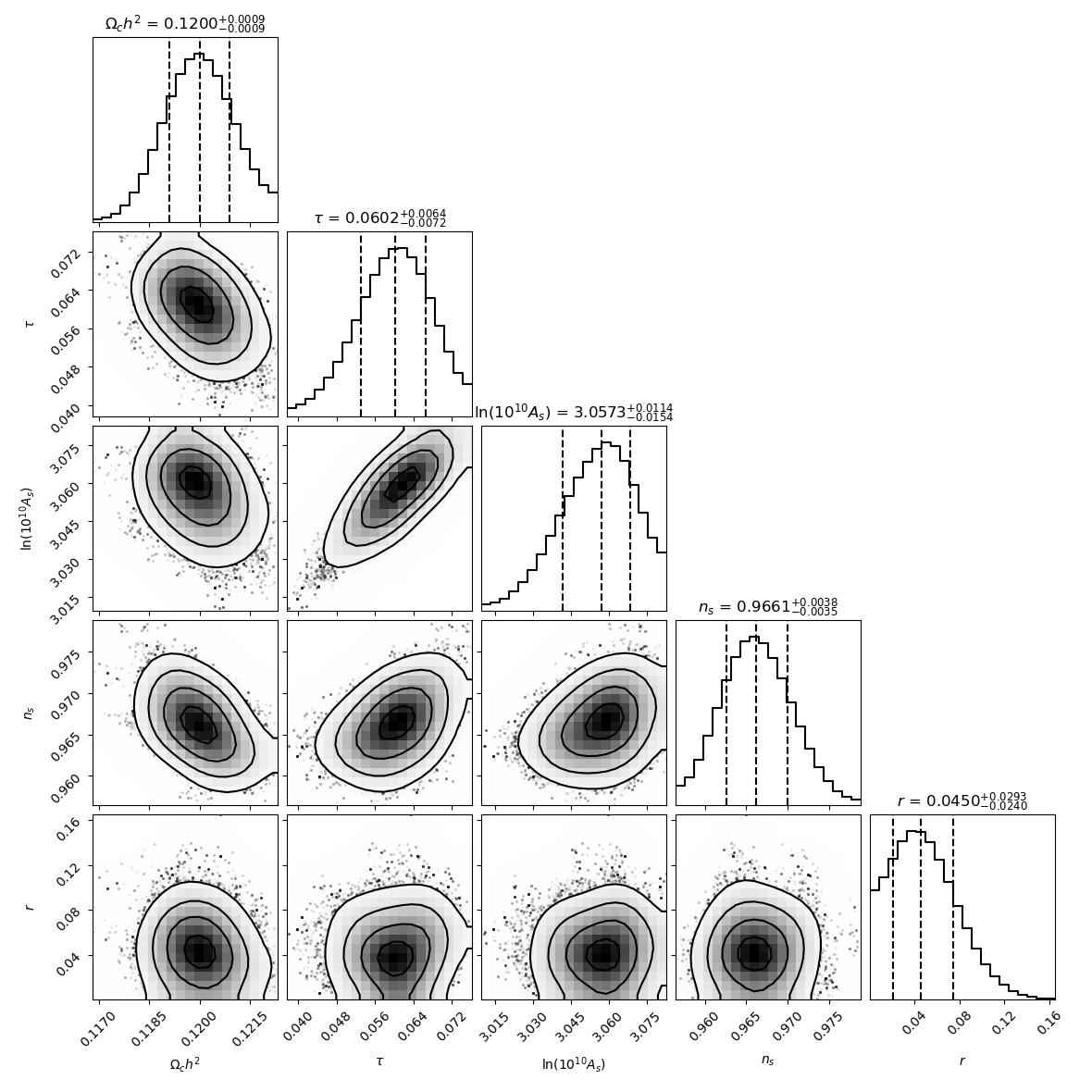}
    \caption{Cosmological parameter constraints from the ``EE+cosmo'' run. Contours represent confidence levels of 68.3\%, 95.4\%, and 99.7\%, respectively, from the inside out.}
    \label{fig:EEcosmo_contours2}
\end{figure}

Fig.~\ref{fig:EEcosmo_contours2} shows the marginalized constraints on $r$ and other cosmological parameters. For better readability we skipped $\theta_{\rm MC}$ and $\Omega_bh^2$ that are very well determined by the Planck+BAO prior. We do not find very strong degeneracy between $r$ and other cosmological parameters. Compared to the ``p10'' case, the alteration of the best-fit $r$ is mainly due to that $EE$ power spectrum prefers slightly difference foreground parameters.

\section{Nulling noise bias by crossing two data splits}
\label{app:null}
In the main text, we debais noise power spectrum bias by estimating the noise bias from 50 ancillary noise simulations for the cILC and GLS methods. However, in recent studies~\citep{Tristram:2020wbi,tristramImprovedLimitsTensortoscalar2022}, when constraining $r$, they calculate cross-correlations of two dataset maps to circumvent noise bias and potential instrumental errors in the angular power spectrum. The use of cross-spectra to nullify noise bias arises from concerns regarding biases: any systematic effect not characterized by noise simulations could substantially bias $r$, especially since the $BB$ spectrum is noise-dominated. Due to the half observation time, the uncertainty of cross-spectra between two data splits is approximately $\sqrt{2}$ times larger than that of auto-spectra, as indicated by the following Eqs.,
\begin{eqnarray} 
\sigma(\hat{{D}}_\ell^{BB,d_1\times d_2})&=&\sqrt{\frac{1}{(2\ell+1)\Delta\ell f_{\rm sky}}} \left[({D}_\ell^{BB,\rm CMB}+{N}_\ell^{BB,d_1}/B_\ell^2)({D}_\ell^{BB,\rm CMB}+{N}_\ell^{BB,d_2}/B_\ell^2)+({D}_\ell^{BB,\rm CMB})^2\right]^{1/2}\,,
\\
\sigma(\hat{{D}}_\ell^{BB,d})&=&\sqrt{\frac{2}{(2\ell+1)\Delta\ell f_{\rm sky}}}({D}_\ell^{BB,\rm CMB}+{N}_\ell^{BB,d}/B_\ell^2)\,,
\end{eqnarray}
where $\hat{D}_\ell^{BB,d_1\times d_2}$ is the unbiased estimate of the cross-spectrum of data splits 1 and 2, and $\hat{D}_\ell^{BB,d}$ is the estimated auto-spectrum of the full mission data combining two data splits. For two half-mission maps that equally split the horns in the focal plane (thus having uncorrelated systematic errors), we assume their noise power spectra $N_\ell^{BB,d_1}$ and $N_\ell^{BB,d_2}$ are the same, as large as twice the full-mission noise power spectrum $N_\ell^{BB,d}$ so that $\sigma(\hat{D}_\ell^{BB,d_1\times d_2}) \approx \sqrt{2}\sigma(\hat{D}_\ell^{BB,d})$ given the theoretical $BB$ spectrum $D_\ell^{BB,\rm CMB}$ much smaller than the noise spectrum. Despite the expansion of uncertainties, the robustness of the estimated $BB$ power spectrum against biases might be the primary concern to obtain the unbiased estimate of $r$. Therefore, we compute the cross-spectra of two half-mission maps cleaned by cILC and GLS, and fit $r$ from the cross-power spectra.

\begin{figure}[htpb]
    \centering
    \includegraphics[width=0.55\linewidth]{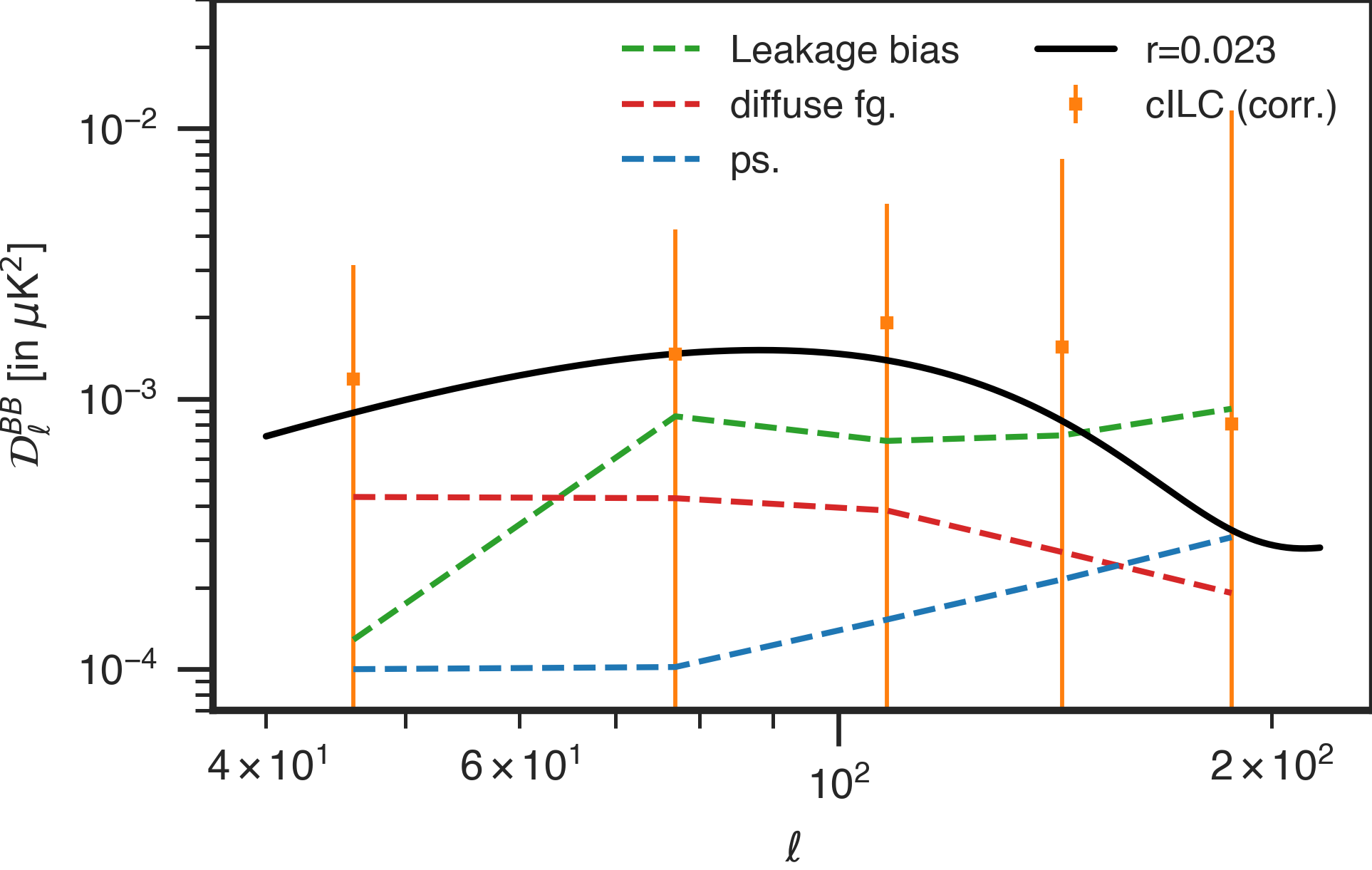}
    \includegraphics[width=0.35\linewidth]{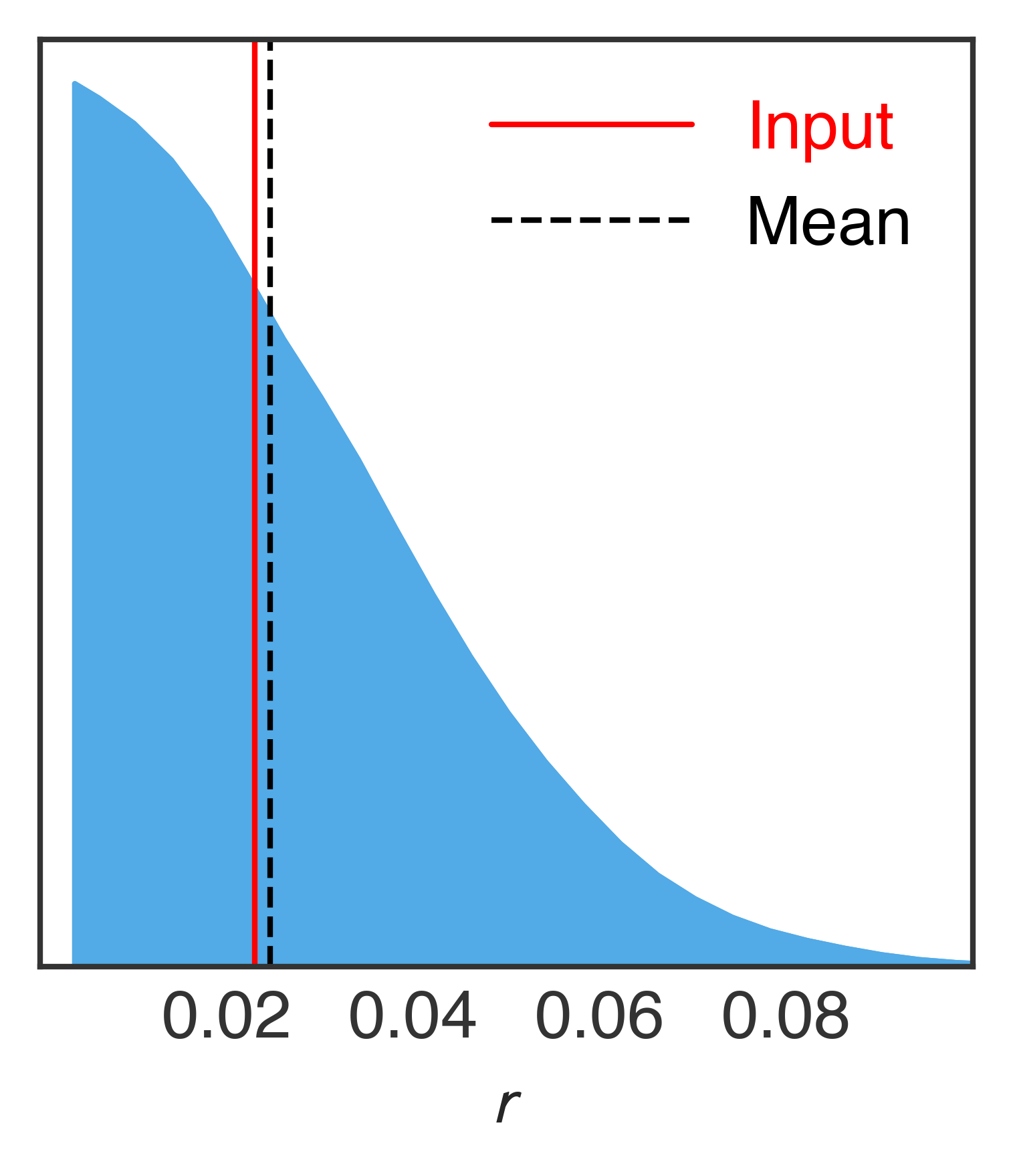}
    \caption{Cross power spectra (left) and posterior distribution (right) of $r$, estimated from the cross power spectrum computed by the cILC pipeline. Left: The orange bars depict the mean and error bars of the cross BB power spectrum between two half-mission maps, corrected for leakage bias. Dashed curves represent the leakage bias (green), foreground residual (red), and point source residual (blue). Right: The solid red line represents the input $r$-value of 0.023 for DC1 data, while the dashed black line illustrates the mean value ($r=0.025$) of the posterior distribution (blue shaded).}
    \label{fig:cross-cilc}
\end{figure}

\begin{figure}[htpb]
    \centering
    \includegraphics[width=0.55\linewidth]{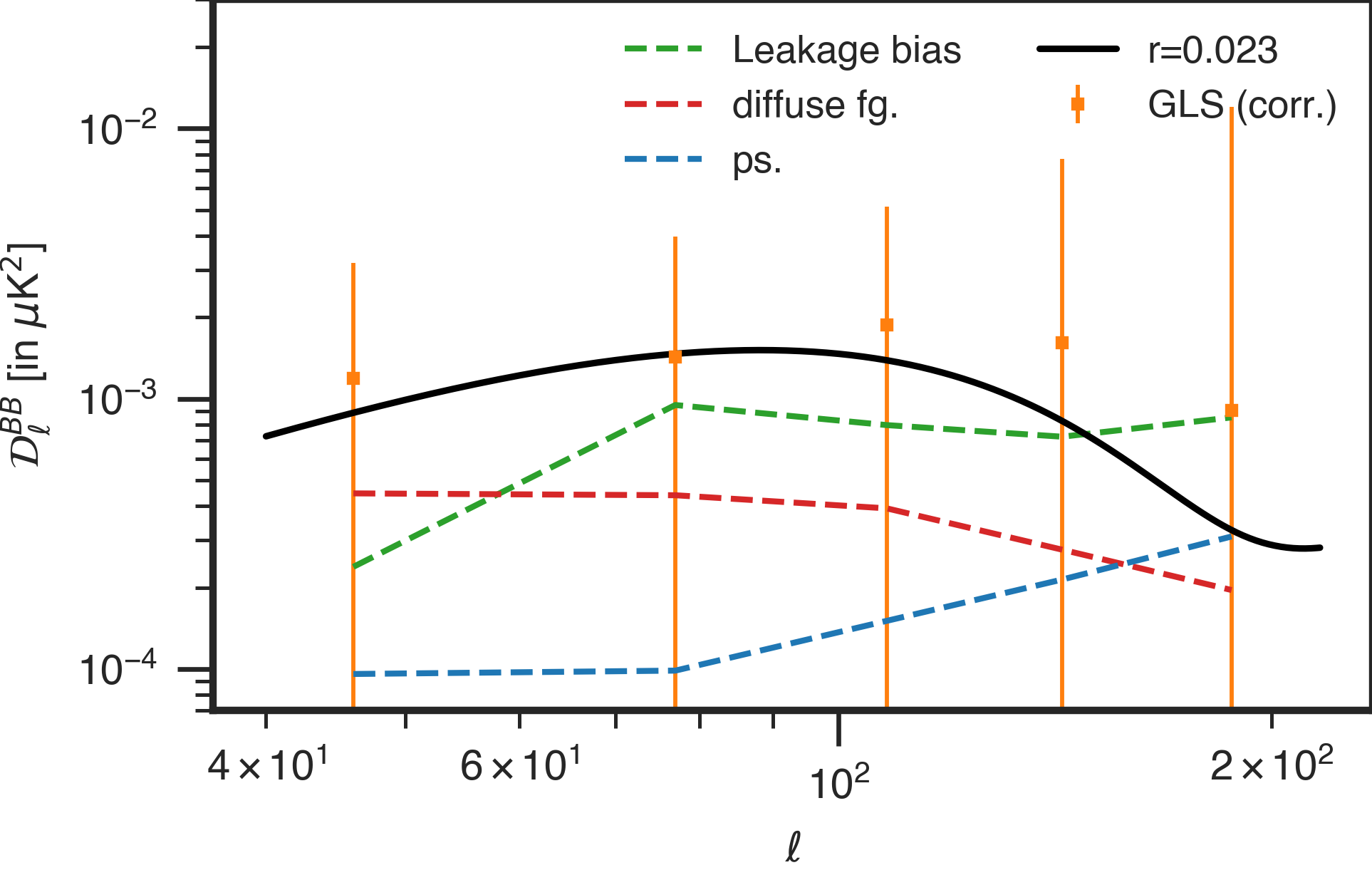}
    \includegraphics[width=0.35\linewidth]{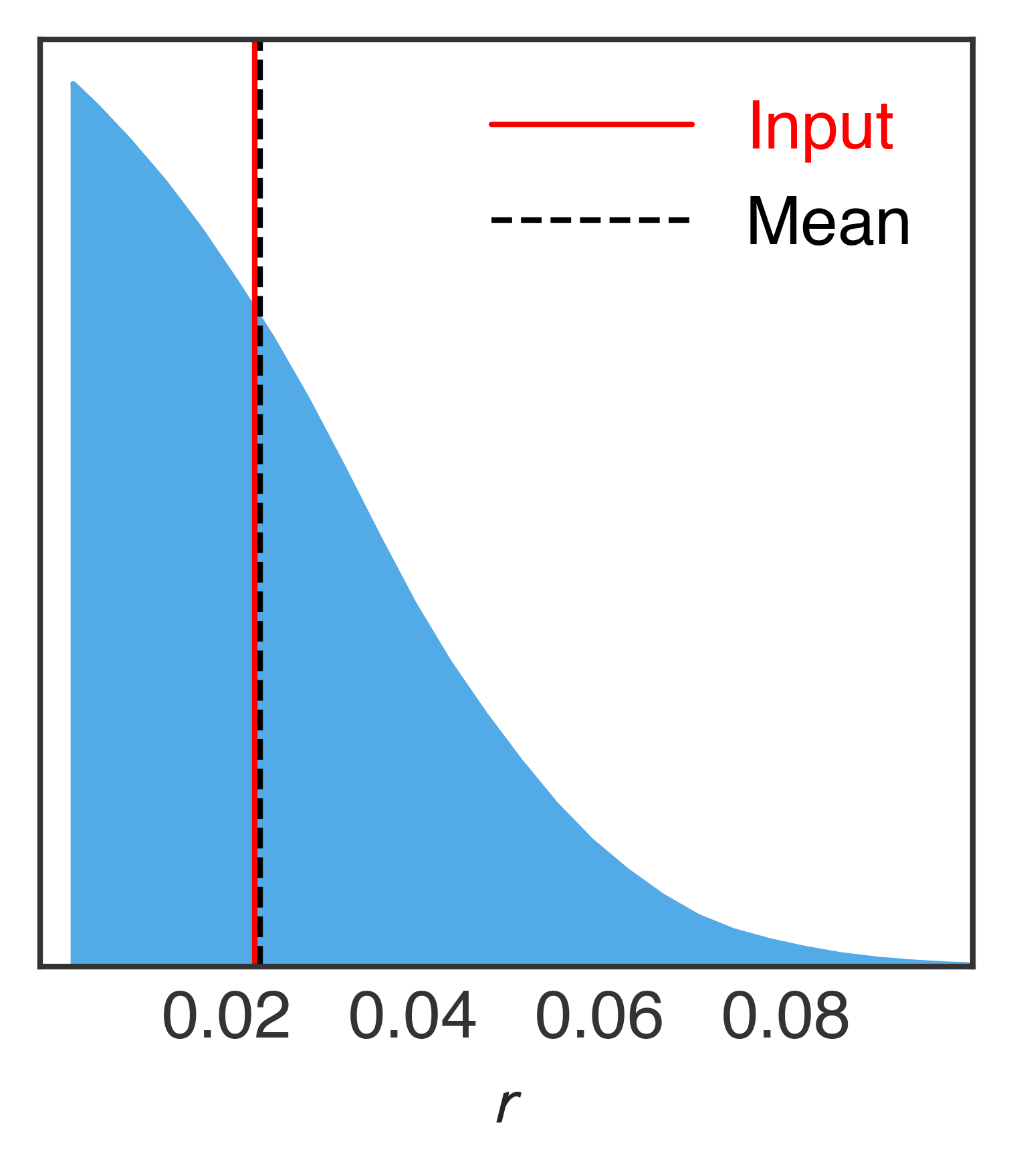}
    \caption{Same as Fig.~\ref{fig:cross-cilc}, but with the GLS pipeline on the data splits. The posterior $r$ is $r=0.023 \pm 0.018$.}
    \label{fig:cross-gls}
\end{figure}

\begin{figure}[htpb]
    \centering
    \includegraphics[width=0.55\linewidth]{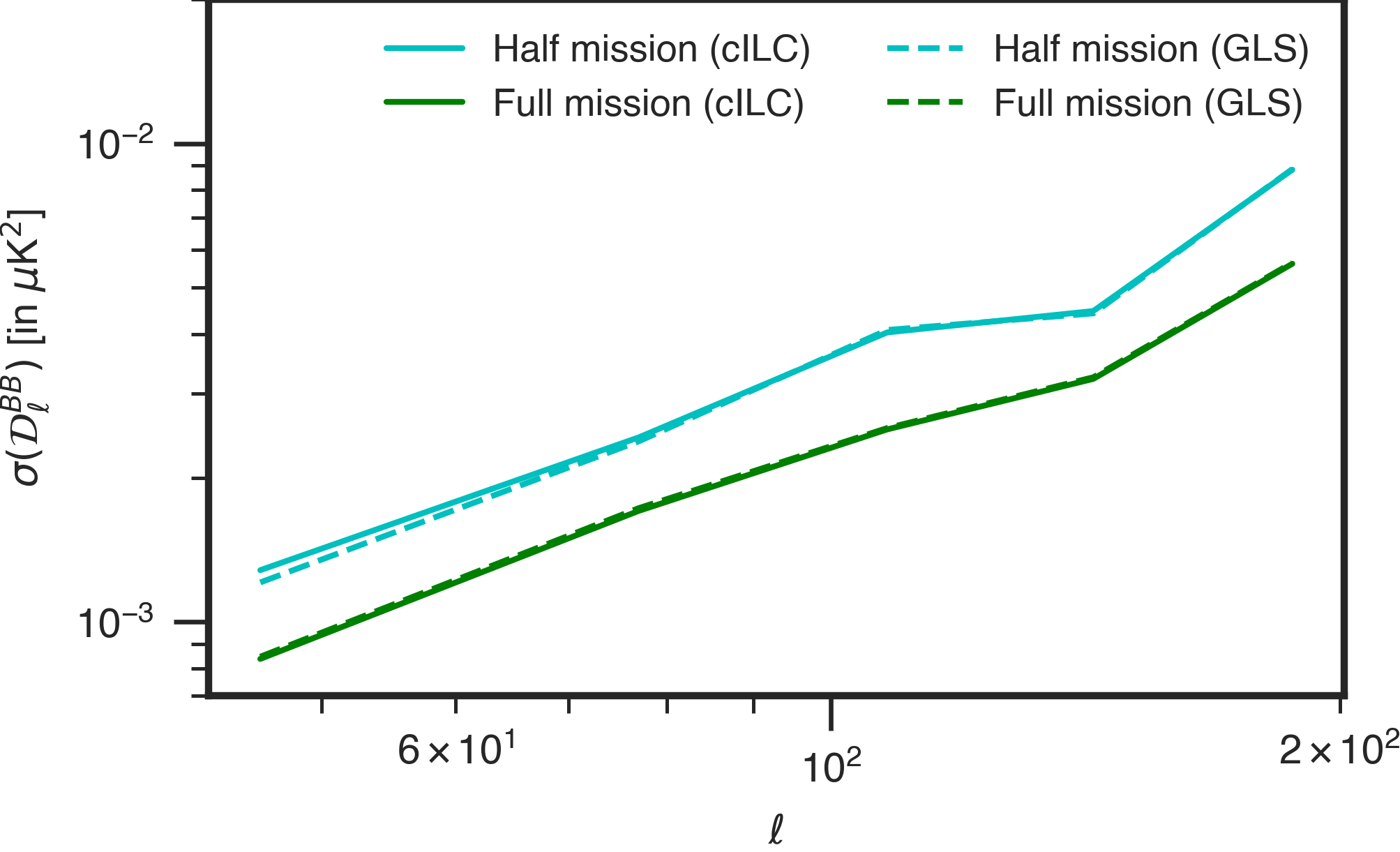}
    \caption{Uncertainties of the power spectrum for half mission and full mission cases using the cILC (solid) and GLS (dashed) pipelines. The auto-spectrum uncertainty for the full mission case is indicated by green curves, while the cross-spectrum uncertainty for the half mission case is represented by cyan curves. Their ratio agrees with the expected value of $\sqrt{2}$.}
    \label{fig:cross-eDl}
\end{figure}

The half-mission maps are created by adding the same CMB and foreground realization to two independent noise simulations, with noise levels $\sqrt{2}$ times larger than that of the full mission. The foreground cleaning pipeline (cILC or GLS) remains unchanged across the data splits, where we apply the same cILC/GLS weights obtained from the full-mission maps to clean the foregrounds more thoroughly. We execute the cILC and GLS pipeline on 25 pairs of half-mission maps, and the mean cross-power spectra are depicted in the left plot of Fig.~\ref{fig:cross-cilc} (cILC) and Fig.~\ref{fig:cross-gls} (GLS). Following a likelihood analysis on the cross-power spectra, we derive estimated $r$ values: $r=0.025\pm 0.018$ for cILC and $r=0.023\pm 0.018$ for GLS, with their posterior distribution displayed in the right plot. In each case, the mean $r$ values are in good agreement with the truth, indicating unbiased estimation for these pipelines. Note that the constraint on $r$ for GLS does not exhibit higher uncertainty compared to the auto-spectrum in the full mission case (see Tab.~\ref{tab:fg_res}). This occurs because the uncertainty of $r$ tends to decrease as the mean value of $r$ approaches zero under a positive $r$ prior. We compare the uncertainties of the estimated auto- and cross-power spectra for both the full mission and the half mission, respectively, as shown in Fig.~\ref{fig:cross-eDl}. The ratio between the uncertainties of the two cases, $\sigma(\hat{{D}}\ell^{BB,d_1\times d_2})/\sigma(\hat{{D}}\ell^{BB,d})$ averaged over $\ell\in[40,200]$, is 1.50 (cILC) or 1.47 (GLS), nearly consistent with $\sqrt{2}$.

\section*{Acknowledgments}

This work is supported by the National Key R\&D Program of China Grant (No. 2020YFC2201600, 2021YFC2203102), National Science Foundation of China (11903030, 12150610459, 11653003, 12273035), Guangdong Basic and Applied Basic Research Foundation (2024A1515012309).   We also wish to acknowledge the Beijing Super Cloud Center (BSCC) and Beijing Beilong Super Cloud Computing Co., Ltd (http://www.blsc.cn/) for providing HPC resources that have significantly contributed to the research results presented in this paper. Some of the results in this paper have been derived using the \texttt{HEALPix} package~\citep{Healpix2005} and the \texttt{emcee} python package~\citep{2013PASP..125..306F}.  

%

\bibliographystyle{aasjournal}
\bibliography{references}

\end{document}